\documentclass[aps,prd,superscriptaddress,showpacs,preprintnumbers,a4paper,
floatfix]{revtex4}
\usepackage{graphicx,color}
\usepackage{color}
\usepackage{amssymb}
\usepackage{amsmath}

\usepackage[active]{srcltx} 

\newcommand{\vect}[1]{\ensuremath{{\bm{#1}}}}
\newcommand{\bm}[1]{\mbox{\boldmath $#1$}}
\newcommand{\be}{\begin{equation}}
\newcommand{\ee}{\end{equation}}
\newcommand{\bea}{\begin{eqnarray}}
\newcommand{\eea}{\end{eqnarray}}

\newcommand{\bfk}{\mbox{\boldmath $k$}}

\newcommand{\bfp}{\mbox{\boldmath $p$}}

\newcommand{\bfP}{\mbox{\boldmath $P$}}

\newcommand{\qup}{q^\uparrow}

\def\lsim{\mathrel{\rlap{\lower4pt\hbox{\hskip1pt$\sim$}}\raise1pt\hbox{$<$}}}
\def\gsim{\mathrel{\rlap{\lower4pt\hbox{\hskip1pt$\sim$}}\raise1pt\hbox{$>$}}}
\def\nostrocostruttino#1\over#2{\mathrel{\mathop{\kern 0pt \rlap
{\hbox{$#1$}}} \hbox{\kern-.135em $#2$}}}

\def\kt{k_\perp}
\def\bkt{\bfk_\perp}

\def\pp{p_\perp}

\def\xb{x_{_{\!B}}}

\def\avk{\langle k_\perp ^2\rangle}
\def\avp{\langle p_\perp ^2\rangle}
\def\avPT{\langle P_T^2\rangle}

\def\C{_{_C}}
\def\G{_{_G}}
\def\BM{_{_{B\!M}}}

\newcommand{\xbj}{x_B}
\def\kt{k_\perp}
\def\ktm2{k^2_{\perp{\rm max}}}
\def\bkt{\bfk_\perp}

\newcommand{\ptsmG}{\langle P_T^2\rangle_G}
\def\xb{x_{_{\!B}}}

\begin{document}

\title{Partonic Transverse Motion in Unpolarized \\Semi-Inclusive Deep Inelastic Scattering Processes}
%

\author{M.~Boglione}
\affiliation{Dipartimento di Fisica Teorica, Universit\`a di Torino,
              Via P.~Giuria 1, I-10125 Torino, Italy\\
              INFN, Sezione di Torino, Via P.~Giuria 1, I-10125 Torino, Italy}
\author{S.~Melis}
\affiliation{European Centre for Theoretical Studies in Nuclear Physics and
Related Areas (ECT*) \\
              Villa Tambosi, Strada delle Tabarelle 286, I-38123 Villazzano,
Trento, Italy}
\author{A.~Prokudin}
\affiliation{Jefferson Laboratory, 12000 Jefferson Avenue, Newport News, VA
23606}
%
%
\begin{abstract}
We analyse the role of partonic transverse motion
in unpolarized Semi-Inclusive Deep Inelastic Scattering (SIDIS) processes. 
Imposing appropriate kinematical conditions,
we find some constraints which fix an upper limit to the range of allowed $\kt$ values.
We show that, applying these additional requirements on the partonic kinematics,
we obtain different results with respect to the usual phenomenological
approach based on the Gaussian smearing with analytical integration over an
unlimited range of $\kt$ values. These variations are particularly interesting for some observables,
like the $\langle \cos \phi_h \rangle$ azimuthal modulation of the unpolarized SIDIS cross section or the average transverse momentum of the final, detected hadron. 
\end{abstract}

\pacs{13.88.+e, 13.60.-r, 13.85.Ni}

\maketitle

\section*{INTRODUCTION}

Inclusive and Semi-Inclusive Deep Inelastic Scattering (DIS and SIDIS) processes
are important tools to
understand the structure of nucleons and nuclei. Spin asymmetries in polarized
SIDIS are directly
related to Transverse Momentum Dependent (TMD) parton distribution and
fragmentation functions,
and are the subject of intense theoretical and experimental studies. The usual,
collinear parton
distribution functions depend on the fraction $\xb$ of hadron momentum carried
by the scattering parton and 
on the virtuality of the probe, $Q^2$. TMDs additionally depend on the intrinsic
transverse momentum of
the parton, $\bfk _\perp$, opening invaluable opportunities to unravel the 
three-dimensional partonic  
picture of the nucleon in momentum space.  

At leading twist, the spin structure of a spin-1/2 hadron can be described by eight 
TMDs \cite{Mulders:1995dh,Bacchetta:2006tn,Anselmino:2011ch}. TMDs 
represents  particular physical aspects of spin-orbit correlations at the
partonic level. 
The dependence of the SIDIS cross section on the azimuthal angle, $\phi _h$, of
the
electro-produced hadron with respect to the lepton scattering plane and on the
nucleon polarization azimuthal angle, $\phi _S$,  
allows a term by term separation of the different azimuthal contributions to the
measured unpolarized and polarized cross sections and spin asymmetries.

The unpolarized SIDIS cross-section can be used not only to study the
unpolarized TMD distribution function 
$f_{q/p}(x,\kt)$ and the unpolarized TMD fragmentation function
$D_{h/q}(z,\pp)$, that encode the intrinsic dynamics 
of unpolarized partons, but also the Boer-Mulders distribution and the Collins
fragmentation functions, which carry information about the dynamics of
transversely polarized partons inside hadrons and give rise, for instance, to a
$\cos 2 \phi_h$ modulation of the unpolarized cross section. 
The existence of partonic intrinsic transverse momenta is also unequivocally
signaled, in the unpolarized SIDIS cross section, by a $\cos \phi_h$ modulation,
which is a subleading twist effect, suppressed by one power of $Q$. This
contribution to the unpolarized cross section consists of a purely kinematical
term, the Cahn effect~\cite{Cahn:1978se,Cahn:1989yf}, proportional to the
convolution of unpolarized distribution and fragmentation functions, and a term
proportional to the convolution of the Boer-Mulders and the Collins functions,
together with other twist-3 contributions, as pointed out in
Ref.~\cite{Bacchetta:2006tn}.

Polarized SIDIS experiments, in addition, allow us to explore the other
TMDs which describe the dynamics of polarized and unpolarized partons inside
polarized nucleons, such as the helicity, transversity and Sivers functions.
These will not be considered in this paper.

TMD factorization theorems~\cite{Ji:2004wu} are proven for SIDIS at momentum
scales $Q \gg P_{T}\sim \Lambda_{QCD}$; here large $Q$ values are needed to
allow a perturbative treatment of the underlying partonic subprocess, while the
small scale $P_{T}$ ensures that the observables are sensitive to the intrinsic
parton motion.
Moreover, the smallness of intrinsic quark momenta is also explicitly required,
as this approach is based on a series expansion in terms of the ratio $\kt/Q$
and higher orders are usually neglected.
However, the present knowledge of TMDs and azimuthal asymmetries is mainly based
on the available experimental data from the HERMES and COMPASS Collaborations,  
that operate at relatively low $Q^2$ values: typically, the cut  $Q^2 > 1$
GeV$^2$ is assumed in order to ensure SIDIS kinematics. 

Very often, in phenomenological analysis, the transverse momentum distribution
of the TMDs is assumed to be a Gaussian.
Although very simple, this approximation leads to a successful description of
many sets of data (see for instance
Refs.~\cite{Anselmino:2005nn,Anselmino:2005ea,Anselmino:2005sh,Anselmino:2006rv,
Anselmino:2007td,Anselmino:2008sga,Anselmino:2007fs,
Anselmino:2009st,Schweitzer:2010tt,Avakian:2010ae,Boglione:2011zw}).
However, as the quality and the amount of data has significantly improved
lately, 
a more detailed picture of the partonic description is now becoming more and
more necessary.
In particular, the latest COMPASS analysis~\cite{Rajotte:2010ir} of the SIDIS
unpolarized cross section suggests that the average transverse momentum of the
detected hadrons could sensitively depend on the parton momentum fractions $x_B$
and $z_h$ 
and even on flavor, although similar conclusions cannot be presently inferred
from HERMES~\cite{Airapetian:2009jy,Giordano:2008zzc} and JLAB
data~\cite{:2008rv}, as we will discuss in more details in Section
\ref{res-unp}.
In Ref.~\cite{Anselmino:2005sh}, a simple Gaussian model was used to
successfully describe the Cahn $\cos\phi_h$ azimuthal modulation measured by the
EMC Collaboration~\cite{Ashman:1991cj} and allowed the extraction of the
unpolarized TMD's Gaussian widths. Later, it has been realized that the
predictions of Ref.~\cite{Anselmino:2005sh} largely overestimated the more
recent data from HERMES~\cite{Giordano:2010zz} and COMPASS~\cite{Sbrizzai:2009fc} 
which have meanwhile become available. 
On the contrary, a large Cahn effect now seems to be required in order to
describe the $\cos 2\phi_h$ asymmetries~\cite{Barone:2009hw}, at least at the
COMPASS kinematics. 
Furthermore, 
a global phenomenological analysis of data led by Schweitzer, Teckentrup and
Metz~\cite{Schweitzer:2010tt} suggests 
that the average transverse momenta $\langle k_\perp ^2\rangle$, 
used as a free parameter in the Gaussian, depends on $\sqrt{s}$, the center of
mass energy of the target and the incoming lepton.

This complex scenario thus requires an accurate critical analysis of the work
done so far, to try and shed some light on these apparent controversies.
The detailed study of the COMPASS Collaboration on the $P_T$ distribution of
their SIDIS measurements~\cite{Rajotte:2010ir}, 
pointing to a considerable deviation from the expected behavior, prompted us to a reanalysis
of the approximations used in our $\kt$ integration. 
In this paper,  we examine the kinematical regions where order
$\mathcal{O}(k^2_{\perp}/Q^2)$ and higher can be safely neglected
and we realize that not all the kinematical domains
of the existing and planned experiments satisfy the basic criteria
$k_{\perp}\simeq P_T\simeq\Lambda_{QCD}\ll Q$.
Inspired by the parton model, and still adopting a Gaussian model for our TMDs, 
we bound the integration range of transverse momenta $k_{\perp}$ and 
we observe, in some kinematical regions, remarkable deviations
from the predictions obtained from the common TMD approach, based on the
Gaussian parametrization integrated over the full $\kt$ range, $[0,\infty]$.
We show that some kinematical ranges, typically
low $\xb$ or equivalently low $Q^2$ regions, are not safely controlled
by the present phenomenological model, while 
bounds on $k_{\perp}$ can prevent uncontrolled large $k_{\perp}/Q$
contributions.
This leads, for instance, to a better description of some observables like the
$\langle\cos\phi_h\rangle$ asymmetry and introduces
some interesting effects in the $\langle P_T^2\rangle$ behaviors.
However, simple parton model considerations are not sufficient
to fully describe the present, in some sense puzzling, data: soft gluon
emissions, higher twist contributions and QCD evolution of the TMDs can play a
role.

Higher order contributions to cross sections and asymmetries are difficult to
estimate; here, we will consider only those coming from purely kinematical
corrections to twist-2 TMDs and show that they are not negligible in most of the
present experimental setups. This makes the extraction of twist-3 TMDs from
existing data troublesome. 

Future experimental data from an Electron Ion
Collider~\cite{Deshpande:2005wd,Horn:2009cu}, where the $Q^2$ range would be
easily adjustable in order to estimate higher order contributions, will
definitely help to disentangle them from leading twist contributions.

\section{SIDIS Kinematics \label{sect-kin}}

\begin{figure}[t]
\includegraphics[width=0.5\textwidth]{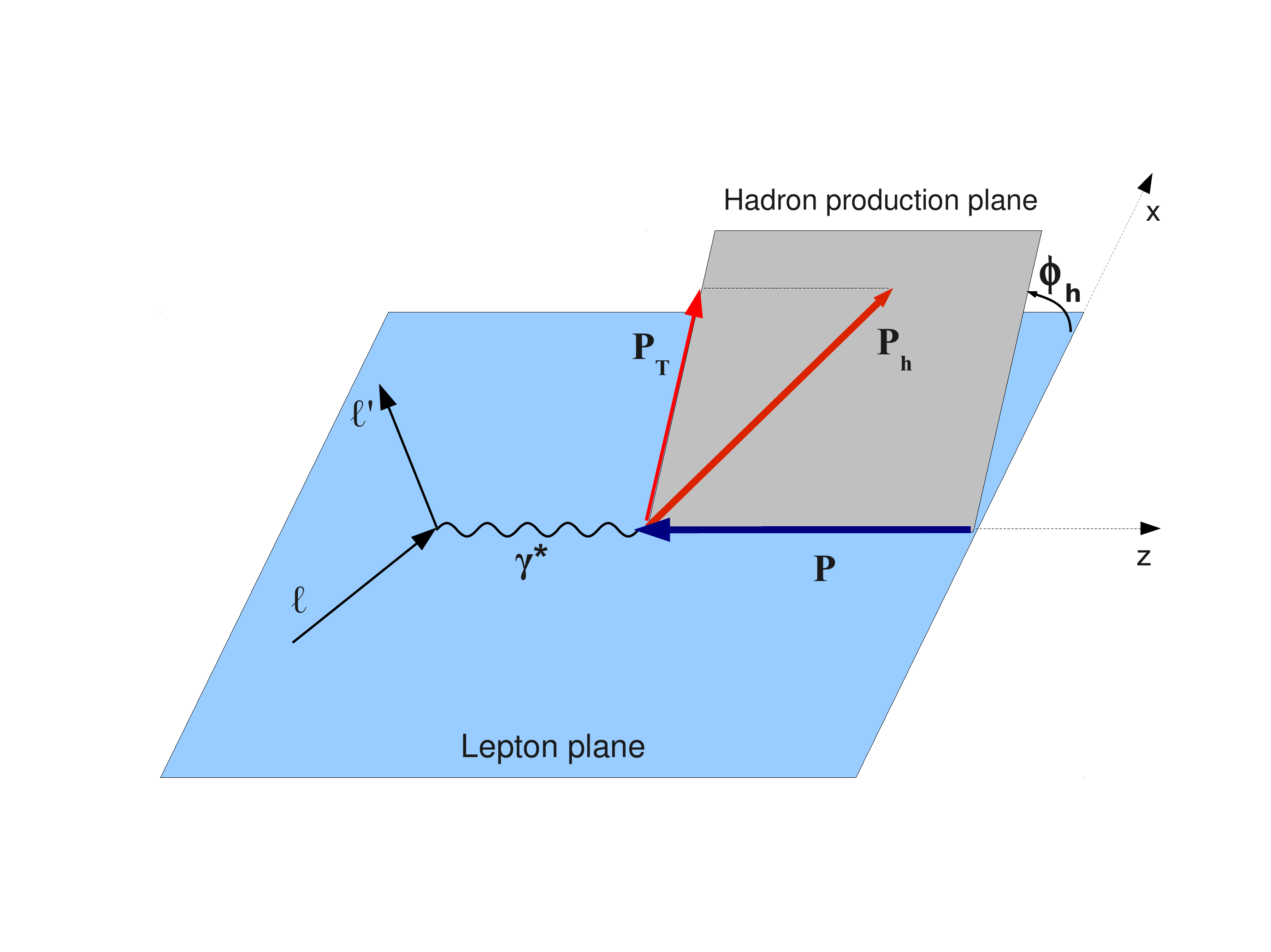}
\caption{\label{sidisfig}
Kinematical configuration and conventions for SIDIS processes in the $\gamma^*
-p$ c.m. frame. 
The initial and final lepton momenta define the $\hat{x}$-$\hat{z}$ plane, while
the detected hadron momentum and the $\hat{z}$ axis determine the hadron
production plane, at an angle $\phi_h$.}
\end{figure}

Let us give a brief review of $\ell p\to \ell^{\prime} h + X$ 
Semi-Inclusive Deep Inelastic Scattering kinematics.
We study the SIDIS process in the $\gamma^* - p$ c.m. frame,
where $\gamma^*$ denotes the virtual photon.
Following the Trento conventions~\cite{Bacchetta:2004jz},
we take the virtual photon momentum $q$ along the $+\hat{z}$ direction
and the proton momentum $P$ in the opposite direction, as shown in
Fig.~\ref{sidisfig}.
The leptonic momenta define a plane that coincides with our
$\hat x$-$\hat z$ plane. The detected hadron has momentum $P_h$, 
its transverse component is denoted by $\bfP_T$ and $\phi_h$ is its azimuthal
angle:
together with $\hat z$ they identify the hadron production plane,
Fig.~\ref{sidisfig}.
%
%
%
%
\begin{figure}[b]
\includegraphics[width=0.5\textwidth]{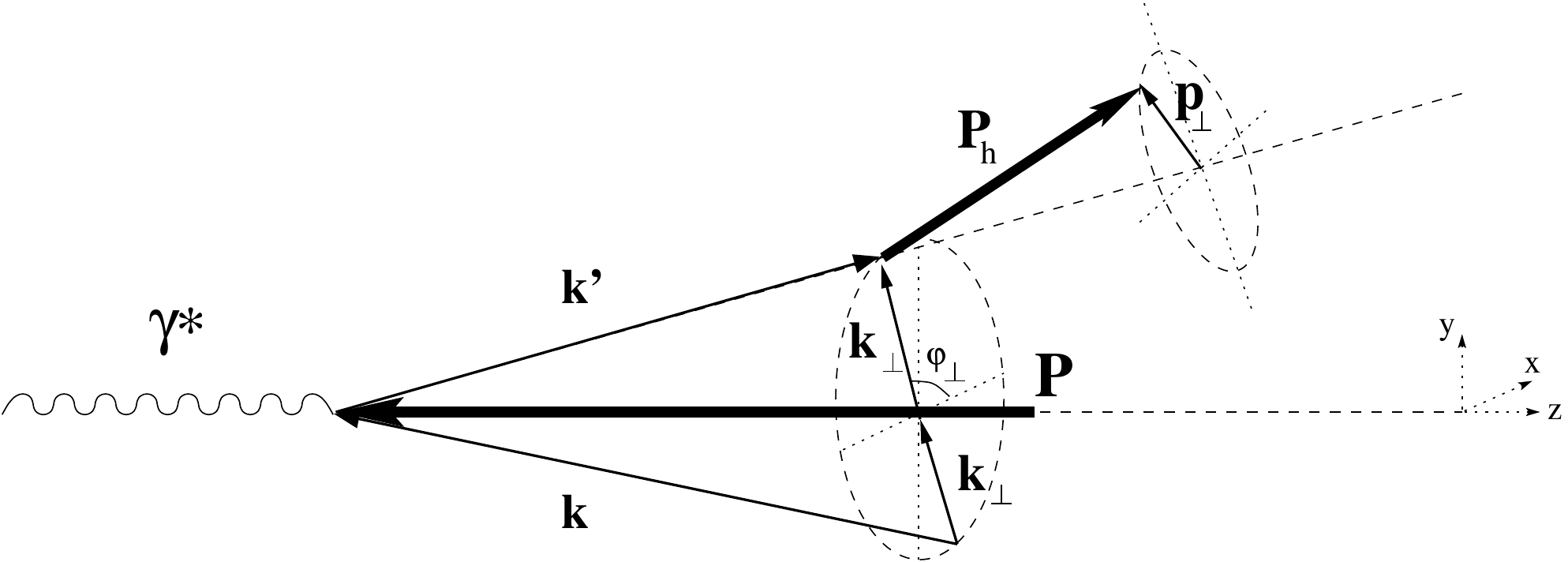}
\caption{\label{kinfig}Kinematical configuration and conventions for the SIDIS
partonic sub-process: $\bfk$ and $\bfk^\prime$ are the initial and final quark
momenta, and $\bfk_{\perp}$ is their transverse component. $P_h$ is the final,
detected hadron momentum, with a $\bfp _\perp$ component, transverse with
respect to the fragmenting quark direction $\bfk^\prime$.
}
\end{figure}
%
We adopt the usual SIDIS variables neglecting the lepton,
the proton and the final hadron masses:
\begin{eqnarray}
s = (P + \ell)^2 \quad\quad  Q^2 = -q^2 \quad\quad
(P+q)^2 = W^2 \simeq \frac{1-\xbj}{\xbj}\,Q^2
\nonumber \\
\xbj = \frac {Q^2}{2P \cdot q} \simeq \frac{Q^2}{W^2 + Q^2}
\quad\quad y = \frac{P \cdot q}{P \cdot \ell} \simeq \frac{Q^2}{\xbj s}
\quad\quad z_h = (P\cdot P_h)/(P\cdot q)
\> \,,\label{kin}
\end{eqnarray}
where $\ell$ is the incoming lepton momentum.
The proton and the virtual photon momenta can be written
in the $\gamma^* -p$ c.m. frame, as functions of the invariants $W$ and $Q$ in
this way:
\begin{equation}
q = \frac{1}{2}\left ( W - \frac{Q^2}{W},0,0, W + \frac{Q^2}{W}\right )\>
\qquad 
P = P_0(1,0,0,-1) \> \qquad 
P_0 = \frac{1}{2}\left ( W + \frac{Q^2}{W} \right ) \> \cdot
\end{equation}

In the parton model the virtual photon scatters off a on-shell quark.
The quark momentum $k$ can be written, in the $\gamma^*-p$ c.m. as:
\begin{eqnarray}
k&=&
\left(x P_0+\frac{k_{\perp}^2}{4xP_0},\bfk_{\perp},-x
P_0+\frac{k_{\perp}^2}{4xP_0}
\right) \\
\end{eqnarray}
where $x=k^-/P^-$ 
is the quark light-cone momentum fraction (see Appendix \ref{Sudakov} for more
details)
and $\bfk_{\perp}$ is the quark intrinsic transverse momentum, see
Fig.~\ref{kinfig}.
The final emitted quark has  momentum $k^{\prime}=q+k$. Its on-shell condition
\begin{equation}
k'^2 = 2 q \cdot k - Q^2 = 0
\end{equation}
implies~\cite{Anselmino:2005nn}
\begin{equation}
x = \frac{1}{2} \, \xbj\left ( 1+ \sqrt{1+\frac{4\kt^2}{Q^2}}\right ) \cdot
\label{x}
\end{equation}
As indicated in Fig.~\ref{kinfig}, $\bfp _\perp=\bfP _h -(\bfP _h \cdot
\hat{\bfk}^\prime)\,\hat{\bfk}^\prime$ is the transverse momentum of the hadron
$h$ {\it with respect
to the direction} $\bfk ^\prime$  of the fragmenting quark, while $z =
P_h^+/k^{\prime+}$ is the
light-cone fraction of the quark momentum carried by the resulting hadron. The
fragmenting variables $z$ and $\bfp _\perp$ can be expressed in terms of the
usual observed hadronic variables $\bfP _T$ and $z_h$, as shown in Eqs.~(26) and
(28) of Ref.~\cite{Anselmino:2005nn}.

All the kinematical relations given above are exact expressions at all orders in
a $(\kt / Q)$ expansion.
Neglecting terms of order ${\cal O}(\kt^2/Q^2)$, they considerably simplify and
we find:
\begin{eqnarray}
  x\simeq x_B\,, \qquad z\simeq z_h\,\qquad \bfp_{\perp}\simeq \bfP_T-z_h
\bfk_{\perp}\,.
\end{eqnarray}

Transverse Momentum Dependent distributions depend on the kinematical variables
$x$ and $k_\perp$ defined above. Let's consider, for instance, the unpolarized
distribution function $f_{q/p}(x,k_\perp)$, which gives the number density of
unpolarized
quarks inside an unpolarized proton; this function is usually normalized  in
such a way that 
\be
\int d^2 k_\perp \, f_{q/p}(x,k_\perp) = f_{q/p}(x) \, ,
\label{normalization}
\ee
where $f_{q/p}(x)$ is the usual, collinear parton distribution function at some
given scale $Q^2$. The same logic holds for
the unpolarized TMD fragmentation function, $D_{h/q}(z,\pp)$.

Very often, in phenomenological analysis, a Gaussian dependence of the TMDs is
assumed, adopting the following  parametrizations:
\be
f_{q/p}(x,k_\perp) = f_{q/p}(x) \, \frac{1}{\pi \langle\kt^2\rangle} \,
e^{-{\kt^2}/{\langle\kt^2\rangle}}
\label{partond}
\ee
and
\be
D_{h/q}(z,p _\perp) = D_{h/q}(z) \, \frac{1}{\pi \langle p_\perp^2\rangle}
\, e^{-p_\perp^2/\langle p_\perp^2\rangle}\,,
\label{partonf}
\ee
where $f_{q/p}(x)$ and $D_{h/q}(z)$ can be taken
from the available fits of world data,
while $\langle\kt^2\rangle$ and $\langle p_\perp^2\rangle$ are free parameters
to be extracted from dedicated fits.

Note that the functions defined in Eqs.~\eqref{partond} and \eqref{partonf} obey
the normalization condition of Eq.~\eqref{normalization}, so that
\be
 \int_{0}^{2\pi}\! \!d\varphi \int_{0}^{\infty} \! d k_{\perp}\,k_{\perp} \,
f_{q/p}(x,k_\perp) = f_{q/p}(x)\,.
\ee 
The Gaussian parametrization in principle allows any value of $k_{\perp}$ from
zero to infinity.
However, the integrand is weighted by the Gaussian $k_{\perp}$ distribution, 
so that $\kt$ values larger than the Gaussian width, $\langle \kt ^2 \rangle$,
are strongly suppressed. 
Typical values of $\langle k_{\perp}^2\rangle$ are of a few hundreds MeV. Thus
if $Q^2$ is large
with respect to $\langle k_{\perp}^2\rangle$, the Gaussian represents an
effective model
that prevents large $k_{\perp}/Q$ contributions to the cross section.
However in many low energy SIDIS experiments, like
HERMES and COMPASS, $\langle Q^2\rangle\simeq2$ $\textrm{GeV}^2$ and the
experimental cut $Q^2>1$ $\textrm{GeV}^2$ is used: thus at low $x$ we have 
$Q^2 \simeq 1$ $\textrm{GeV}^2$.
Therefore, in these particular cases, the Gaussian smearing is not
sufficient to cut away large  $k_{\perp}/Q$ contributions.
Notice that these considerations are still valid for any phenomenological
parametrization that
does not impose any cut to the range of allowed $k_{\perp}$ values.
Hence the necessity to explore whether it is possible to find a physical
picture that allows us to put some further constraints on the partonic intrinsic motion.

\section{Physical partonic cuts \label{kcut}}

In order to find some constraints on the partonic intrinsic motion, we adopt the
simple picture of nucleons provided by the parton model.
Although ``non-physical'' in many aspects, the parton model can be seen as 
good approximation and 
a toy--model to understand some physical QCD features.
In particular, it gives  kinematical limits on the transverse momentum size,
which can be obtained by requiring the 
energy of the parton to be less than the energy of the parent hadron and by
preventing the parton to move backward with respect to the parent hadron direction ($k_z < 0$).
The energy bound implies:
\begin{eqnarray}
x P_0+\frac{k_{\perp}^2}{4xP_0}\le P_0 &\Rightarrow &k_{\perp}^2\le
4x(1-x)P_0^2\nonumber\\
&\Rightarrow & k_{\perp}^2\le\frac{x(1-x)}{x_B(1-x_B)}Q^2\,\cdot \label{cut1}
\end{eqnarray}
Inserting Eq.~(\ref{x}) in Eq.~(\ref{cut1}) and solving one finds:
\begin{equation}
k_{\perp}^2\le(2-\xb)(1-\xb)Q^2 \,\, \,\, , \, \,\, { 0 < \xb  <  1}\,.
\label{cutenergy}
\end{equation}
Requiring the parton to move in the forward direction
with respect to the parent hadron gives:
\begin{eqnarray}
(\boldsymbol{P}\cdot \boldsymbol{k})>0 &\Rightarrow&  k_{\perp}^2\le 4 x^2
P_0^2\nonumber\\
&\Rightarrow&k_{\perp}^2\le\frac{x^2}{x_B(1-x_B)}Q^2\label{cut2}\,.
\end{eqnarray}
Using Eq.~(\ref{x}) and solving we find
\begin{equation}
\kt ^2\le \frac{\xb(1-\xb)}{(1-2\xb)^2}Q^2 \, \, , \, \, { \xb  <  0.5}\,.
\label{cutdirection}
\end{equation}
Notice that these are exact relations, which hold at all orders in $(\kt/Q)$.
These constraints are obtained assuming that partons are on-shell 
(the parton off-shellness should be  very small, at least in the region
where we will apply the parton model).

\begin{figure}[t]
\includegraphics[width=0.5\textwidth ]{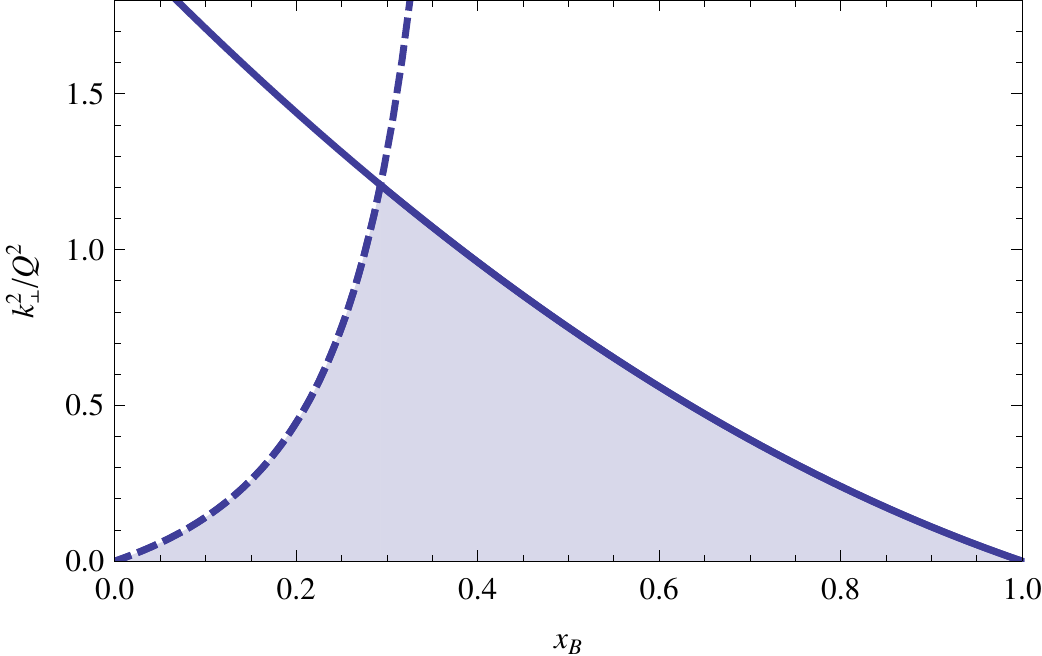}
\caption{
$\kt^2/Q^2$ phase space as determined by the bounds of Eqs.~\eqref{cutenergy}
and \eqref{cutdirection}. The allowed region, which fulfills  both bounds, is
represented by the shaded area below the solid line, corresponding to
Eq.~\eqref{cutenergy}
and the dashed line, corresponding to Eq.~\eqref{cutdirection}. Notice that
present data from HERMES and COMPASS experiments span the region
$x_B\lesssim0.3$, where only the momentum bound of Eq.~\eqref{cutdirection}
plays a role.
}
\label{cuts}
\end{figure}

The ratio $\kt^2/Q^2$, as constrained by Eqs.~(\ref{cutenergy})
and~(\ref{cutdirection}),
is shown in Fig.~\ref{cuts} as a functions of $\xb$:
from this plot it is immediately evident that although in principle
Eq.~\eqref{cutdirection} (represented by the dashed line)
gives a stringent limit on $\kt^2/Q^2$ in the region $\xb<0.5$,
it intercepts the bound of Eq.~(\ref{cutenergy}) (solid line) in 
$\xb\simeq0.3$,
where the latter becomes most relevant.
Notice also that present data from HERMES and COMPASS experiments span the
region $x_B\lesssim0.3$,  
where only the momentum bound of Eq.~\eqref{cutdirection} plays a role.

Once the maximum value of $k_\perp$ is bounded by the conditions of
Eqs.~\eqref{cutenergy} and \eqref{cutdirection}, to comply with the 
normalization condition of Eq.~\eqref{normalization}, we set the right
normalization coefficient 
\be
f_{q/p}(x,\kt) =  f_{q/p}(x) \, \frac{1}{1-e^{-(k_\perp^{\rm max})^2/\avk}}
 \, \frac{e^{-{\kt^2}/{\avk}}}{\pi \avk}\,,
\ee
where $(k_\perp^{\rm max})^2$ denotes the maximum value of $k^2_\perp$ for each
given values of $\xb$ and $Q^2$ as required by  
Eqs.~\eqref{cutenergy},\eqref{cutdirection}, 
so that 
\be
f_{q/p}(x) = \int_{0}^{2\pi} \!\! d \varphi  \int_{0}^{k_\perp^{\rm max}}
\!\!k_\perp \,d k_\perp \, f_{q/p}(x,\kt)\,.
\label{normalization1}
\ee

This normalization will allow us to reproduce correctly  all inclusive
cross-sections, such as the DIS cross-section, for which
collinear results are known. Note that if $Q\rightarrow \infty$
then $k_\perp^{max} \rightarrow \infty$ and one 
obtains the usual normalization of Eq.~\eqref{normalization}.

%

\section{Unpolarised SIDIS cross section \label{Xsect}}

According to Refs.~\cite{Mulders:1995dh,Anselmino:2011ch, Bacchetta:2006tn}
the unpolarized differential cross section for the SIDIS process,
$\ell \, + \, p  \to \ell^\prime \, h \, X$ can be written as
\be
\frac{d\sigma^{\ell + p \to \ell^\prime h X}}
{d\xb \, dy \, dz_h \, d^2 \bfP_T} =
\frac {4 \pi \, \alpha^2}{\xb s y^2} \bigg\{
\frac{1+(1-y)^2}{2} F_{UU} + (2-y)\sqrt{1-y} \, \cos\phi_h \,
F_{UU}^{\cos\phi_h} + (1-y) \, \cos2\phi_h \, F_{UU}^{\cos2\phi_h}\bigg\}\,, \;
\label{sidis-Xsec-final}
\ee
where the $F$
``structure functions",
which involve the relevant convolutions of distribution and fragmentation
functions
over the intrinsic transverse momenta, are defined within a TMD factorization
scheme, at order ${\cal O}(\kt/Q)$, as
\bea
F_{UU} &=& \sum_{q} e_q^2 \, \int d^2\bkt \, f_{q/p}(x,\kt) \, D_{h/q}(z,\pp)\,,
\label{FUU} \\
F_{UU}^{\cos\phi_h} &=&
2 \sum_{q} e_q^2 \, \int d^2\bkt \; \frac{\kt}{Q} 
\left[(\hat{\bfP}_T \cdot \hat{\bfk}_\perp )\,f_{q/p}(x,\kt)\,D_{h/q}(z,\pp)
\right.\nonumber \\ && \hspace{3.5cm}
+ \left.\frac{P_T-z_h \kt \,(\hat{\bfP}_T \cdot \hat{\bfk}_\perp )}{2\pp}\,
\Delta f_{\qup/p}(x,\kt)\,\Delta^N  D_{h/q^\uparrow}(z,\pp)\right]\,,
\label{FUUcosphi} \\
F_{UU}^{\cos2\phi_h} &=&
-\sum_{q} e_q^2 \, \int d^2\bkt \!
\left[\frac{ P_T\,(\hat{\bfP}_T \cdot \hat{\bfk}_\perp) - 2 z_h \kt
(\hat{\bfP}_T \cdot \hat{\bfk}_\perp)^2 +
z_h \kt }{2 \pp}\right]\!\Delta f_{\qup/p}(x, \kt)\,\Delta^N  D_{h/q^\uparrow}
(z,\pp) \,.
\nonumber \\
\label{FUUcos2phi}
\eea
Although three different azimuthal modulations are simultaneously at work in the
total unpolarized SIDIS cross section of  Eq.~\eqref{sidis-Xsec-final},
to extract single effects one can introduce the following azimuthal moments
\be
\langle \cos \phi_h \rangle=
\frac{
\int d\phi_h \,
d\sigma^{\ell p \to \ell^\prime h X} \,\cos \phi_h }
{
\int d\phi_h\,
d\sigma^{\ell p \to \ell^\prime h X} }\;, \hspace*{1.6cm}
\langle \cos  2\phi_h \rangle=
\frac{
\int d\phi_h \,
d\sigma^{\ell p \to \ell^\prime h X} \,\cos 2\phi_h }
{
\int d\phi_h\,
d\sigma^{\ell p \to \ell^\prime h X} } \; ,
\label{val-med}
\ee
which isolate the contributions of $F_{UU}^{\cos\phi_h}$ and
$F_{UU}^{\cos2\phi_h}$ from the traditional, collinear term $F_{UU}$.
As we showed in Ref.~\cite{Anselmino:2011ch}, the $F$
structure functions in Eqs.~\eqref{FUU}, \eqref{FUUcosphi} and
\eqref{FUUcos2phi} coincide with
those defined in Ref.~\cite{Bacchetta:2006tn}, 
considering only leading twist TMDs. In the phenomenological study we will
perform in what follows, 
we will not consider higher twist dynamical contributions
\cite{Bacchetta:2006tn} to the structure functions, therefore we shall keep in
mind that strong deviations of measured quantities
from the predictions obtained in this simple model will signal the presence of
higher twist contributions. 

In Eqs.~(\ref{FUU}) and (\ref{FUUcosphi}), $f_q(x, k_\perp)$ is the unpolarized
TMD distribution function and $D_q^h(z,p _\perp)$ is the unpolarised TMD
fragmentation function discussed in Section~\ref{sect-kin}, while $\Delta
f_{\qup/p}(x, \kt)$ is the Boer-Mulders distribution function, related to 
the number density of transversely polarized quarks inside an unpolarized
proton, and $\Delta^N D_{h/q^\uparrow} (z,\pp)$ is the Collins fragmentation
function which, in turn, is related to the number density of transversely polarized quarks
fragmenting into a spinless hadron.
Other common notations used for the Boer-Mulders and Collins functions  are:
\bea
&&\Delta f_{\qup/p}(x, \kt)  
= - \frac {\kt}{M} \, h_{1}^\perp (x, \kt) \label{b-m}\,,\\
&&\Delta^N  D_{h/q^\uparrow}(z,p_\perp)  = 
\frac{2p_{\perp}}{z M_h}H_{1}^{\perp}(z,p_{\perp})\,,
\label{D-prop}
\eea
where $M$ and $M_h$ are the masses of the initial proton and of the final
hadron, respectively.

Summarizing, Eq.~(\ref{FUU}) corresponds to the usual, collinear
$f_{q/p}\,\otimes  D_{h/q}$ 
contribution to the unpolarized SIDIS cross section;
in the $\cos\phi _h$ azimuthal modulation of Eq.~(\ref{FUUcosphi})
the first term is the Cahn effect~\cite{Cahn:1978se,Cahn:1989yf},
signaling the existence of transverse momenta in the partonic scattering
even when considering only unpolarized partons,
while the second term corresponds to
the Boer-Mulders $\otimes$ Collins contribution
of transversely polarized quarks to the unpolarized cross section;
finally, Eq.~(\ref{FUUcos2phi}) gives the $\cos 2\phi _h$
azimuthal modulation of the unpolarized SIDIS cross section.
An additional contribution to the $\cos\phi _h$ modulation can be generated by 
 ``dynamical'' higher twist contributions from twist-3 functions~\cite{Bacchetta:2006tn}.
Instead, the $\cos 2\phi _h$ azimuthal moment in Eq.~(\ref{FUUcos2phi}) is not suppressed by $1/Q$ 
and does not receive any  dynamical or kinematical twist-3 contribution~\cite{Bacchetta:2006tn}, 
but can be affected by twist-4 contributions, like the $\cos 2\phi_h$ Cahn effect discussed in
Section~\ref{rescos2}.

Therefore from this azimuthal moment one could extract the Boer-Mulders and
Collins TMD's, in all those kinematical ranges
in which higher orders in the $\kt/Q$ expansion can safely be neglected.
As a matter of facts, as we will see in Section~\ref{kcut},
higher orders of $\left(k_\perp/Q\right)^n$, $n\geq2$,
could  potentially be important when analysing HERMES,
COMPASS and JLab experimental data, and should not be neglected. 
Therefore twist-4 contributions could be important in these kinematics.
One of these contributions, the ``twist-4'' Cahn effect, Eq.\eqref{twist4},
has been analyzed phenomenologically in Ref~\cite{Barone:2009hw}, see
Section~\ref{rescos2}.

For the Boer-Mulders and Collins functions,
we assume parametrizations similar to those in Eqs.~\eqref{partond} and
\eqref{partonf}, with an extra multiplicative
factor $\kt$ or $\pp$, respectively, to give them the appropriate behavior
in the small $\kt$ or $\pp$ region~\cite{Anselmino:2008sga,Anselmino:2011ch}:
\bea
\Delta f_{\qup/p}(x,\kt) &=&   \Delta f_{\qup/p}(x)\;
\sqrt{2e}\,\frac{\kt}{M\BM} \; \frac{e^{-\kt^2/\avk \BM}}{\pi\avk}
\,,\label{BM-dist}
\eea
\bea
\Delta^N  D_{h/q^\uparrow}(z,\pp) &=& \Delta^N  D_{h/q^\uparrow}(z)\;
\sqrt{2e}\,\frac{\pp}{M_h} \;
\frac{e^{-\pp^2/\avp \C}}{\pi\avp}\,,\label{Coll-frag}
\eea
with
\be
\avk \BM= \frac{\avk \, M^2\BM}{\avk + M^2 \BM}\, ,
\qquad
\avp \C=\frac{\avp \, M_h^2}{\avp +M_h^2}\,\cdot \label{Coll-frag2}
\ee
The $x$-dependent function $\Delta f_{\qup/p}(x)$ and the $z$-dependent
function 
$\Delta^N  D_{h/q^\uparrow}(z)$  in Eqs.~(\ref{BM-dist}) and (\ref{Coll-frag})
are 
not known, and should be determined phenomenologically by fitting the available
data 
on azimuthal asymmetries and moments; the $\kt$ and $p_\perp$ dependent terms
and 
their normalization are chosen in such a way that positivity 
bounds~\cite{Bacchetta:1999kz} are fulfilled automatically.

At this stage, the usual procedure would be to perform an analytical
$\kt$ integration of Eqs.~(\ref{FUU}-\ref{FUUcos2phi}) over the 
range $[0,\infty]$,  
using the parametrizations in Eqs.~(\ref{partond}), (\ref{partonf}),
(\ref{BM-dist}) and (\ref{Coll-frag}) with the appropriate normalization of Eq.~\eqref{normalization}, and to re-express all the $F$ structure functions in terms of the Gaussian parameters:
\bea
F_{UU} & = & \sum_{q} \, e_q^2 \,f_{q/p}(\xb)\,D_{h/q}(z_h)
\frac{e^{-P_T^2/\avPT\G}}{\pi\avPT\G} \label{g-FUU}\\
F_{UU}^{\cos2\phi_h} & = & -e\,P_T^2 \, \sum_{q} \, e_q^2 \,
\frac{ \Delta f_{\qup/p}(\xb)}{M \BM}\,
\frac{\Delta^N  D_{h/q^\uparrow}(z_h)}{M_h} \,
\frac{e^{-P_T^2/\avPT \BM}}{\pi\avPT ^3 \BM}
\, \frac{z_h\,\avk ^2 \BM \avp ^2 \C}{\avk \avp }
\label{g-FUUcos2phi} \\
F_{UU}^{\cos\phi_h} & = & -2\,\frac{P_T}{Q}\, \sum_{q} \, e_q^2 \,
f_{q/p}(\xb) \, D_{h/q}(z_h) \, \frac{e^{-P_T^2/\avPT\G}}{\pi\avPT \G ^2}\, z_h
\avk\,
\nonumber \\
&& +  2e\,\frac{P_T}{Q}\,\sum_{q} \, e_q^2 \,
\frac{ \Delta f_{\qup/p}(\xb)}{M \BM}\, \frac{\Delta^N  D_{h/q^\uparrow}(z_h)}
{M_h} \, \frac{e^{-P_T^2/\avPT \BM}}{\pi\avPT ^4 \BM} \,
\label{g-FUUcosphi} \\
&& \hspace{1.5cm} \times
\frac{\avk^2 \BM \avp ^2 \C}{\avk \avp} \Big[z_h^2 \avk \BM \Big(P_T^2 -
\avPT \BM \Big) + \avp \C \avPT \BM \Big]\,,
\nonumber
\eea
where
\bea
&&\avPT\G=\langle p_{\perp}^2\rangle+z^2_h \langle
k_{\perp}^2\rangle\,,\label{ptsm-gauss}\\
&&\langle P_T^2\rangle \BM = \langle p_{\perp}^2\rangle\C + z^2_h \langle
k_{\perp}^2\rangle\BM\,.\label{ptsm-gauss-bm}
\eea
We stress that the analytical integration which leads to
Eqs.~(\ref{g-FUU})--(\ref{g-FUUcosphi}) is performed over the full range of
$\bfk_{\perp}$ values:
\begin{equation}
\int \!\! d^2 \bfk_{\perp} \Rightarrow  \int_{0}^{2\pi} \!\!d\varphi 
\int_{0}^{\infty} \!\!d k_{\perp}\,k_{\perp}\,,
\label{eq:int0inf}
\end{equation}
and that the expressions~(\ref{ptsm-gauss}) and (\ref{ptsm-gauss-bm}) which
relate the $\avPT$ to  $\langle p_{\perp}^2\rangle$, $\langle k_{\perp}^2\rangle$ 
and $z_h$ are due to such choice. 
In other words, Eqs.~(\ref{ptsm-gauss}) and
(\ref{ptsm-gauss-bm}) are a direct consequence not only of the assumption of 
Gaussian $\kt$ and $\pp$ distributions of the TMDs, but also of the choice of
the $\kt$ integration range. 
Many phenomenological analysis on the TMDs are
based on the Gaussian assumption and
overlook any issue regarding the limits of integration over $k_{\perp}$.
However, great attention to this should be payed when analysing 
data from JLAB, HERMES and COMPASS experiments, where the average $Q^2$ is not
so large ($\sim 2\textrm{  GeV}^2$).
Therefore in some kinematical ranges 
it could happen that the $(k_{\perp}/Q)$ values accessed are not small.
One immediately visible signal of this is the Cahn effect in both azimuthal
moments, at twist-3 in $\langle \cos \phi_h \rangle$
and at twist-4 in $\langle \cos 2\phi_h \rangle$, which are directly proportional
to 
$k_{\perp}/Q$ and $k_{\perp}^2/Q^2$ respectively, and are found to be (phenomenologically) large.

\subsection{Impact of the partonic cuts on the $F_{UU}$ term of the unpolarized
cross section and on $\langle P_{T}^2\rangle$\label{res-unp}}

In this section we will show the impact of the $\kt$ cuts presented in
Eqs.~(\ref{cutenergy}) and (\ref{cutdirection})
on the calculation of the SIDIS unpolarized cross section
and on the average transverse momenta of the final detected hadron,
$\langle P_{T}^2\rangle$.
%
%
\begin{figure}[b]
\includegraphics[width=0.24\textwidth,angle=-90]{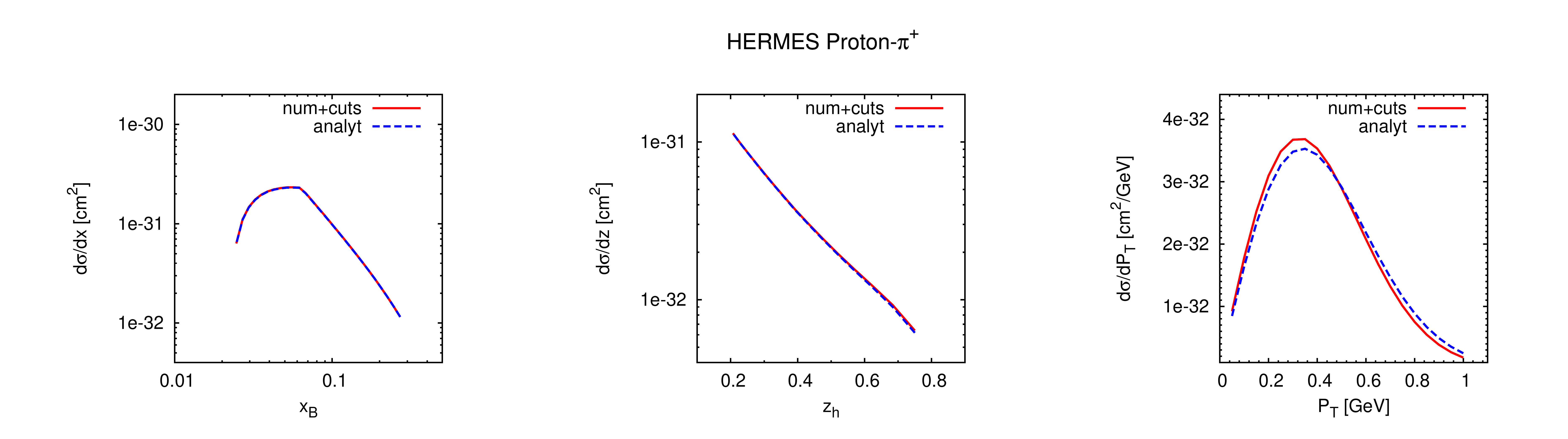}
\\
\includegraphics[width=0.24\textwidth,angle=-90]{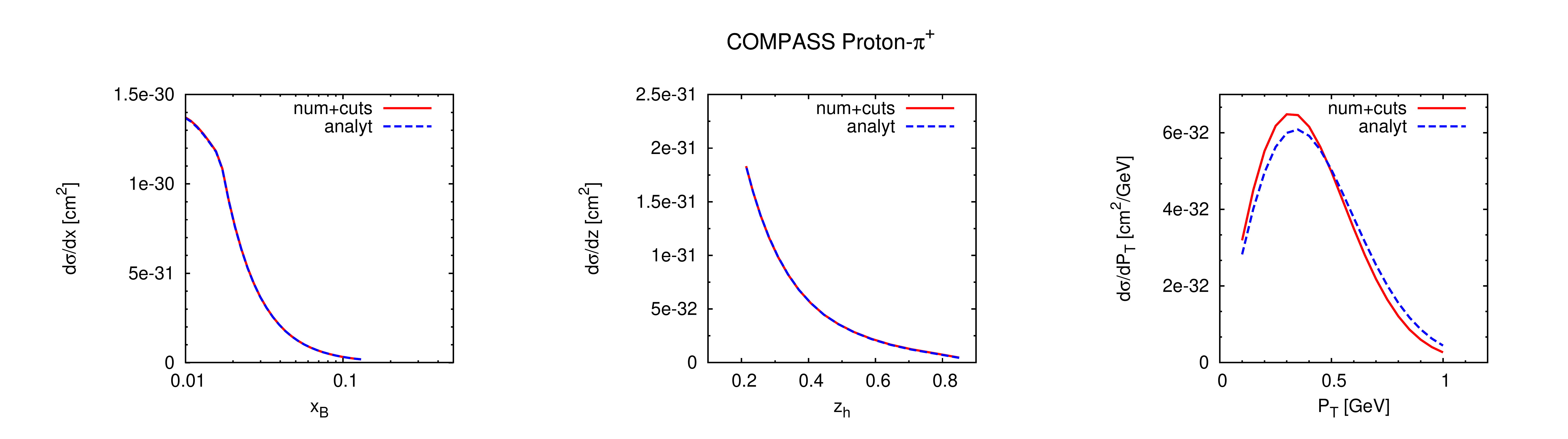}
\caption{Unpolarized cross section for $\pi^+$ production at HERMES (upper
panel) and COMPASS experiment on a proton target (lower panel), as a function of
$\xb$ (left plot), $z_h$ (central plot) and $P_T$ (right plot).
The (red) solid line corresponds to the unpolarized cross section calculated
according to Eq.~(\ref{sidis-Xsec-final}) with a numerical $\kt$ integration
implementing the $\kt$-cuts of Eqs.~(\ref{cutenergy}) and (\ref{cutdirection}).
The dashed (blue) line is the unpolarized cross section calculated according to
Eqs.~(\ref{g-FUU})--(\ref{g-FUUcosphi}) resulting from an analytical $\kt$
integration from zero to infinity.
We do not show the analogous cross section corresponding to the COMPASS
experiment on deuteron target as the effects of the $\kt$ cuts are very similar
to those for COMPASS on proton target.}
\label{xsec-pip}
\end{figure}
%
%
Figure~\ref{xsec-pip} shows the first term of the SIDIS unpolarized cross
section, proportional to $F_{UU}$,
calculated for HERMES and COMPASS kinematics  (the detailed experimental cuts are
reported in Appendix~\ref{exp-cuts}) 
for $\pi^+$ production, integrated over all variables but one, $\xb$, $z_h$ and
$\bfP_T$. 

We use the unpolarized integrated PDF's given in Ref.~\cite{Gluck:1998xa} and the 
unpolarized fragmentation functions of Ref.~\cite{deFlorian:2007aj}.
For the Gaussian widths of the unpolarized distribution and fragmention TMDs we
use the values extracted 
in Ref.~\cite{Anselmino:2005nn}:  $\langle k_\perp^2\rangle = 0.25$ GeV$^2$ and
$\langle p_\perp^2\rangle=0.20$ GeV$^2$.
The  solid (red) line, denoted in the legend as ``{\it num+cuts}'',
corresponds to the unpolarized cross section
calculated according to Eqs.~(\ref{sidis-Xsec-final}),
(\ref{FUU})--(\ref{FUUcos2phi})
with a numerical $\kt$ integration over the range obtained 
implementing the $\kt$-cuts of Eqs.~(\ref{cutenergy}) and (\ref{cutdirection}).
The dashed (blue) line,  indicated as ``{\it analyt}'', is the unpolarized cross
section calculated according to
Eqs.~(\ref{g-FUU})--(\ref{g-FUUcosphi}), resulting from a $\kt$ analytical
integration over the range $[0,\infty]$.
These plots clearly show that, 
as far as the $\xb$ and $z_h$ distributions are concerned,  
there is no difference between the two calculations.
Instead, a slight modification can be observed
in the $P_T$ distribution (see the upper and lower right panels).
%
\begin{figure}[t]
\includegraphics[width=0.24\textwidth,angle=-90]{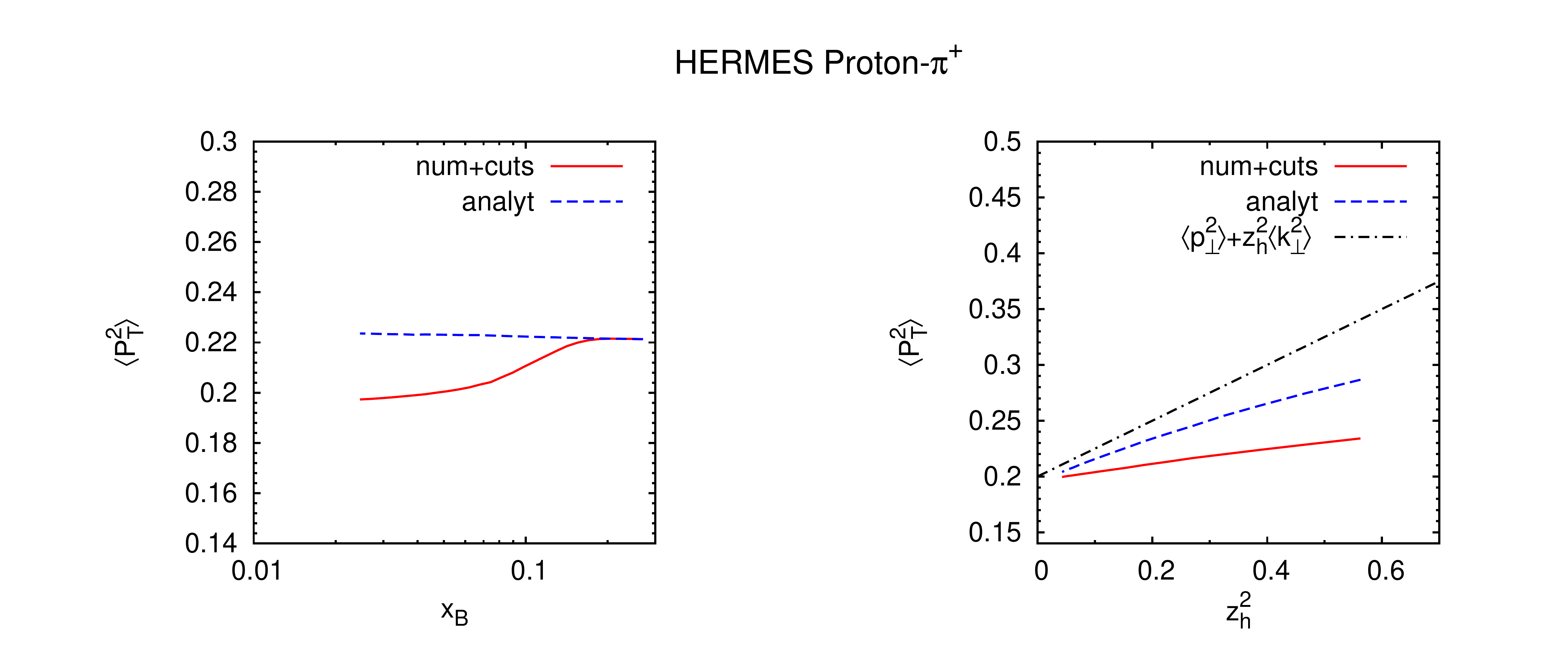}
\\
\includegraphics[width=0.24\textwidth,angle=-90]{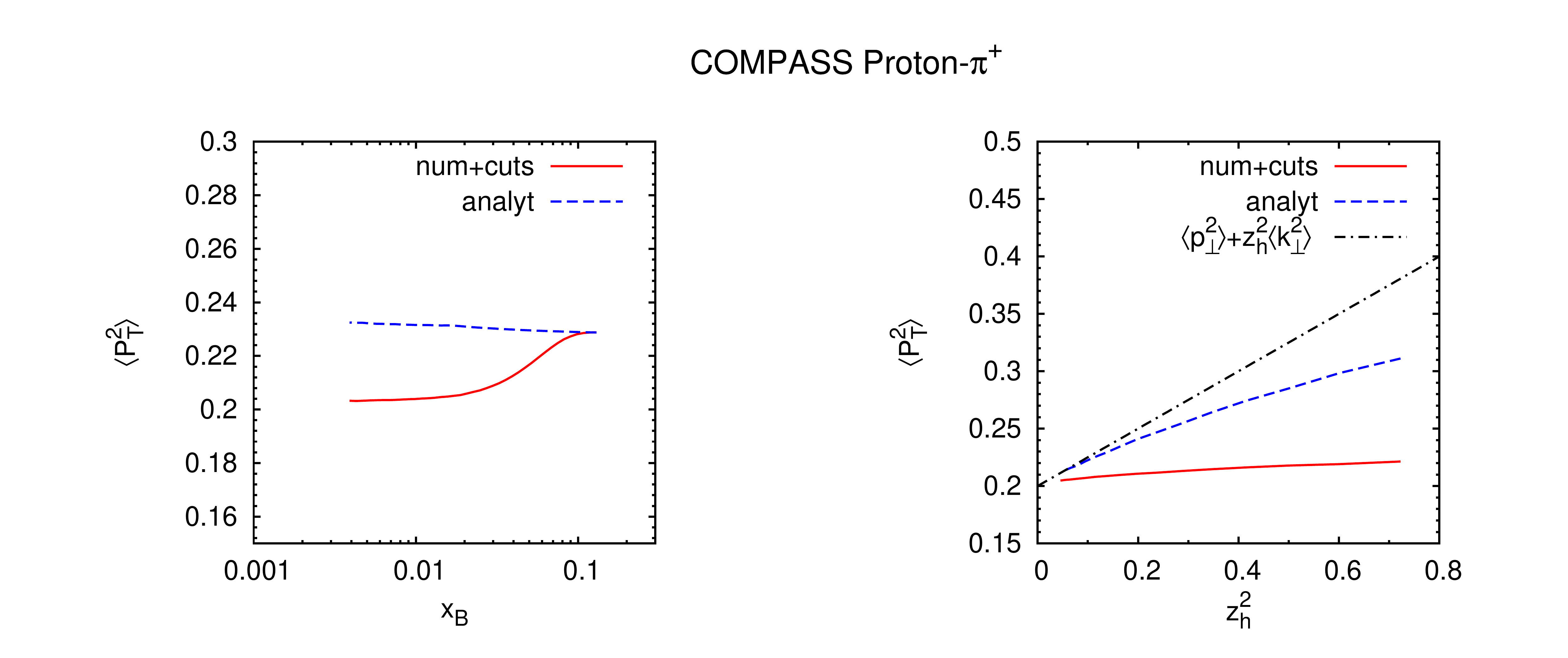}
\caption{$\langle P_{T}^2\rangle$, defined in Eq.~(\ref{ptsmobs}), as a function
of $\xb$ (left plot) 
and of $z_h^2$ (right plot), for $\pi^+$ production at HERMES (upper panel) and
COMPASS (lower panel). 
The solid (red) line corresponds to $\langle P_{T}^2\rangle$ calculated with a
numerical integration implementing Eqs.~(\ref{cutenergy}) and
(\ref{cutdirection}), while 
the dashed (blue) line is $\langle P_{T}^2\rangle$ calculated with an analytical
integration. In both cases we have applied the 
experimental cuts on $P_T$ reported in Appendix~\ref{exp-cuts} . Finally, the dash-dotted (black) line corresponds to
the Gaussian $\ptsmG$ 
. \label{capt-x-ptsm}}
\end{figure}
%

The average hadronic transverse momentum $\langle P_{T}^2\rangle$ of the final,
detected hadron $h$ is defined as:
\begin{equation}
\langle P_{T}^2\rangle=\frac{\int d^2 \boldsymbol{P}_T P_T^2 d\sigma }{\int d^2
\boldsymbol{P}_T d\sigma}\,\cdot\label{ptsmobs}
\end{equation}
Notice that if the integral in Eq.~(\ref{ptsmobs}) is performed over the range
$[0,\infty]$, 
then $\langle P_{T}^2\rangle$
coincides with the Gaussian width of the unpolarized $P_T$ distribution of
Eq.~(\ref{g-FUU}): $\langle P_{T}^2\rangle\equiv \ptsmG$.
The experimental $P_T$ range, however, usually span a finite region between some
$P_T^{min}$ and $P_T^{max}$; 
therefore, in any experimental analysis, one inevitably has
$\langle P_{T}^2\rangle \neq\ptsmG$,
even without considering the cuts in Eqs.~(\ref{cutenergy}) and
(\ref{cutdirection}). Consequently, the relation
$\langle P_{T}^2\rangle\simeq\langle p_{\perp}^2\rangle+z^2_h \langle
k_{\perp}^2\rangle$
holds only approximatively.

\begin{figure}[p]
\includegraphics[width=0.24\textwidth,angle=-90]{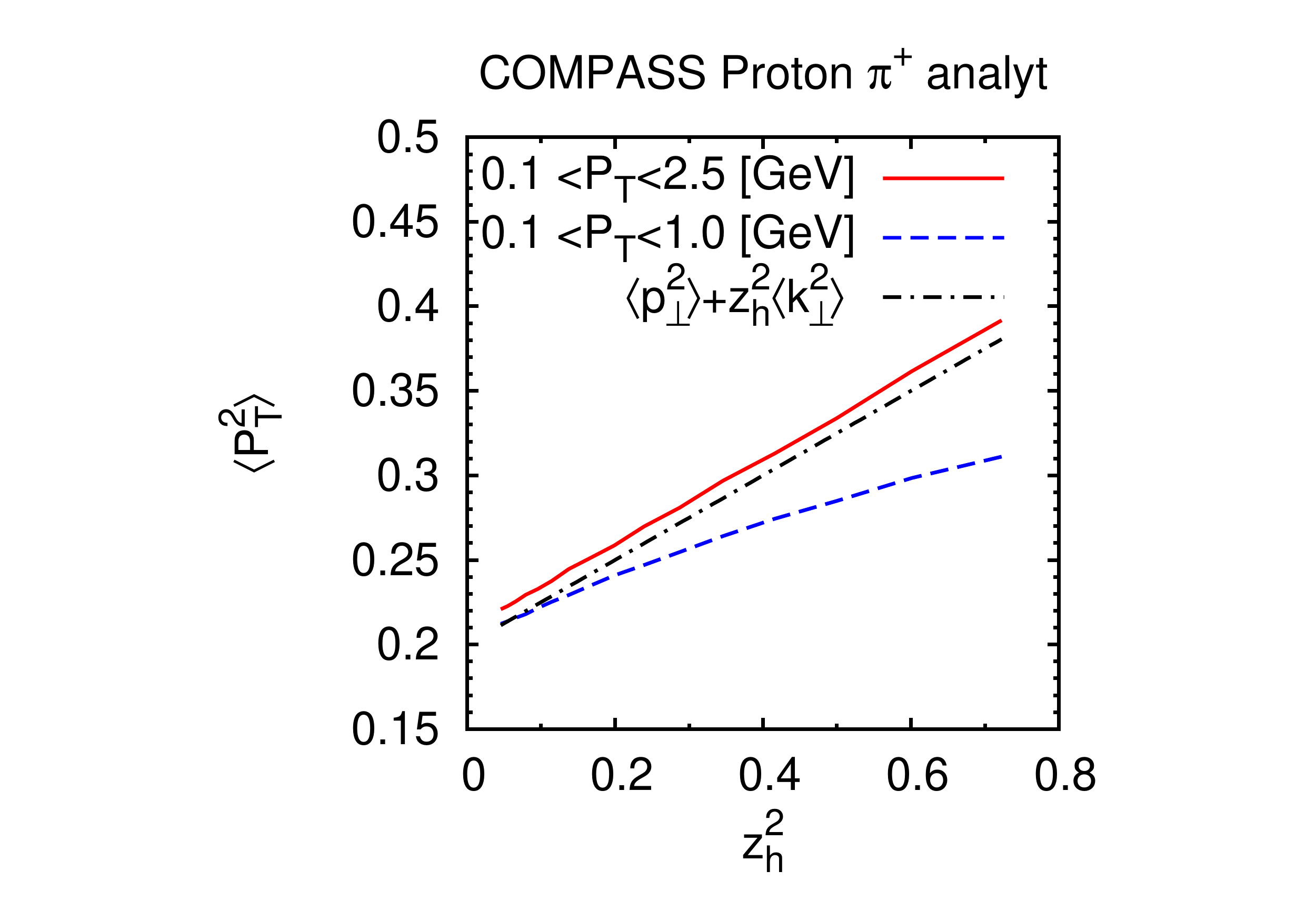}
\includegraphics[width=0.24\textwidth,angle=-90]{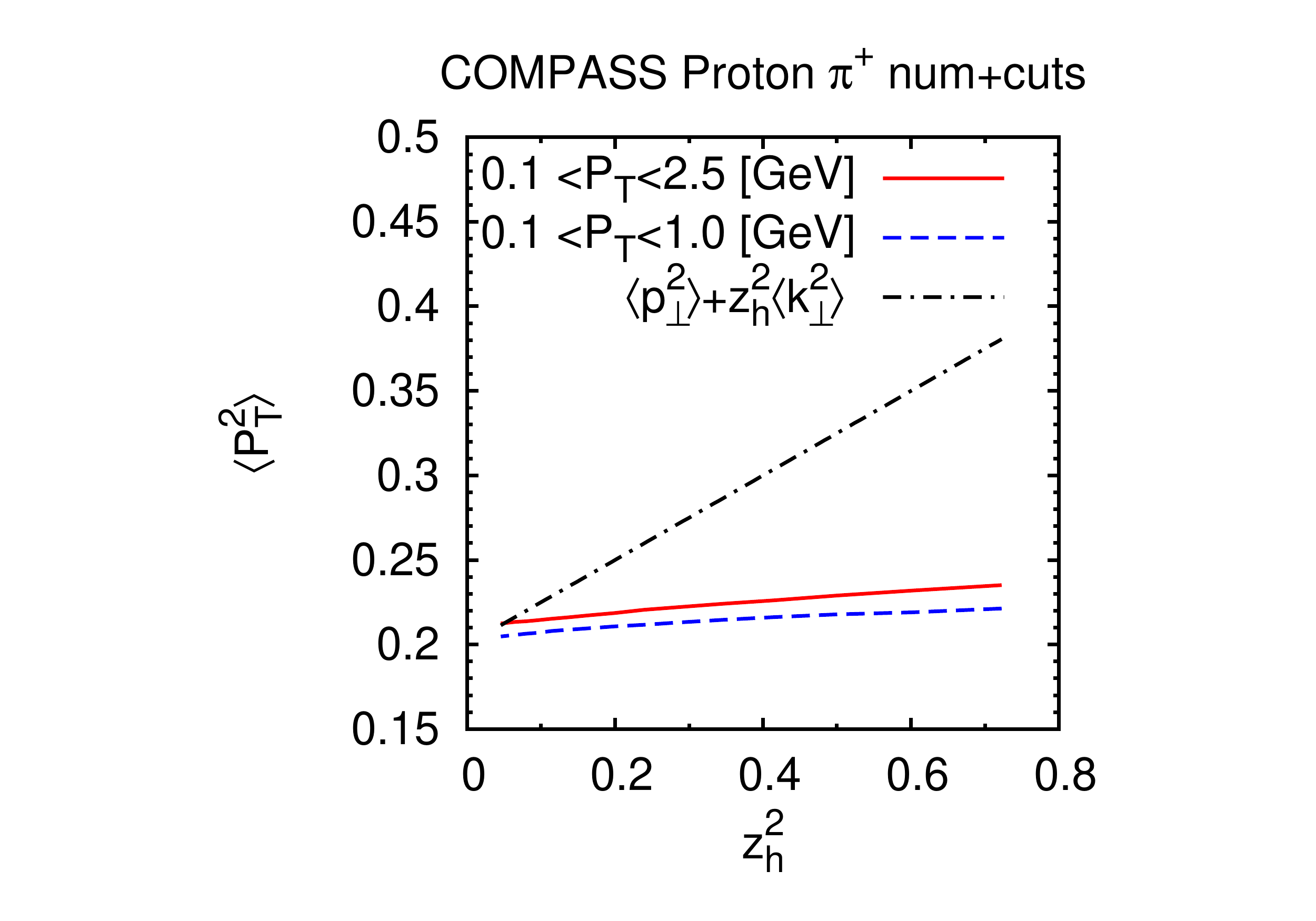}
\caption{$\langle P_{T}^2\rangle$, at COMPASS kinematics as a function of 
$z_h^2$ for different ranges in $P_T$,
calculated analytically (left plot) and numerically implementing
Eqs.~(\ref{cutenergy}) and (\ref{cutdirection}) (right plot). 
\label{capt-z-ptsm-diff_pt_cuts}}
%
\includegraphics[width=0.24\textwidth,angle=-90]
{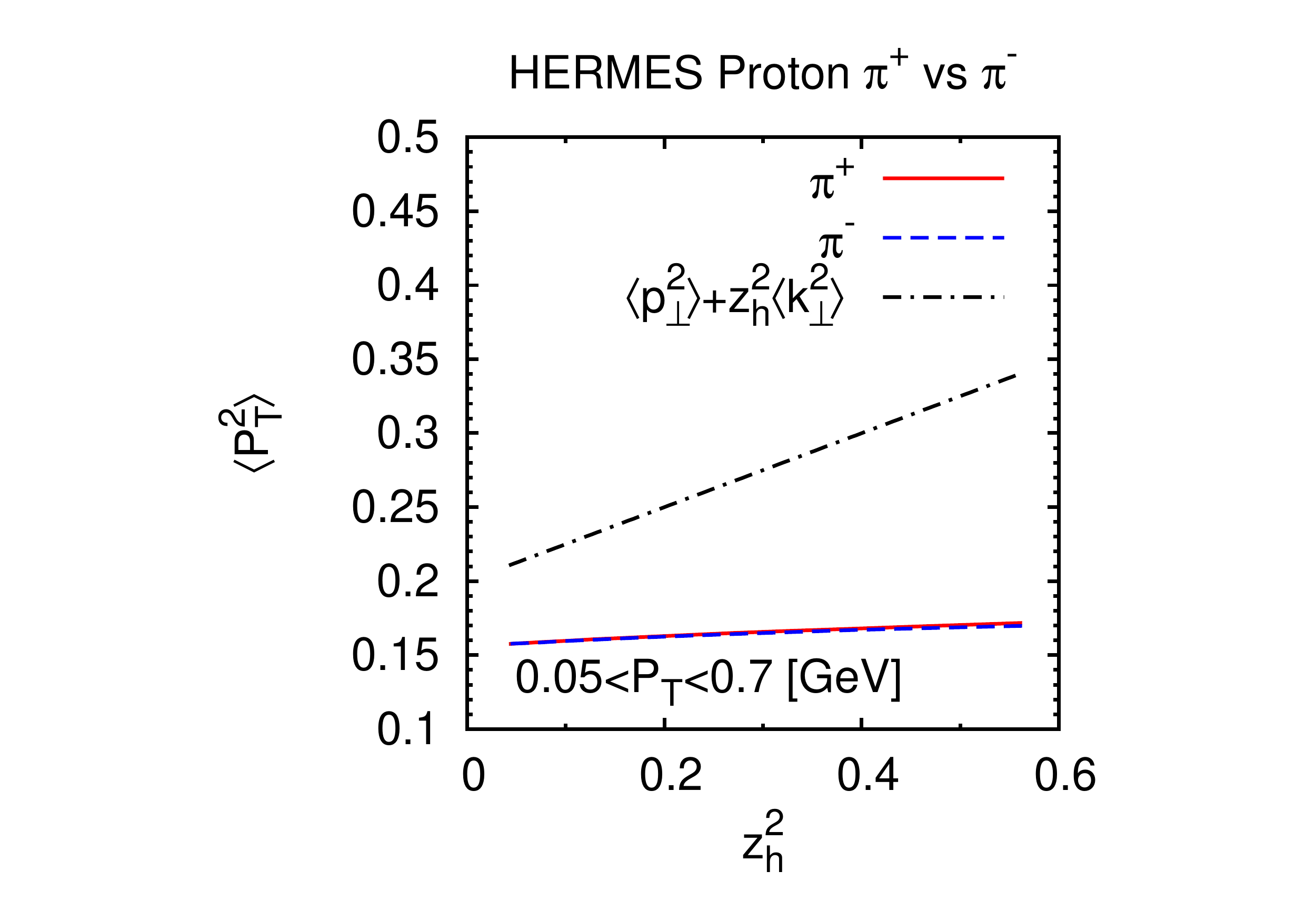}
\includegraphics[width=0.24\textwidth,angle=-90]
{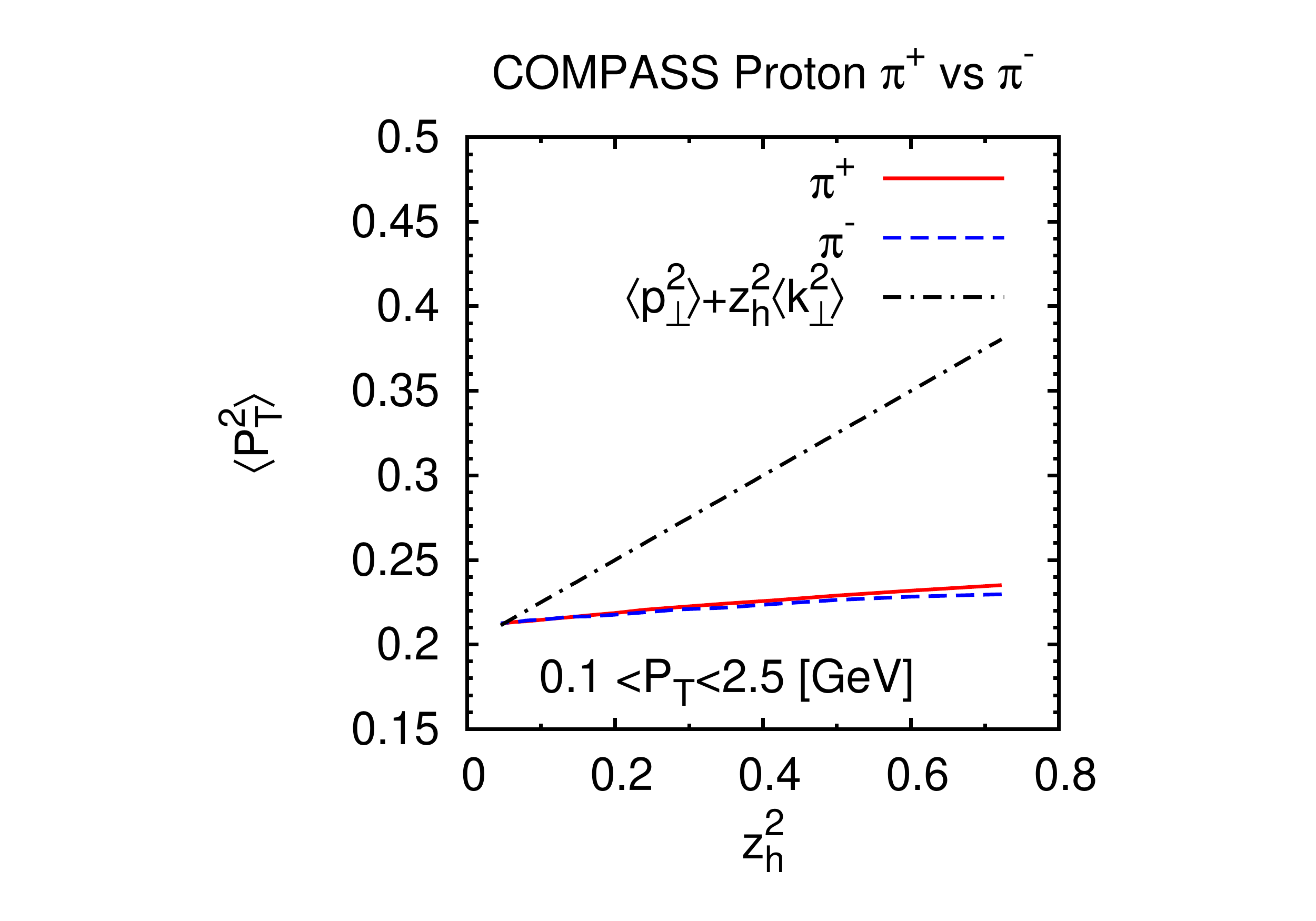}
\caption{$\langle P_{T}^2\rangle$ as function of $z_h^2$ for $\pi^+$ and $\pi^-$
production at HERMES (left plot) and COMPASS(right plot) pion production on a proton target.
\label{zs-pip-pim-ptsm}}
%
\includegraphics[width=0.24\textwidth,angle=-90]{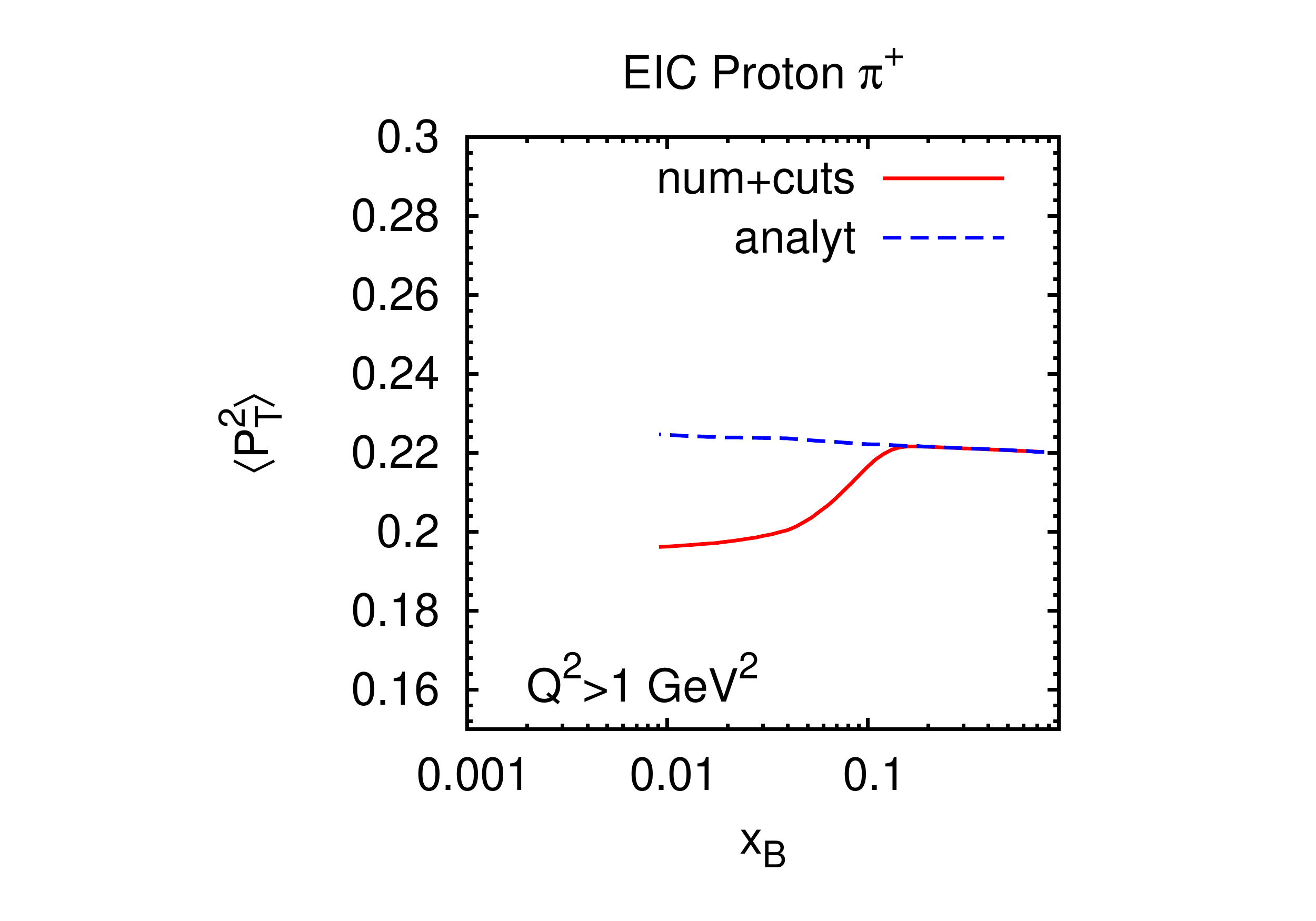}
\includegraphics[width=0.24\textwidth,angle=-90]{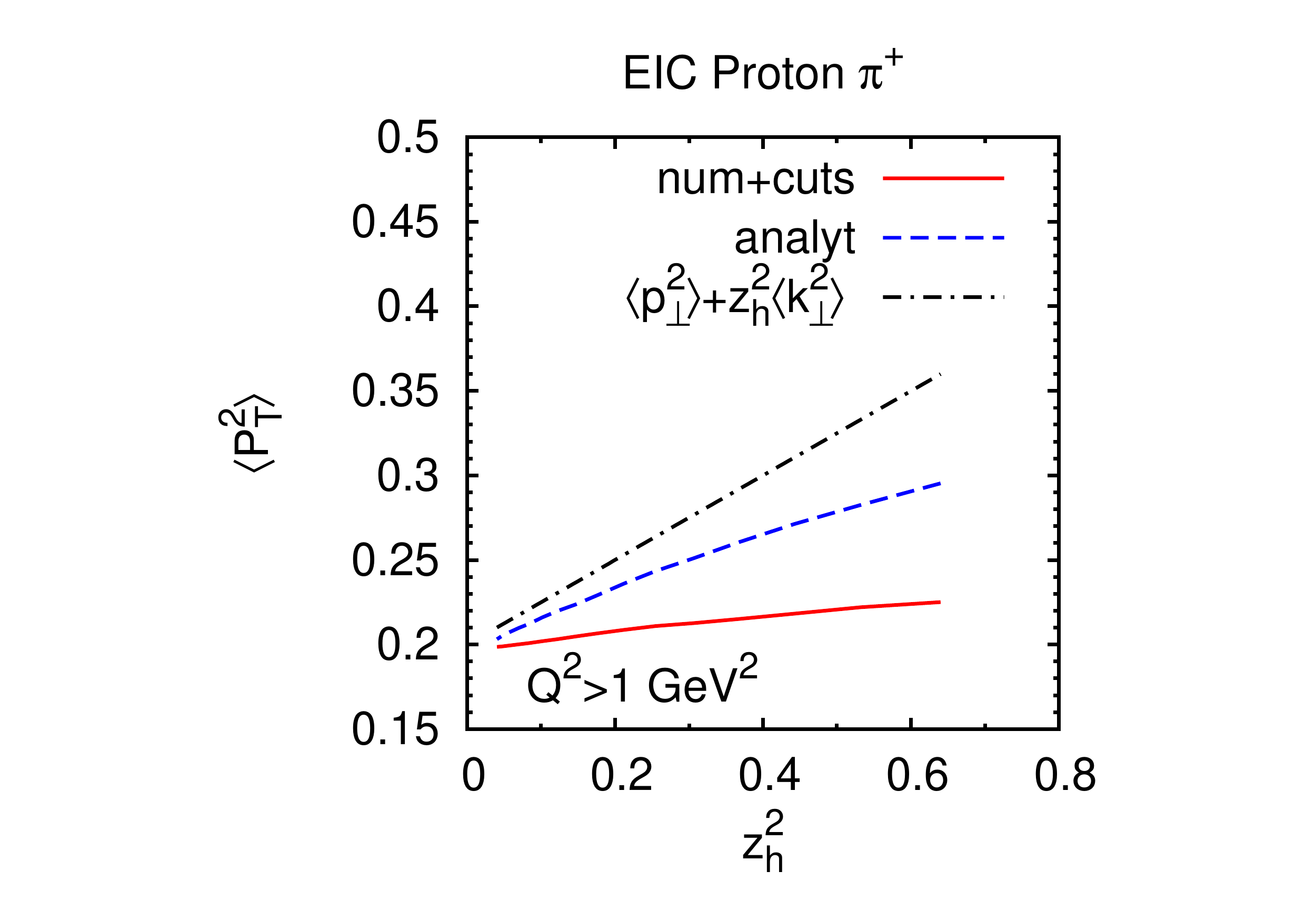}
\\
\includegraphics[width=0.24\textwidth,angle=-90]{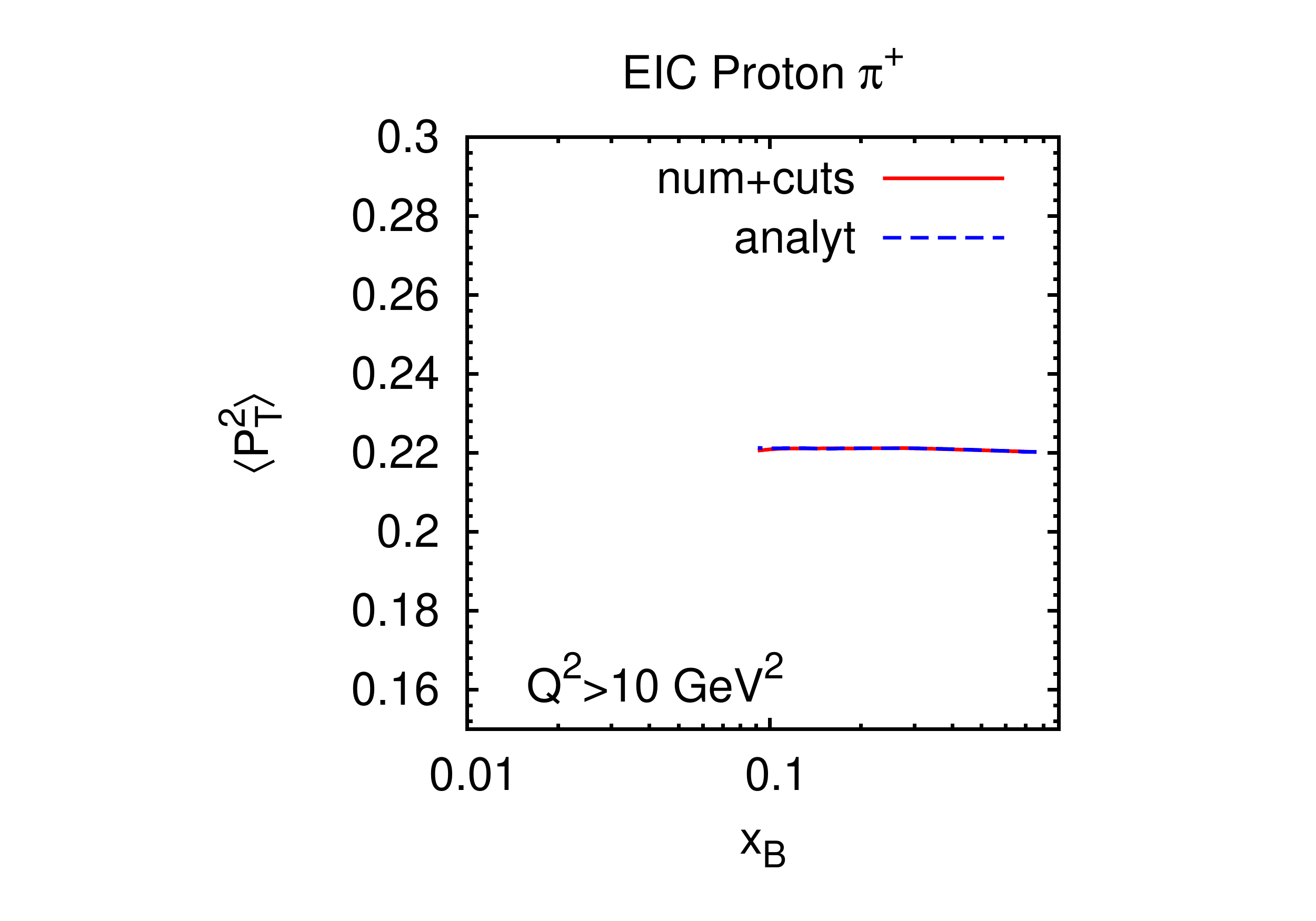}
\includegraphics[width=0.24\textwidth,angle=-90]{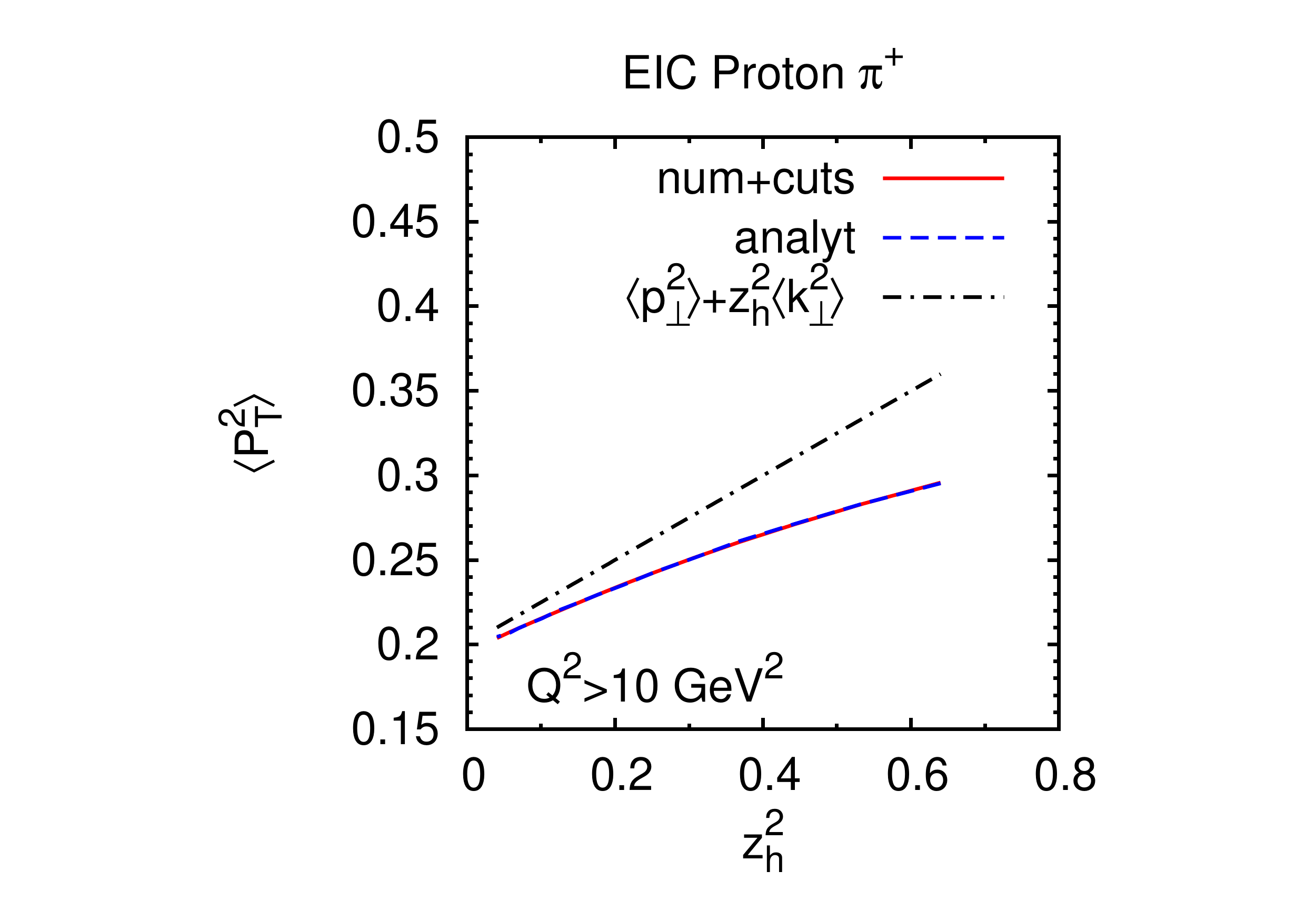}
\caption{$\langle P_{T}^2\rangle$, defined in Eq.~(\ref{ptsmobs}), as a function
of $\xb$ (left plot) 
and of $z_h^2$ (right plot), for $\pi^+$ production at EIC kinematics,
with $Q^2>1\textrm{ GeV}^2$ cut (upper panel) and  $Q^2>10\textrm{ GeV}^2$ cut
(lower panel). 
The solid (red) line corresponds to $\langle P_{T}^2\rangle$ calculated starting
from Eq.~(\ref{sidis-Xsec-final}) 
and then integrating it numerically, implementing
Eqs.~(\ref{cutenergy},\ref{cutdirection}). The dashed (blue) line is 
$\langle P_{T}^2\rangle$ calculated starting from
Eqs.~(\ref{g-FUU})--(\ref{g-FUUcosphi}). }
\label{eic-ptsm-pip}
\end{figure}

Figure~\ref{capt-x-ptsm} shows the average hadronic transverse momentum $\langle
P_{T}^2\rangle$ as a function of $\xb$ and of $z_h^2$ for $\pi^+$ 
at HERMES and COMPASS, respectively.
The  solid (red) lines correspond to $\langle P_{T}^2\rangle$ calculated
according to 
Eq.~(\ref{sidis-Xsec-final}) with a numerical $\kt$ integrations and
implementing Eqs.~(\ref{cutenergy}) and (\ref{cutdirection}). Instead, the
dashed (blue) lines correspond to $\langle P_{T}^2\rangle$ calculated according
to Eqs.~(\ref{g-FUU})--(\ref{g-FUUcosphi}) with an analytical integration. 
In both cases we have taken into account the appropriate experimental cuts on 
$P_T$ reported in Appendix~\ref{exp-cuts}.
Clearly, at low $x$, there is a substantial deviation from the analytical
calculation,
which also affects the value of $\langle P_{T}^2\rangle$ as a function of
$z_h^2$. As far as the $z_h$ dependence is concerned, first of all, 
one can  see that there is a large deviation from the naive formula,
Eq.~(\ref{ptsm-gauss}),
corresponding to the dash-dotted (black) lines,  for both calculations. 
Secondly, although the $z_h^2$-dependence is not linear any more, it seems to be
approaching
an almost constant behavior (as opposed to what COMPASS latest
analysis~\cite{Rajotte:2010ir}
seems to indicate).
Notice that, independently from Eqs.~(\ref{cutenergy}) and (\ref{cutdirection}),
if we naively assume 
$\langle P_T^2\rangle=\langle p_{\perp}^2\rangle +z_h \langle
k_{\perp}^2\rangle$ in any analysis of the data
we would conclude that the effective $\langle k_\perp^2\rangle$
is smaller than
the value $\langle k_\perp^2\rangle = 0.25$ GeV$^2$  that we used as an input in
the Gaussian.
Also $\langle p_\perp^2\rangle\neq0.20$ GeV$^2$, as a consequence of the limits
of integration on $P_T$.
Fig.~\ref{capt-z-ptsm-diff_pt_cuts} shows
how the integration range influences the value of $\langle P_T^2\rangle$ when we
integrate analytically (left plot)
or implementing the cuts in Eqs.~(\ref{cutenergy}) and (\ref{cutdirection})
(right plot).
%
%
%

Finally, different behaviors in $\xb$ could, in principle, imply different
values of $\langle P_{T}^2\rangle
$ for $\pi^+$ and $\pi^-$. We have explored this possibility in order to account
for the slight discrepancy in the $\langle P_{T}^2\rangle
$ corresponding to $\pi^+$ and $\pi^-$ as observed by COMPASS.
Unfortunately the difference predicted by our model for $\pi^+$ and $\pi^-$ is
extremely tiny, even at COMPASS kinematics, as can be appreciated in
Fig.~\ref{zs-pip-pim-ptsm}.  One should keep in mind that the tiny separation
observed here occurs for purely kinematical effects, although larger differences
between the $\pi^+$ and $\pi^-$ average $P_{T}$ could be generated by adopting
different $\kt$ distribution widths for different quark flavours.

In Fig.~\ref{eic-ptsm-pip} we show how a cut in $Q^2$ can change the description
of data.
We can see that cutting at higher $Q^2$ means cutting the lower $\xb$ region,
where the constraint of Eq.~\eqref{cutdirection} strongly applies.
As a consequence, if we apply high $Q^2$ cuts the description of the data
with
or without  $\kt$-cuts 
is the same. 
This means that, provided Eqs.~\eqref{cutenergy} and \eqref{cutdirection} are
right, there is a region of $x_B$ or $Q^2$
where we can safely assume that $(k_{\perp}/Q)$ corrections are small or
negligible and where a phenomenological analysis is safe and unambiguous.


\subsection{Impact of the partonic cuts on the azimuthal moment $\langle
\cos\phi_h\rangle$ \label{rescos}}

At this stage, we are ready to evaluate the effect of the physical partonic cuts
on the $\langle \cos \phi _h\rangle$ and  $\langle \cos 2 \phi _h\rangle$
azimuthal moments,
Eq.~(\ref{val-med}), which represent the most delicate terms of the SIDIS 
unpolarized cross section.

The $\langle \cos \phi _h\rangle$ modulation receives two contributions,
both suppressed by one power of $(k_{\perp}/Q)$, see  Eq.~(\ref{FUUcosphi}).
The Cahn term, which is proportional to the convolution of the unpolarized
distribution
and fragmentation functions, was extensively studied in
Ref.~\cite{Anselmino:2005nn}.
There, EMC measurements~\cite{Ashman:1991cj} on the $\cos\phi _h$ modulation and
of the $P_T$ distribution on the unpolarized SIDIS cross section were used
to determine the Gaussian width of the $\kt$ distribution of the unpolarized
distribution function $f_{q/p}(x,\kt)$. 
The second term is proportional to the convolution of the Boer-Mulders
distribution function and the Collins fragmentation function and was neglected
in Ref.~\cite{Anselmino:2005nn}.
Since then, new and higher statistics experimental data
have  become available~\cite{Giordano:2010zz,Sbrizzai:2009fc}:
it is therefore timely and interesting to evaluate its
net contribution to the  $\langle \cos \phi _h\rangle$ azimuthal moment,
using some reasonable estimate of the Boer-Mulders
and Collins TMDs from the literature~\cite{Barone:2009hw,Anselmino:2007fs}.

\begin{figure}[t]
\includegraphics[width=0.24\textwidth,angle=-90]
{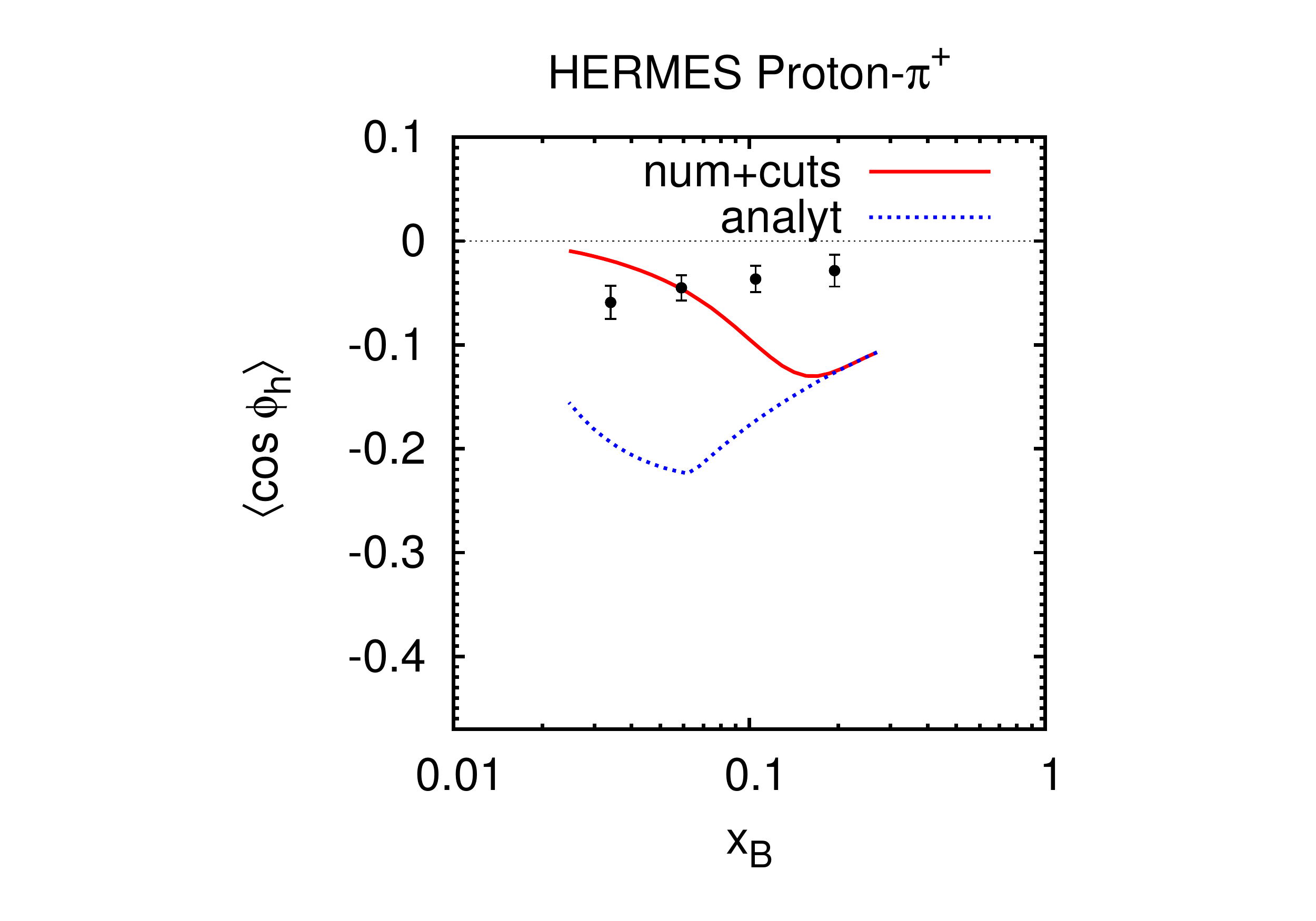}\hspace*{-1.3cm}
\includegraphics[width=0.24\textwidth,angle=-90]
{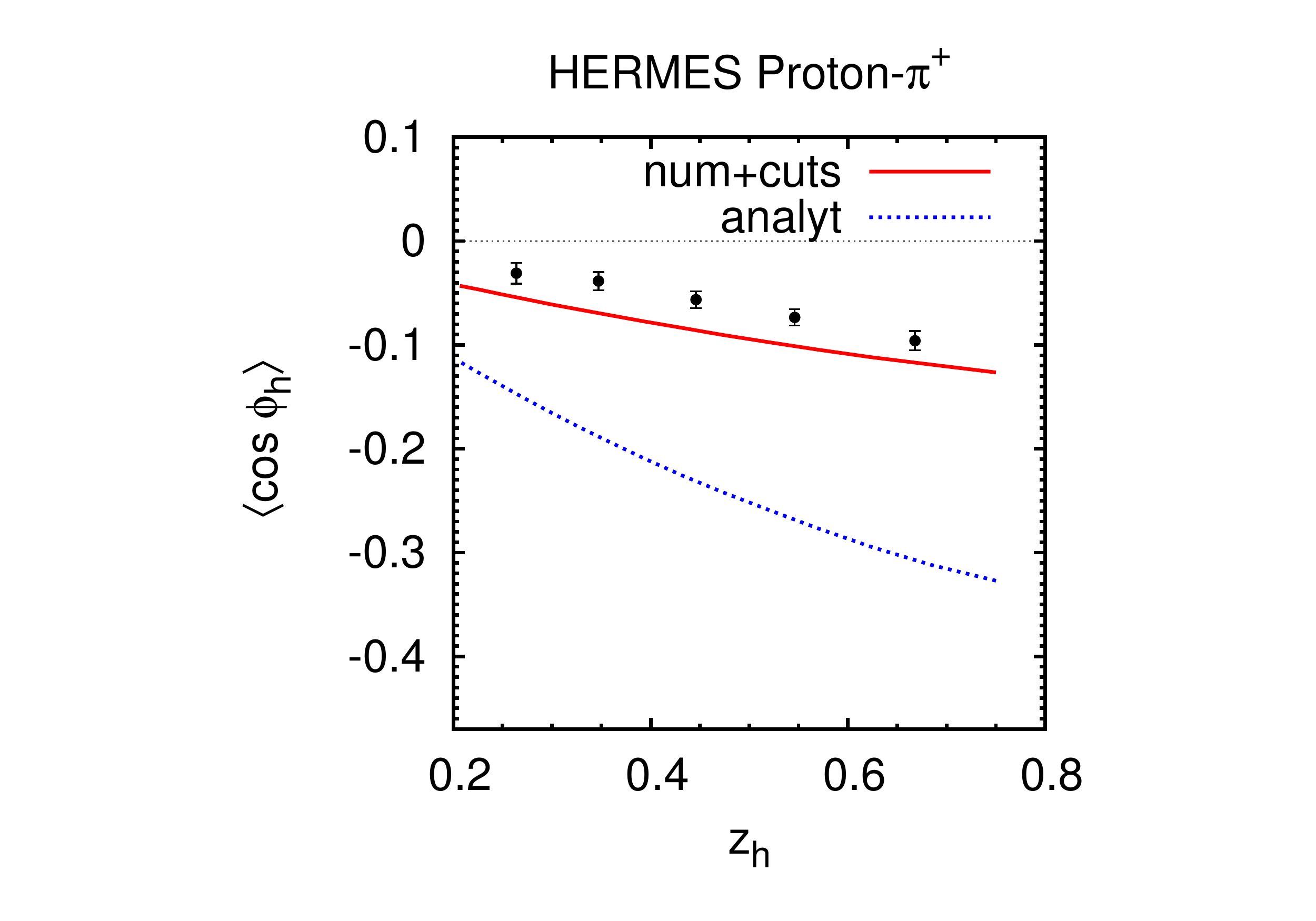}\hspace*{-1.3cm}
\includegraphics[width=0.24\textwidth,angle=-90]
{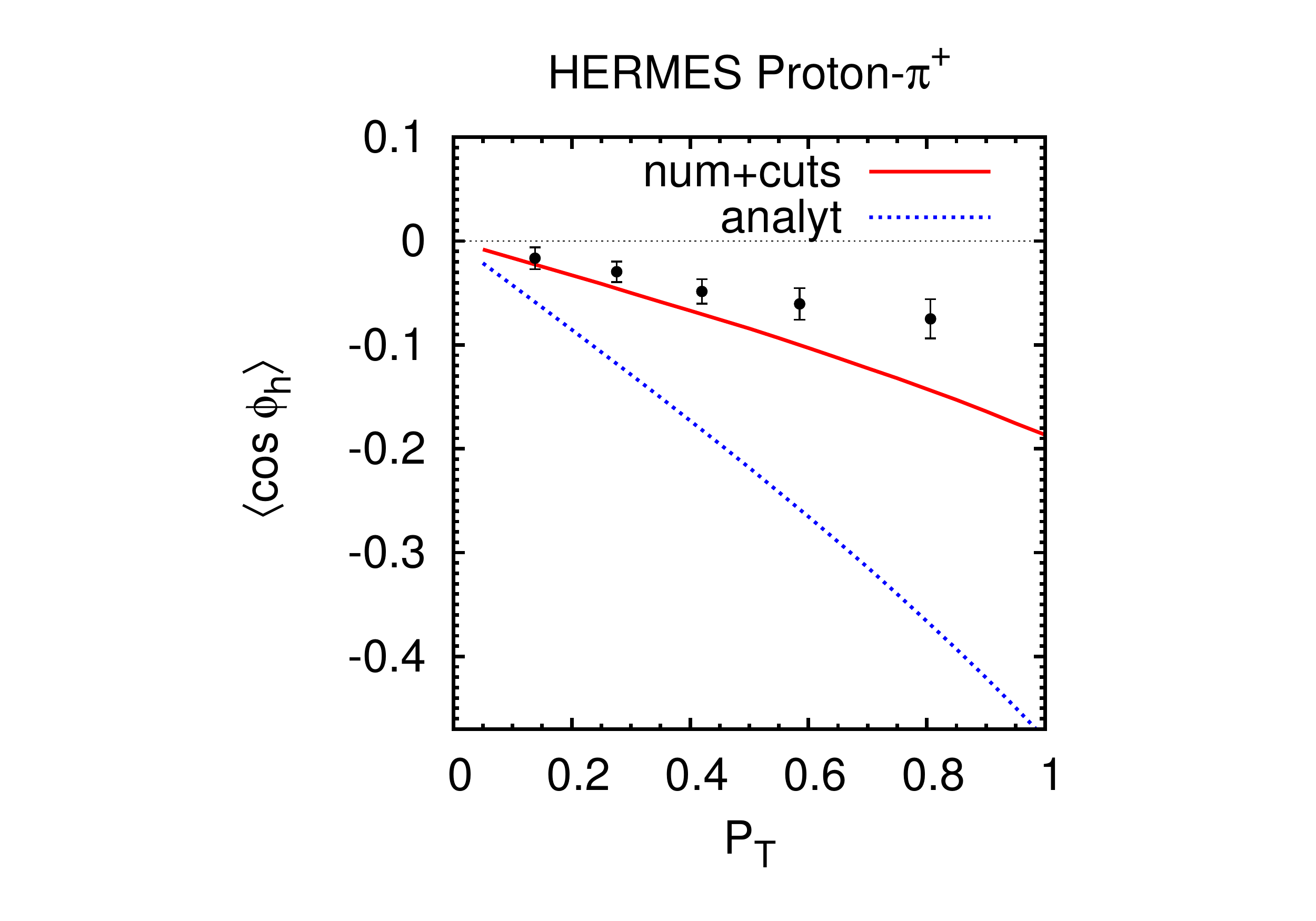}
\\
\includegraphics[width=0.24\textwidth,angle=-90]
{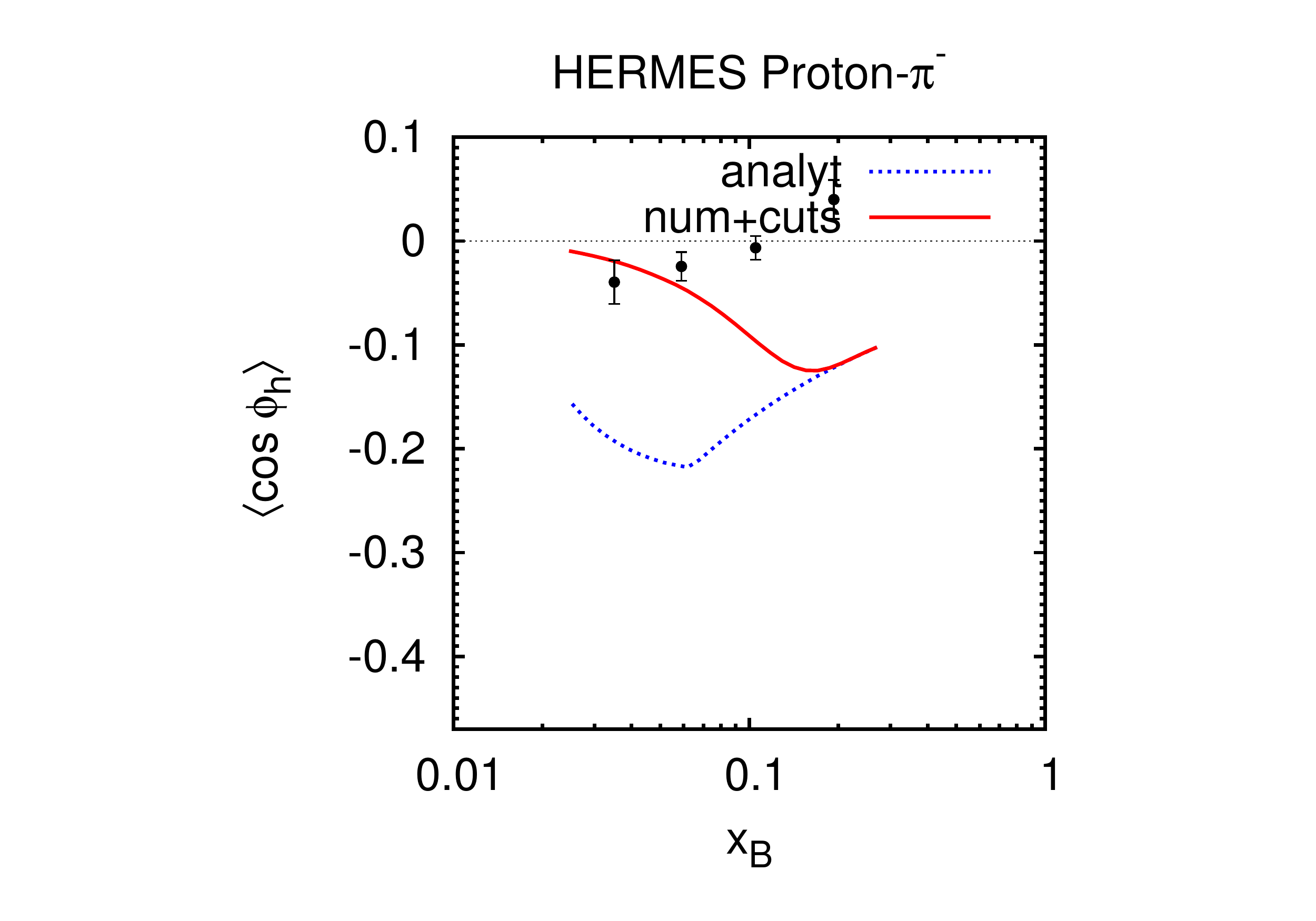}\hspace*{-1.3cm}
\includegraphics[width=0.24\textwidth,angle=-90]
{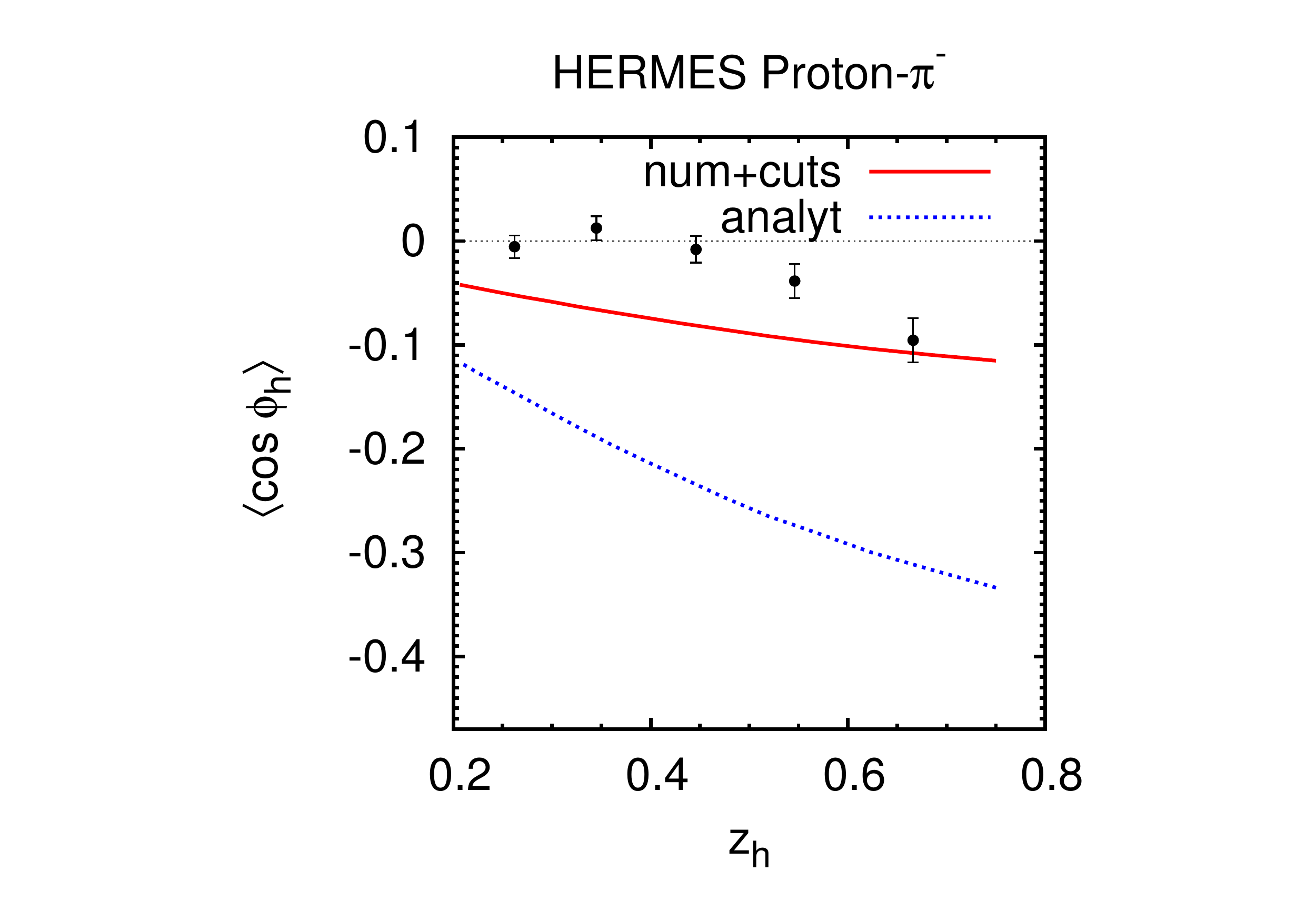}\hspace*{-1.3cm}
\includegraphics[width=0.24\textwidth,angle=-90]
{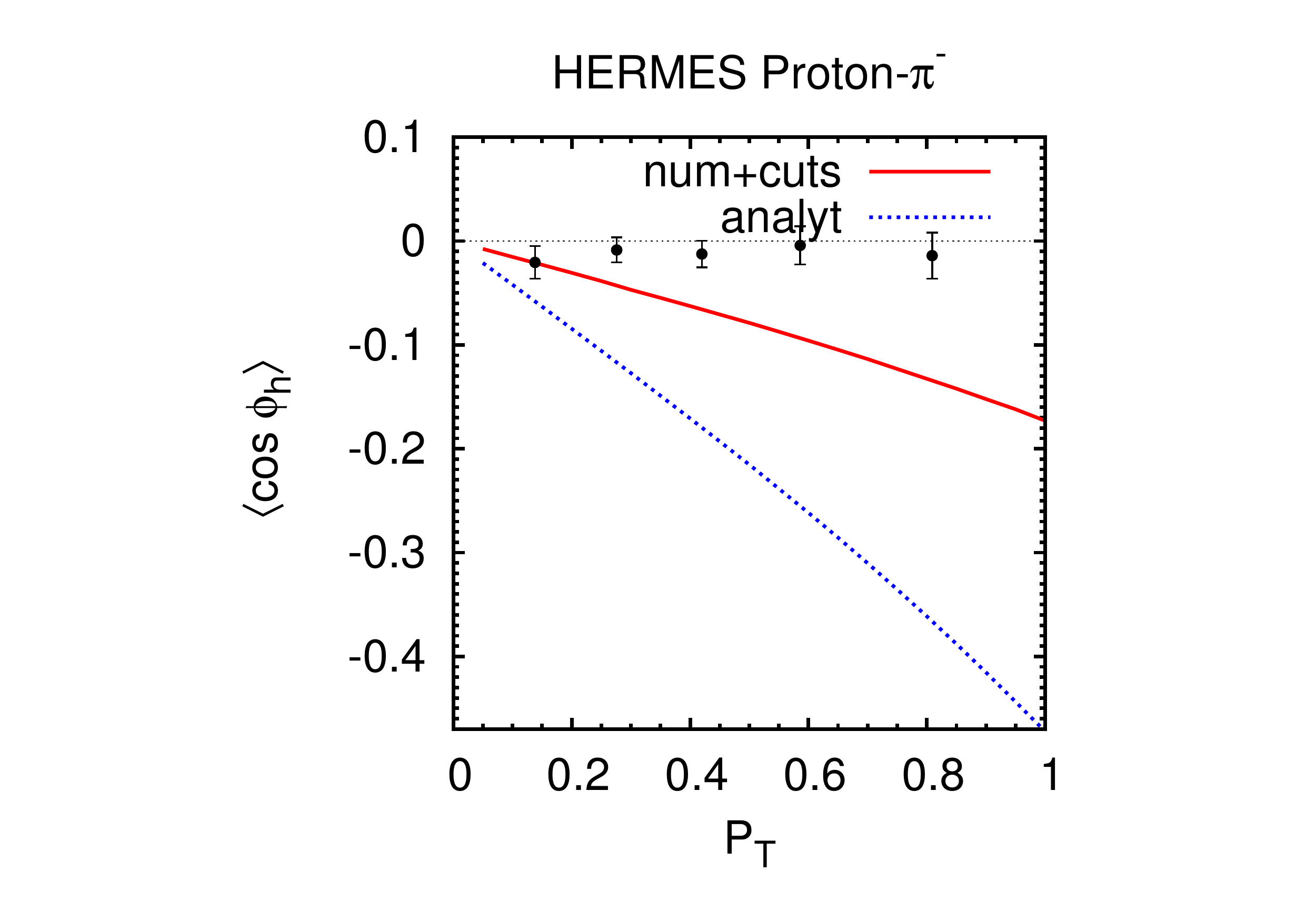}
\\
\includegraphics[width=0.24\textwidth,angle=-90]
{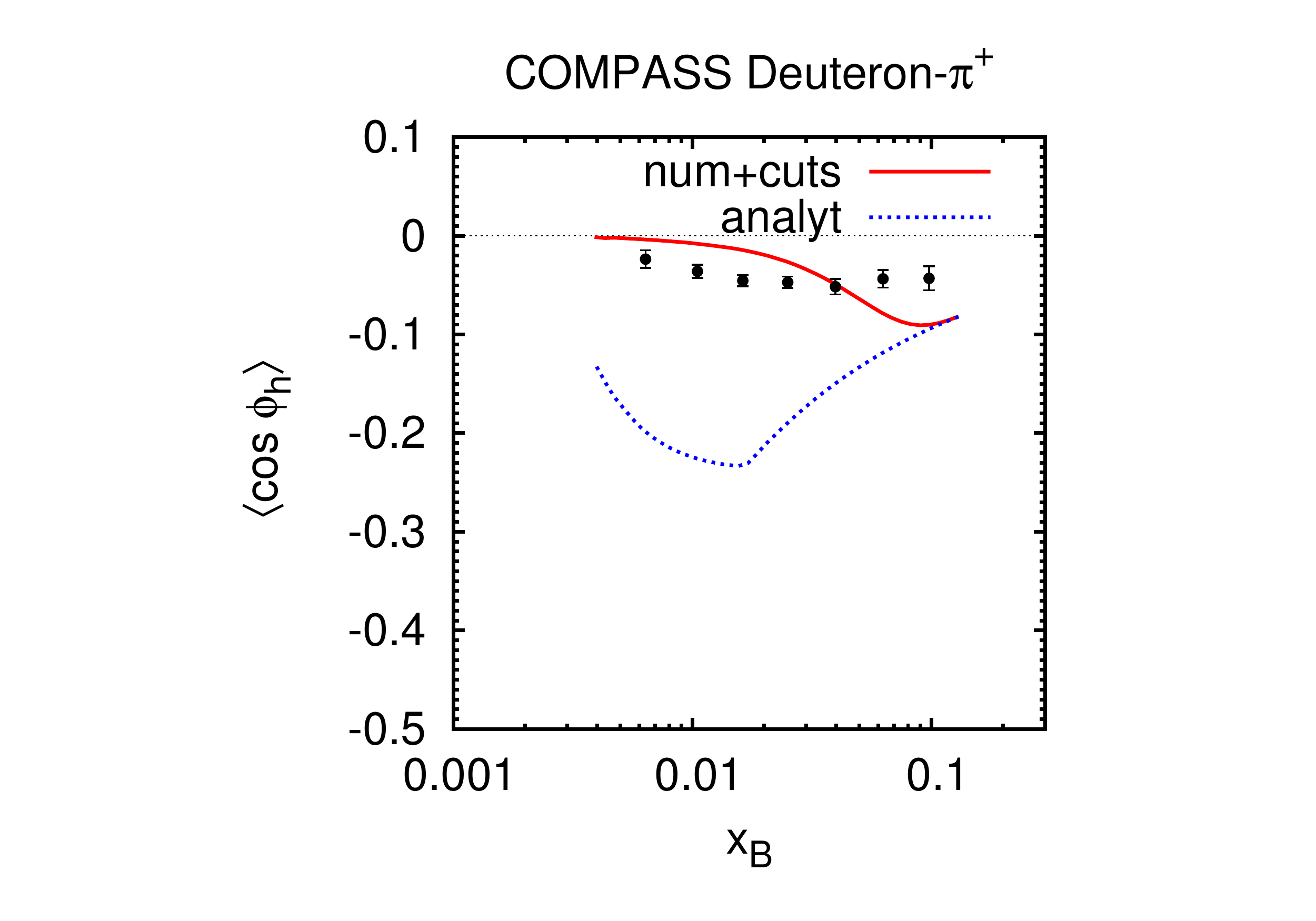}\hspace*{-1.3cm}
\includegraphics[width=0.24\textwidth,angle=-90]
{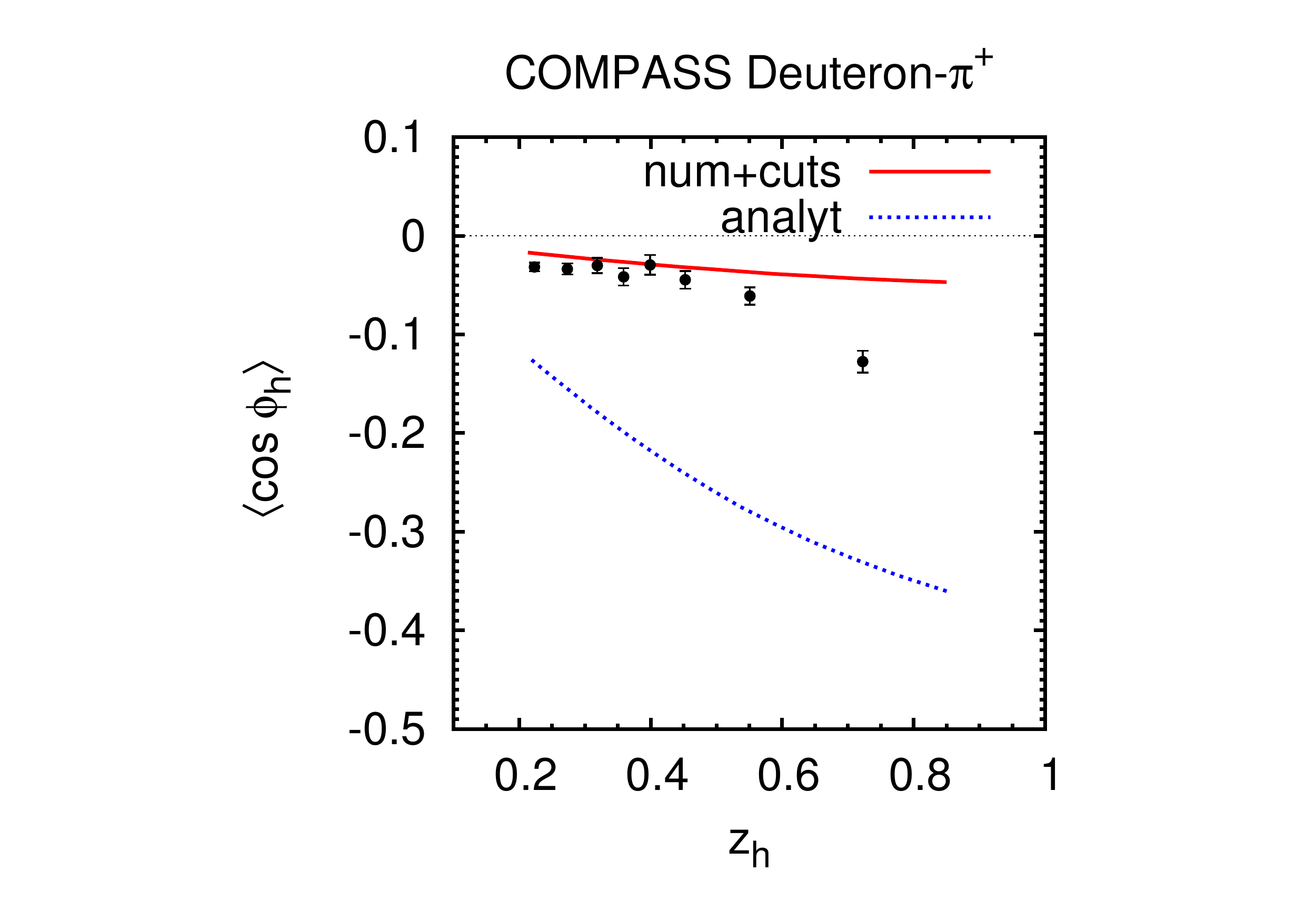}\hspace*{-1.3cm}
\includegraphics[width=0.24\textwidth,angle=-90]
{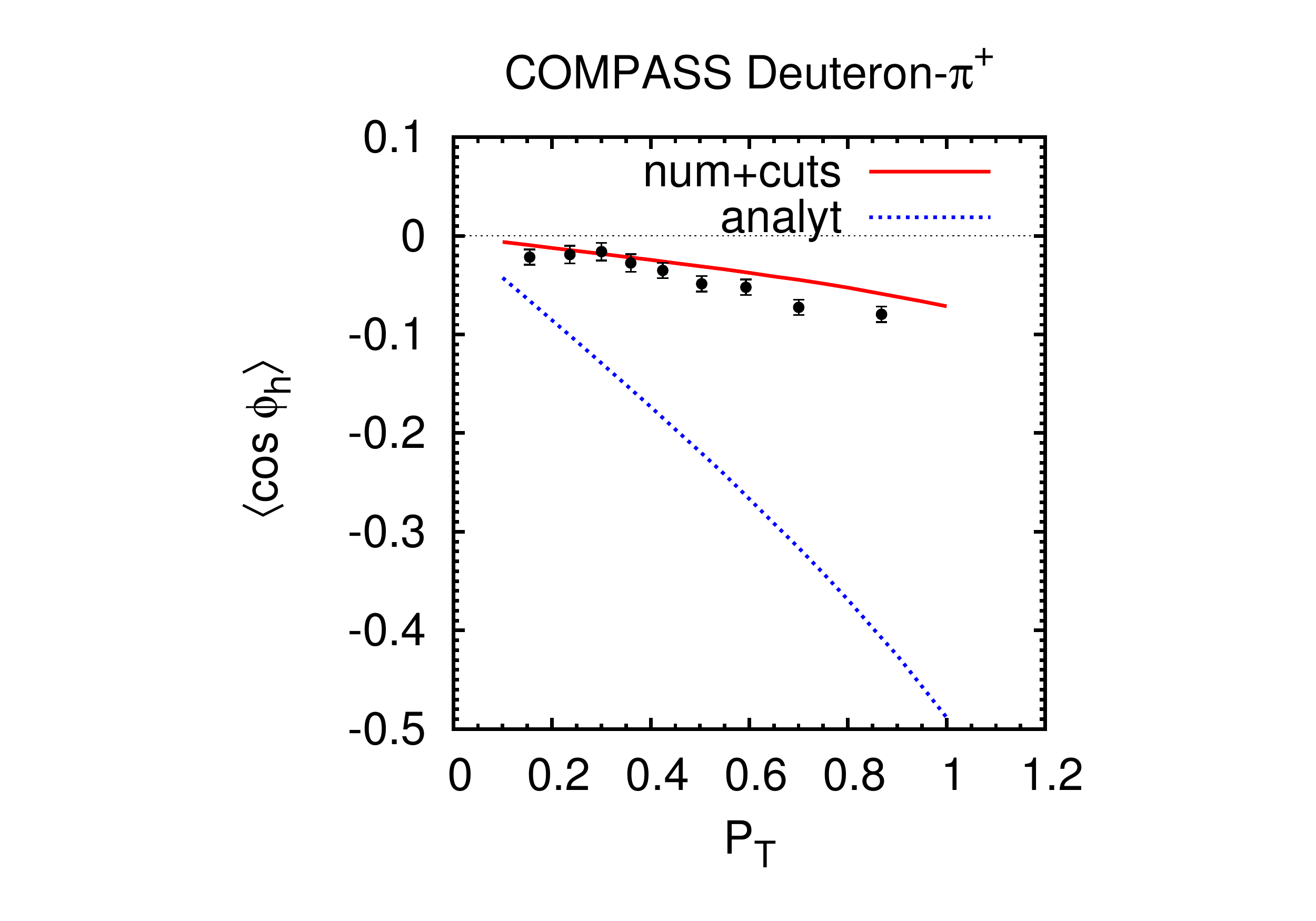}
\\
\includegraphics[width=0.24\textwidth,angle=-90]
{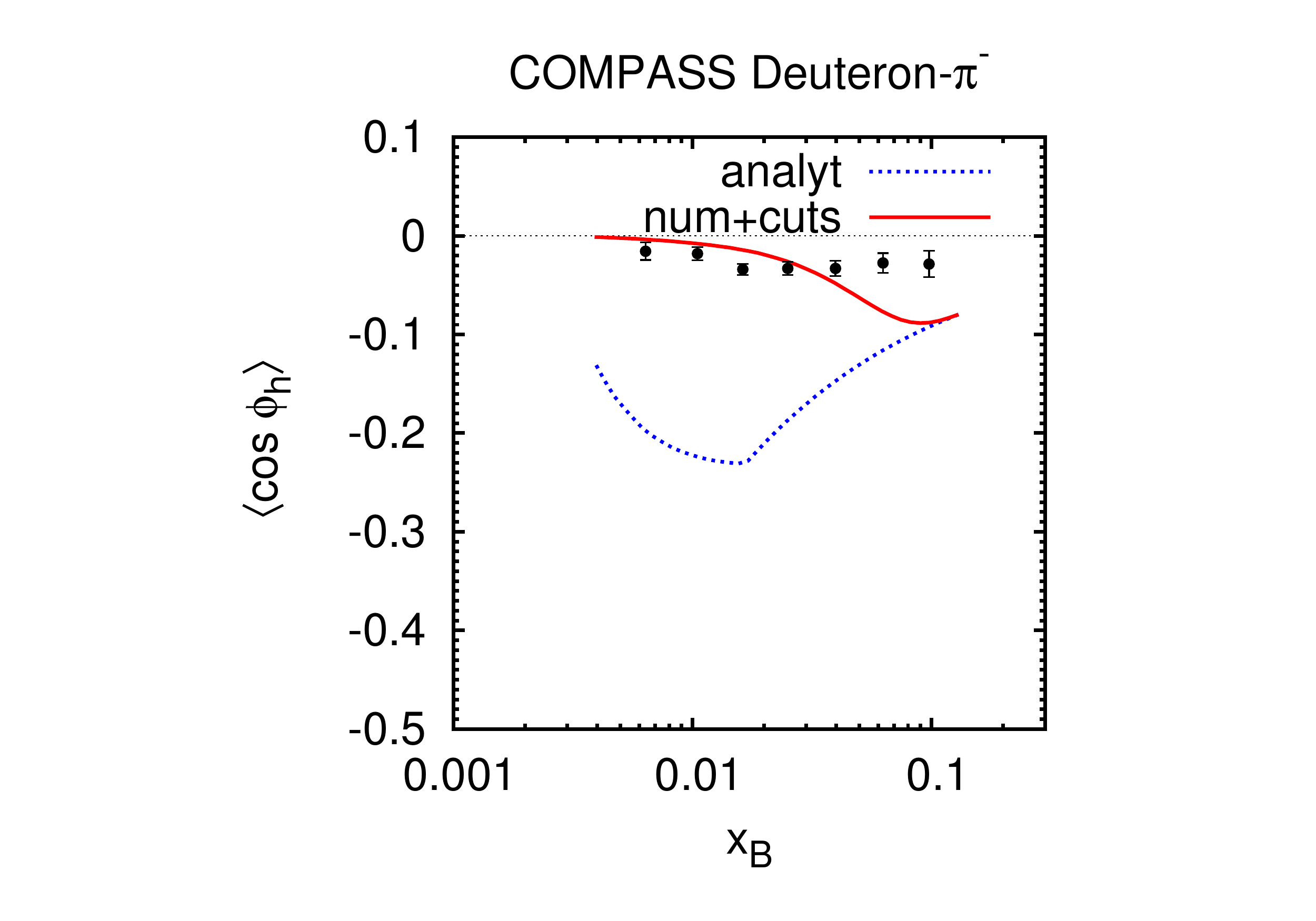}\hspace*{-1.3cm}
\includegraphics[width=0.24\textwidth,angle=-90]
{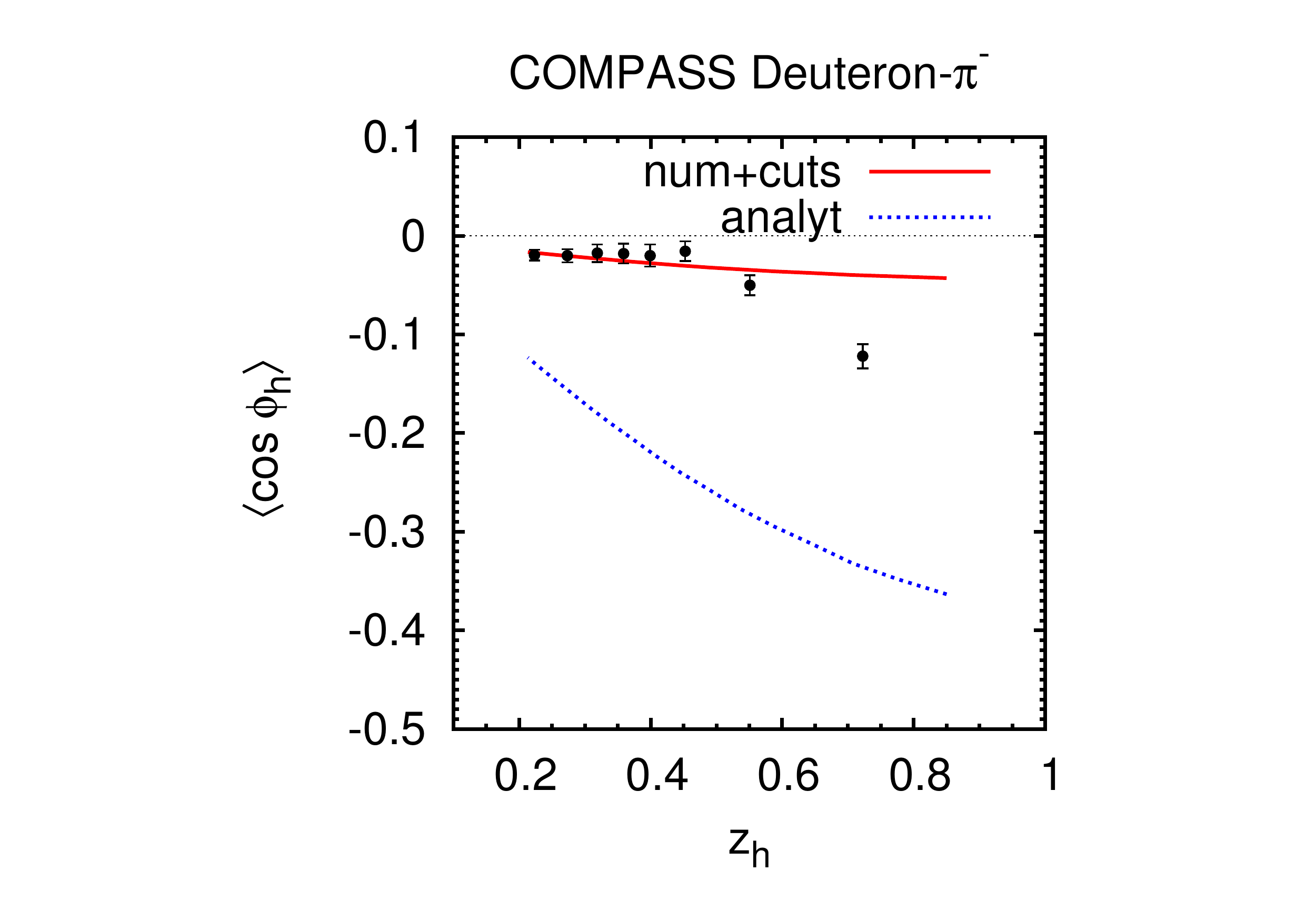}\hspace*{-1.3cm}
\includegraphics[width=0.24\textwidth,angle=-90]
{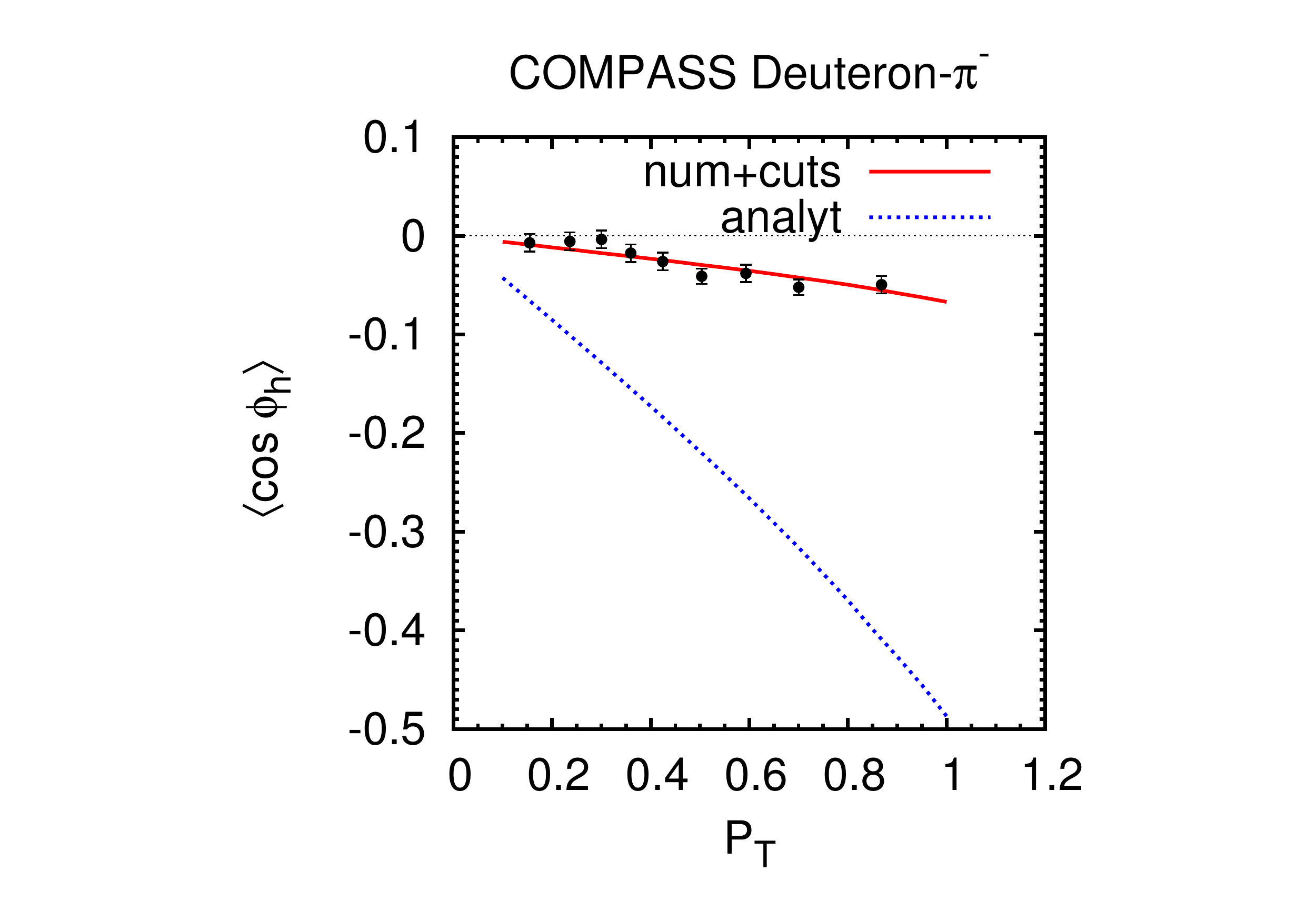}
\caption{\label{hermes-pip-cosphi}
Cahn contribution to $\langle\cos \phi _h\rangle$ for
$\pi ^+$ and $\pi ^-$ production at HERMES and COMPASS kinematics, as a
function of $\xb$ (left plot), $z_h$ (central plot) and $P_T$ (right plot).
The solid (red) line corresponds to $\langle \cos \phi _h\rangle$ calculated
according to Eq.~(\ref{FUUcosphi}) with a numerical $k_\perp$ integration over
the range $[0,k_\perp^{max}]$. 
The dashed (blue) line is $\langle\cos \phi _h\rangle $ calculated according to
Eq.~(\ref{g-FUUcosphi}) obtained by integrating over $k_\perp$ analytically. We
do not show the Boer-Mulders contribution as it is negligible
(see Fig.~\ref{hermes-compass-pip-cosphi-bm}).
The full circles are preliminary experimental data from
HERMES~\cite{Giordano:2010zz} and COMPASS~\cite{Sbrizzai:2009fc}
Collaborations.}
\end{figure}

\begin{figure}[t]
\includegraphics[width=0.24\textwidth,angle=-90]{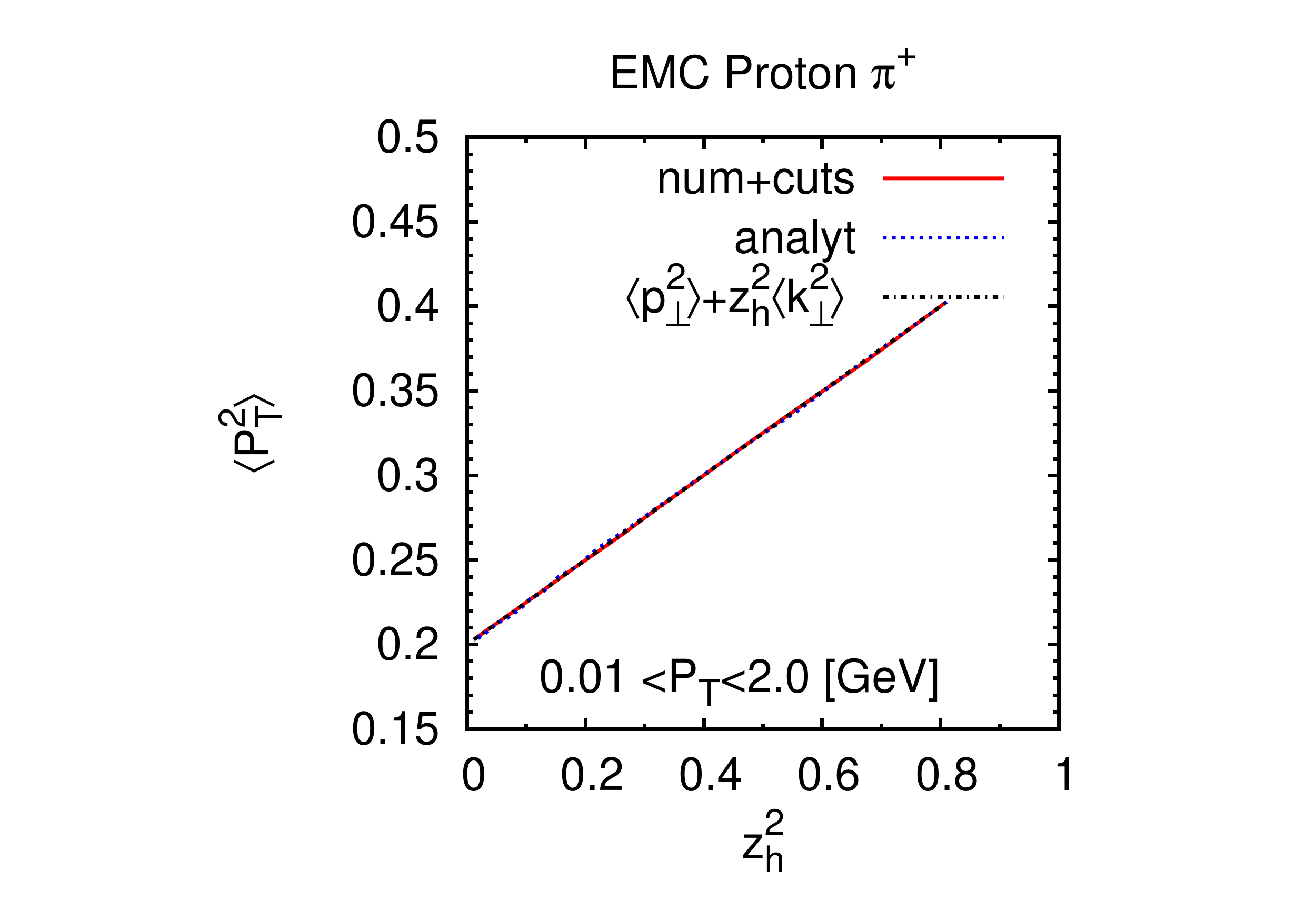}
\hspace*{-1.3cm}
\includegraphics[width=0.24\textwidth,angle=-90]{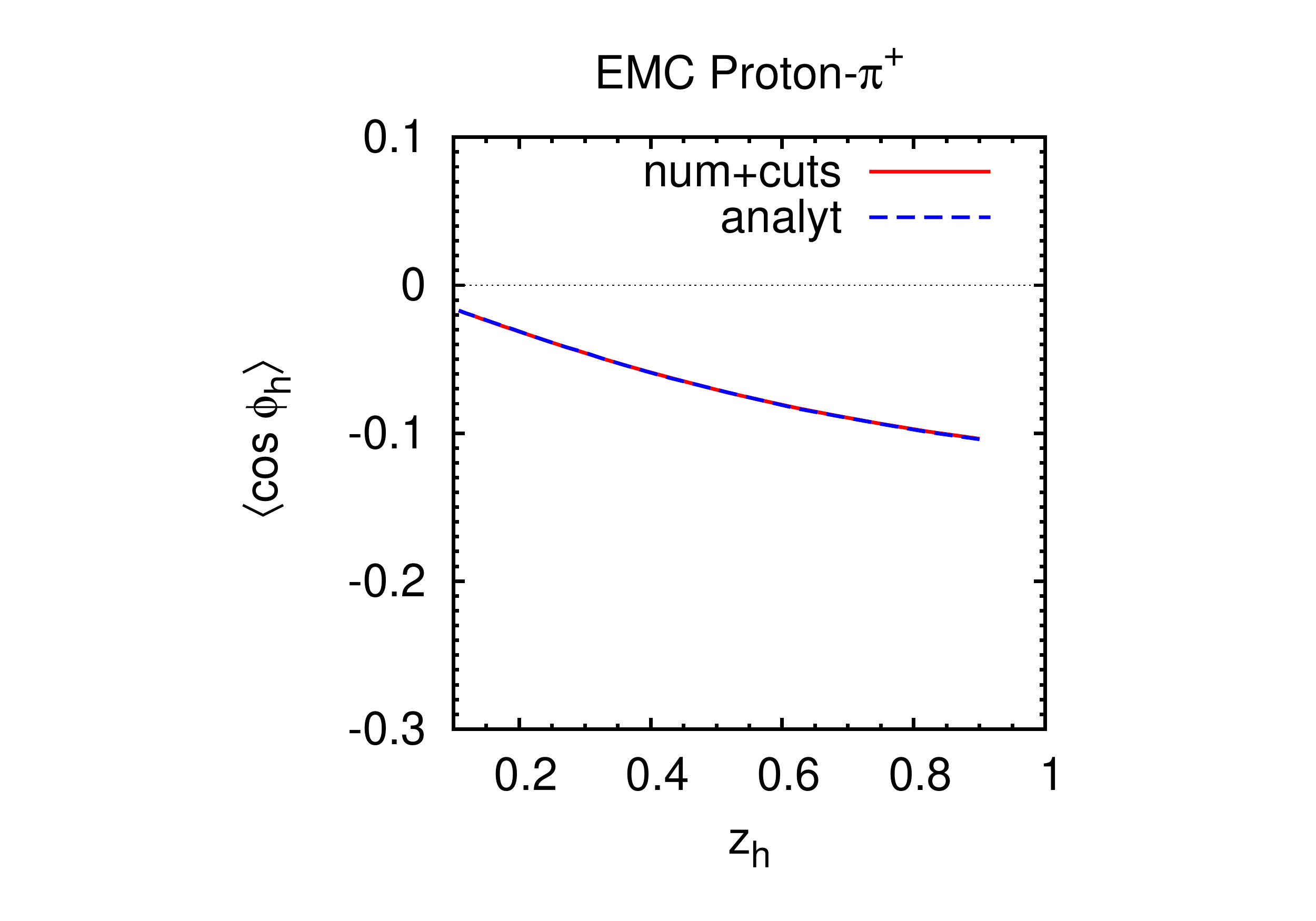}
\hspace*{-1.3cm}
\includegraphics[width=0.24\textwidth,angle=-90]
{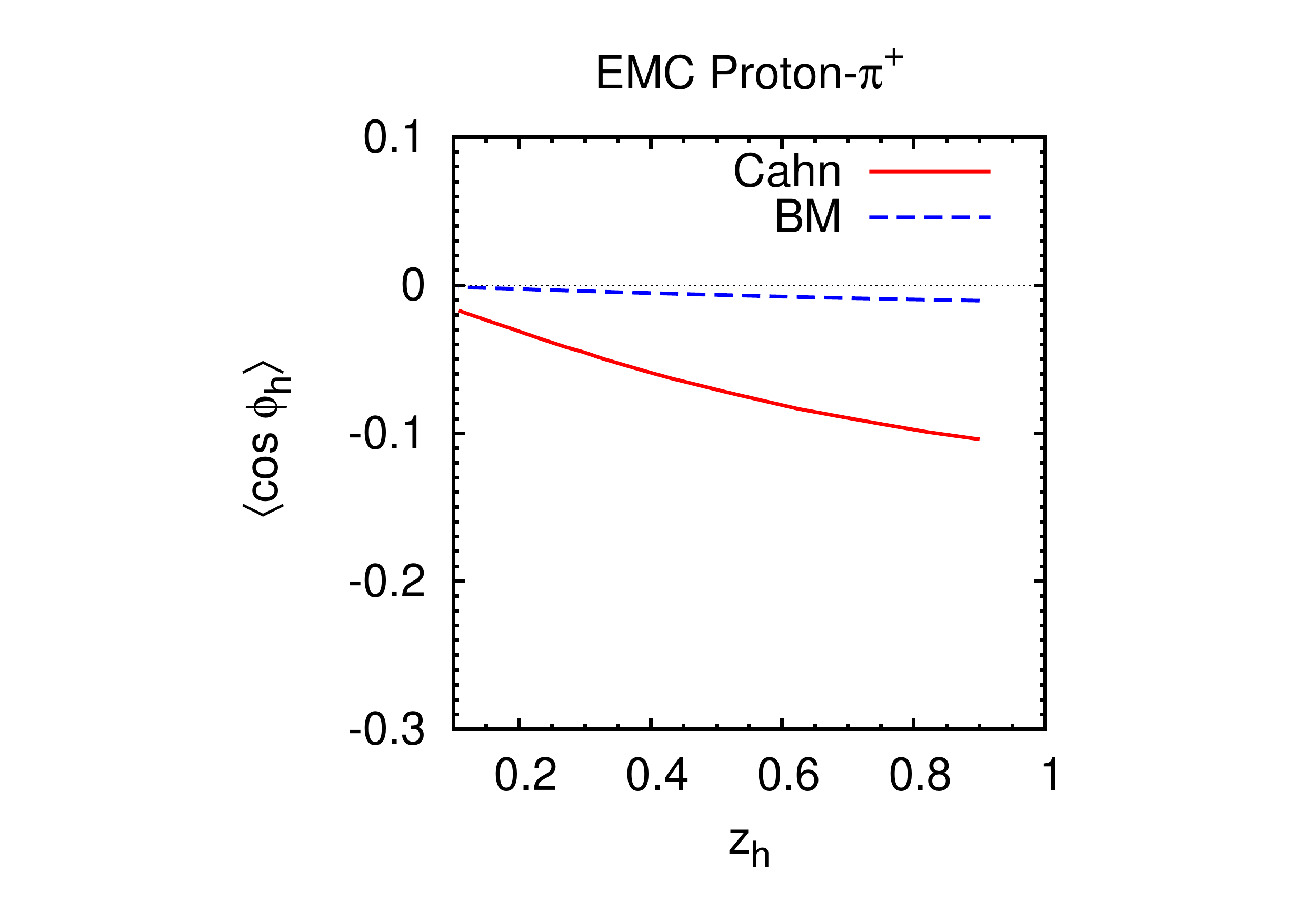}
\caption{\label{emc43-pip-cosphi}
On the left and the central plots, respectively, the $\langle P_T^2 \rangle$
and the $\langle\cos \phi _h\rangle$ azimuthal modulation
for $\pi ^+$ production at EMC.
The solid (red) line corresponds to the calculation performed starting
from Eq.~(\ref{sidis-Xsec-final}) and then integrating it numerically,
implementing Eqs.~\eqref{cutenergy} and \eqref{cutdirection}.
The dashed (blue) line correspond to the analytical calculation.
On the right plot the Cahn (red solid line) and the Boer-Mulder (blu dashed
line) 
contributions to the $\langle\cos \phi _h\rangle$ azimuthal modulation.
}
\end{figure}

Figure~\ref{hermes-pip-cosphi} shows how a large deviation from the analytical
integration results
is obtained by applying the $\kt$ bounds of Eqs.~(\ref{cutenergy}) and
(\ref{cutdirection})
when computing the Cahn effect contribution to $\langle \cos \phi _h\rangle$
corresponding to the HERMES and COMPASS kinematics, whereas for the EMC set up,
see Fig.~\ref{emc43-pip-cosphi}, one can hardly notice any difference. 
The reason is quite obvious: while EMC used high $Q^2 > 5$ GeV$^2$ cuts, HERMES
and COMPASS experiments typically have $Q^2 > 1$ GeV$^2$.

%
%
\begin{figure}[t]
\includegraphics[width=0.24\textwidth,angle=-90]{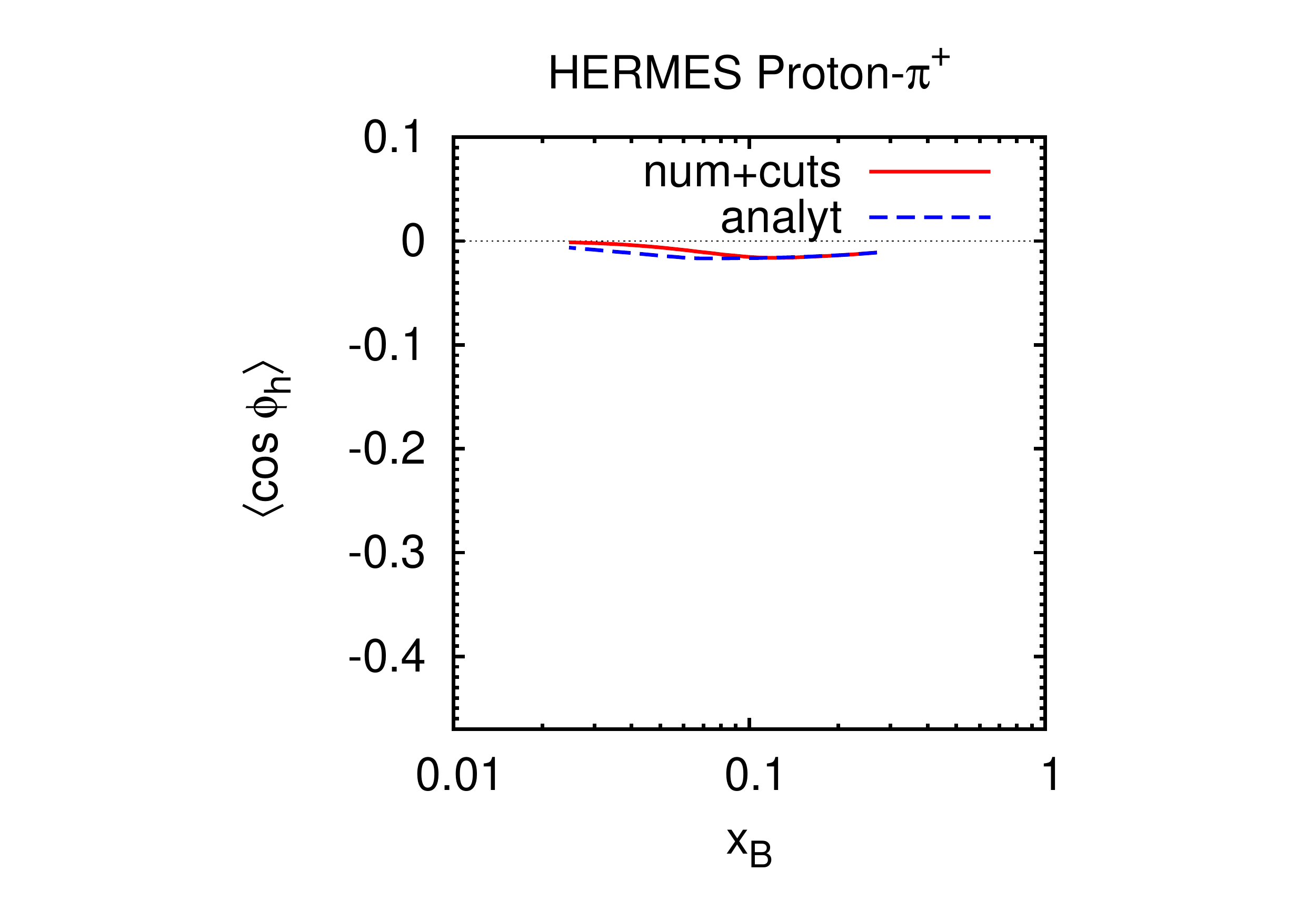}
\hspace*{-1.3cm}
\includegraphics[width=0.24\textwidth,angle=-90]{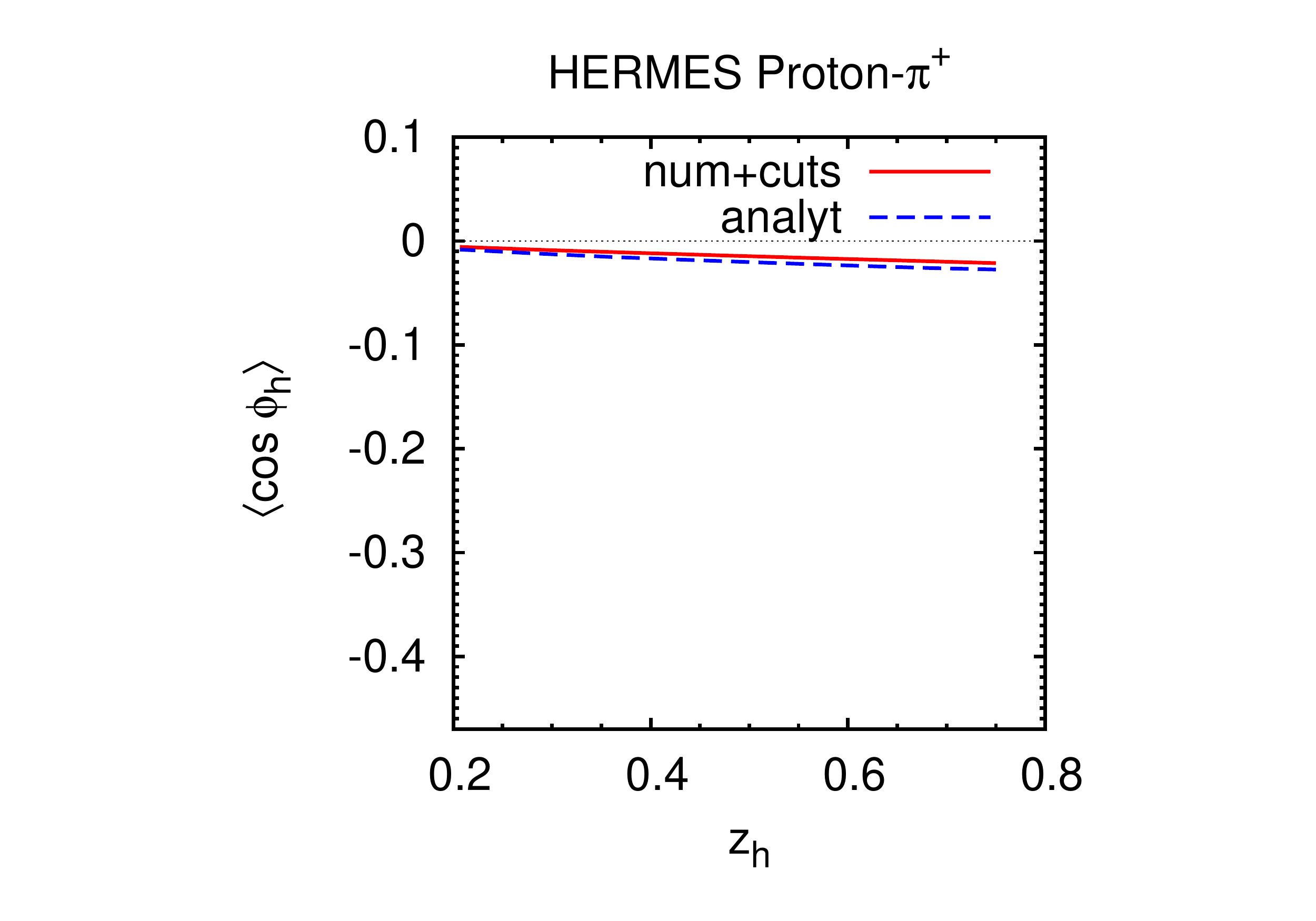}
\hspace*{-1.3cm}
\includegraphics[width=0.24\textwidth,angle=-90]{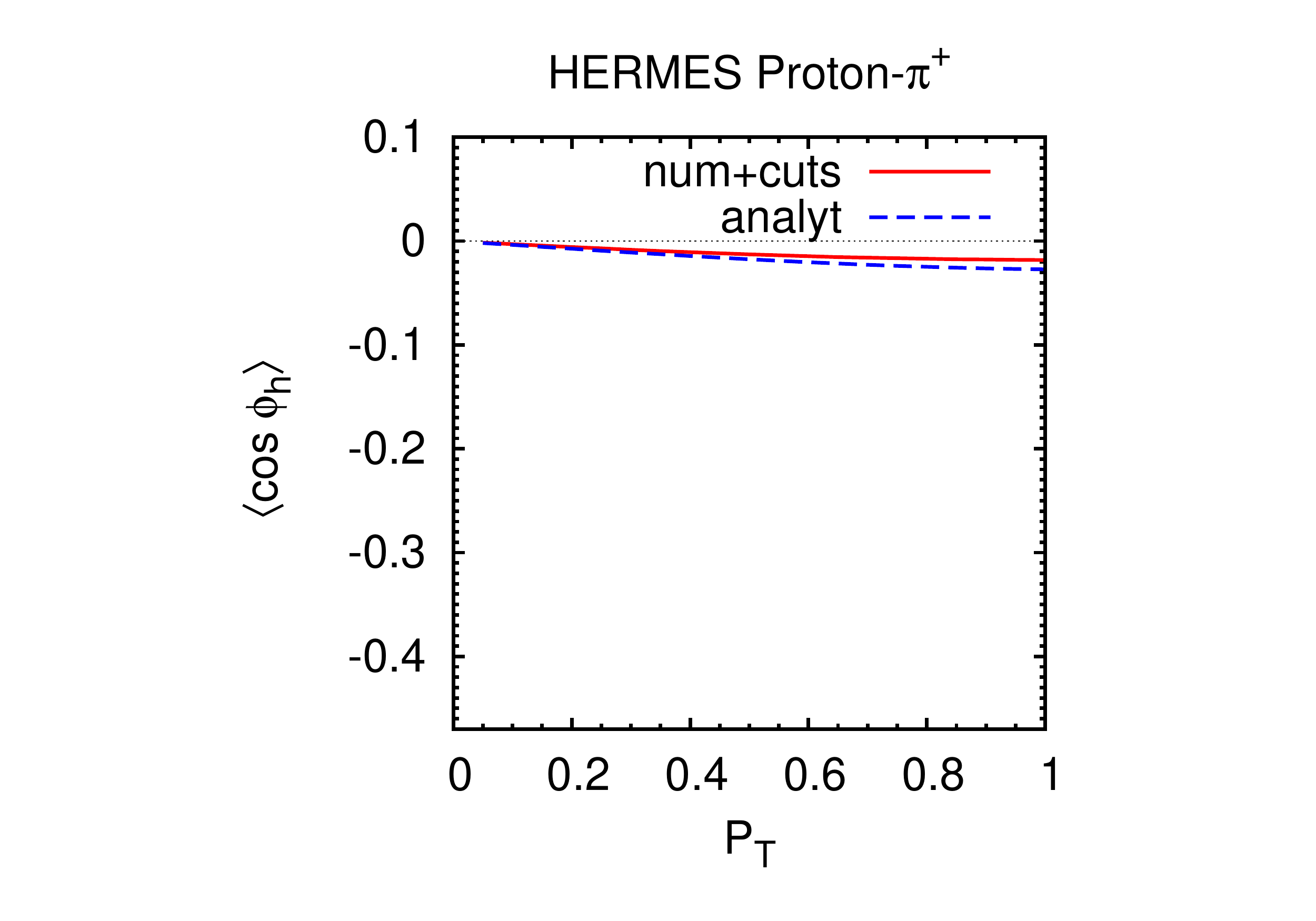}
\\
\includegraphics[width=0.24\textwidth,angle=-90]{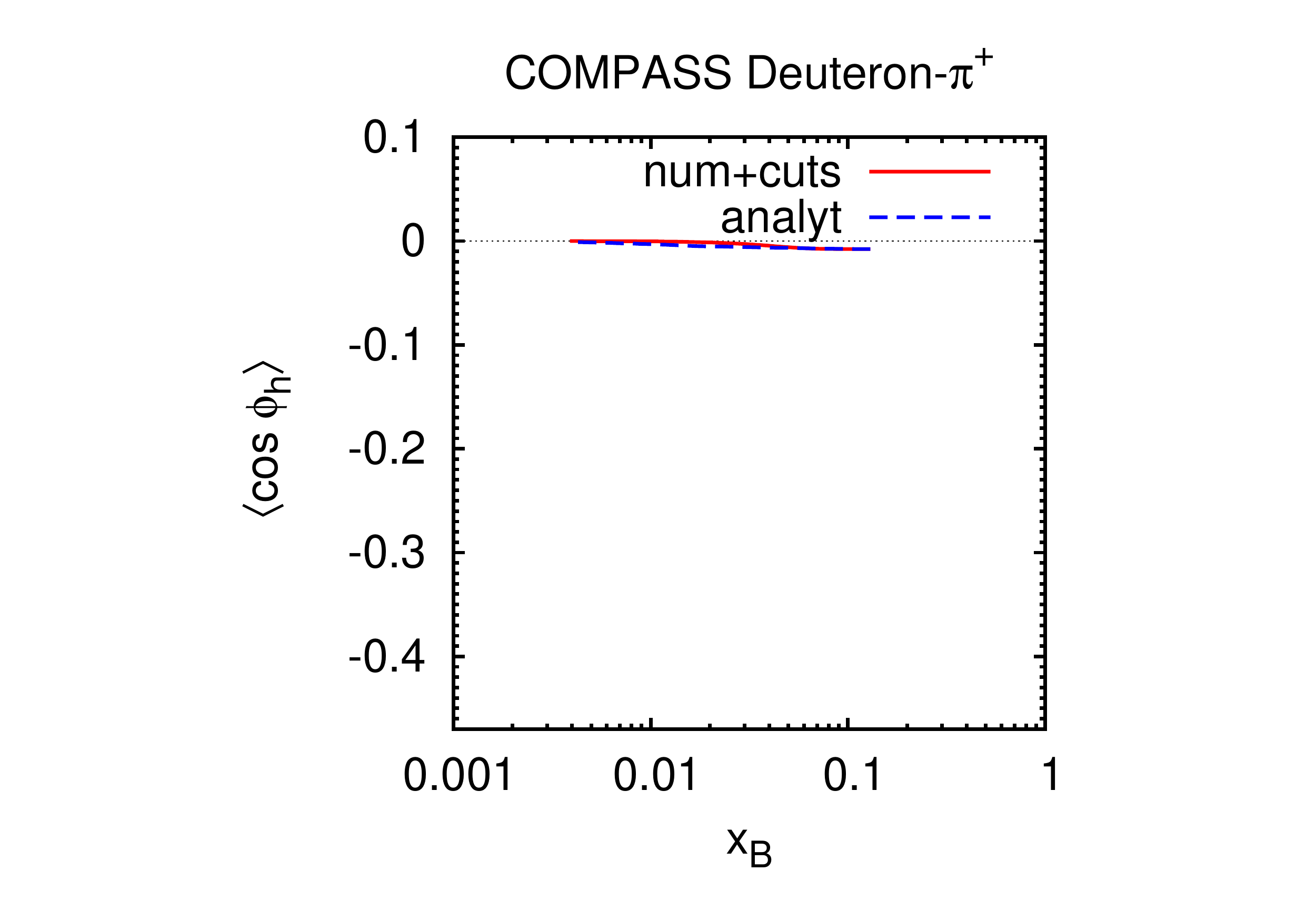}
\hspace*{-1.3cm}
\includegraphics[width=0.24\textwidth,angle=-90]{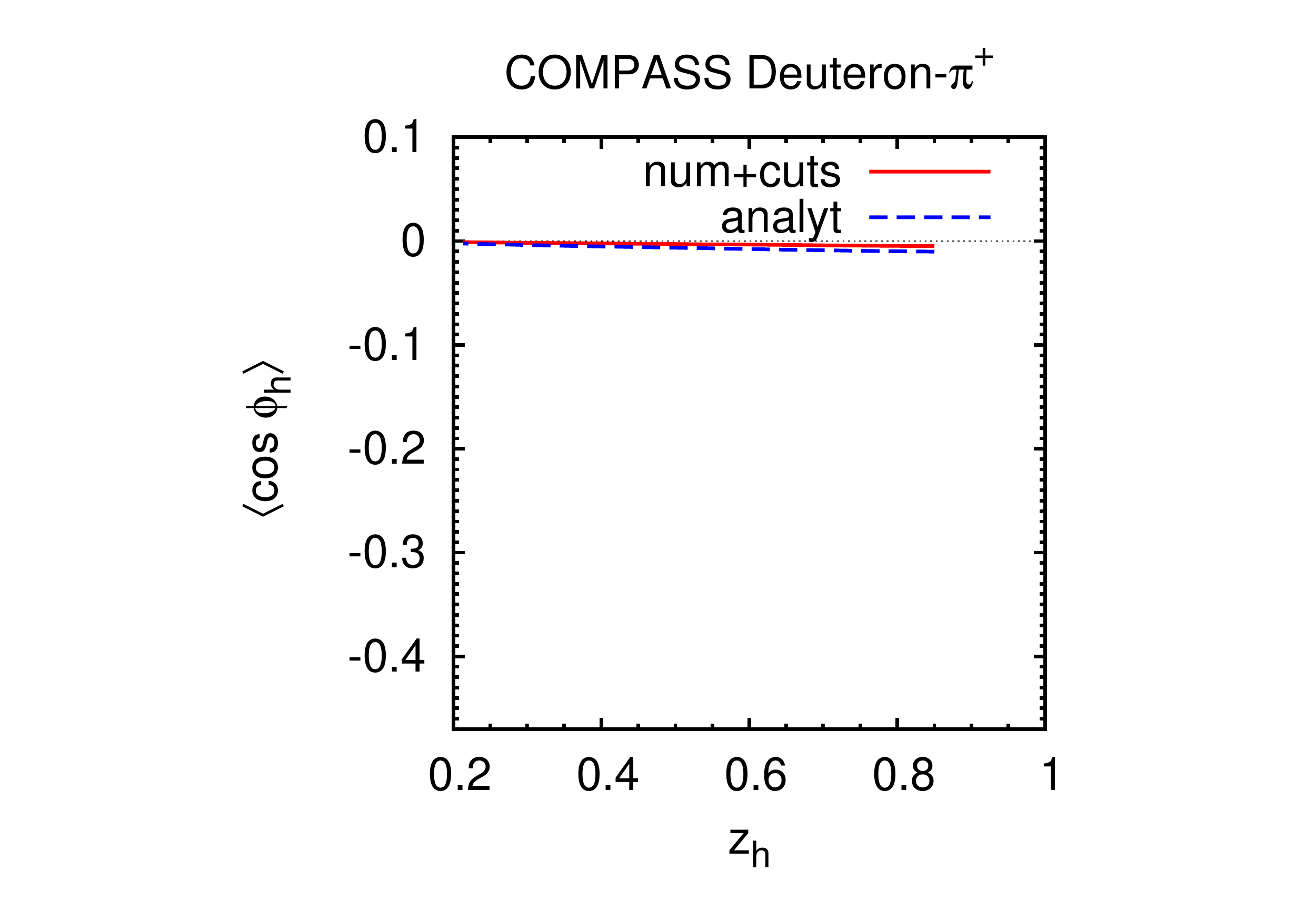}
\hspace*{-1.3cm}
\includegraphics[width=0.24\textwidth,angle=-90]{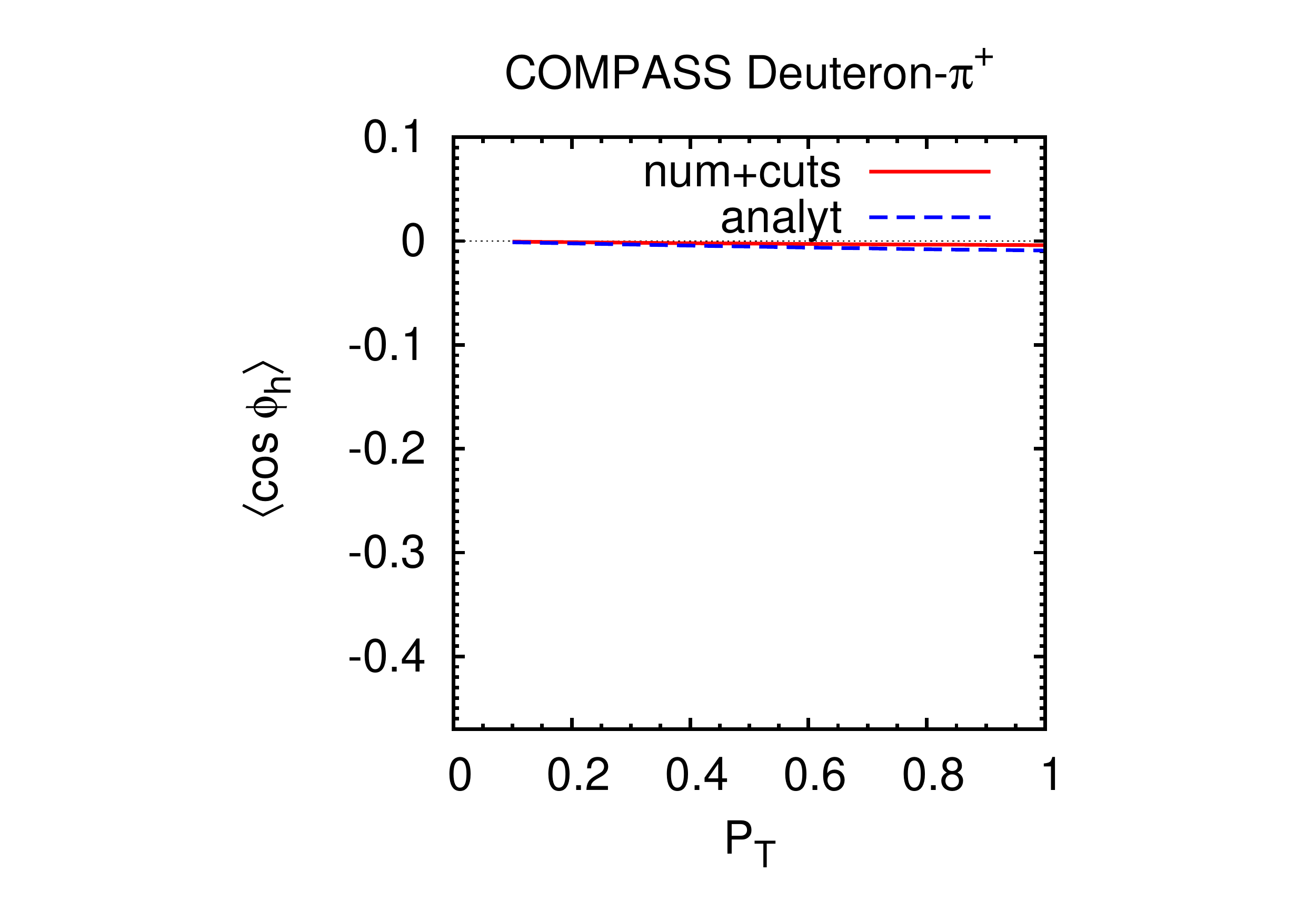}
\caption{Boer-Mulders contribution to the $\langle\cos \phi _h\rangle$ azimuthal
modulation for $\pi ^+$ and 
$\pi ^-$ production at the HERMES and COMPASS kinematics, as a function of $\xb$
(left plot), $z_h$ (central plot) 
and $P_T$ (right plot). The solid (red) line corresponds to $\langle \cos \phi
_h\rangle$ calculated according 
to Eq.~(\ref{FUUcosphi}) with a numerical $k_\perp$ integration over the range
$[0,k_\perp^{max}]$ as given by 
Eqs.~(\ref{cutenergy}) and (\ref{cutdirection}). 
The dashed (blue) line is $\langle\cos \phi _h\rangle $ calculated according to
Eq.~(\ref{g-FUUcosphi}) obtained 
by integrating over $k_\perp$ analytically.\label{hermes-compass-pip-cosphi-bm}}
\end{figure}
%

In Fig.~\ref{hermes-pip-cosphi} our results, obtained with and without $\kt$ -
cuts, are compared to the latest 
HERMES~\cite{Giordano:2010zz} and COMPASS~\cite{Sbrizzai:2009fc} data.
Although still showing a considerable deviation from the experimental data,
our calculation confirms that physical partonic cuts have a quite dramatic
effect in the small $x$ region,
and should therefore be taken into account in any further analysis of these
experimental data.

To evaluate the influence of the partonic cuts on the contribution to  $\langle
\cos \phi_h\rangle$ originating 
from the Boer-Mulders$\otimes$Collins term, 
we use the parametrization of Ref.~\cite{Anselmino:2008jk} for the Collins
function 
while for the Boer-Mulders function we apply the extraction of
Ref~\cite{Barone:2009hw}.
It can be seen from Fig.~\ref{hermes-compass-pip-cosphi-bm} that the
Boer-Mulders contribution
is very tiny (it gives a correction of a few percents over the dominant Cahn
contribution) and is not strongly 
affected by kinematical cuts of Eqs.~\eqref{cutenergy} and \eqref{cutdirection}.

The residual discrepancy between the model prediction and the measurements of
the $\langle \cos \phi_h\rangle$ azimuthal 
moment could indicate that higher twist contributions, from pure twist-3
functions, for example, might be non negligible in this modulation.
More elaborated phenomenological studies including twist-3 TMDs would be
necessary to confirm these observation.

\subsection{Impact of the partonic cuts on the azimuthal moment $\langle \cos
2\phi_h\rangle$ \label{rescos2}}
 
The $\langle\cos 2\phi\rangle$ azimuthal modulation, at twist-2, consists of a term proportional to the Boer-Mulders$\otimes$ Collins, see Eq.~(\ref{FUUcos2phi}). Although it is not affected by any
twist-three corrections, in kinematical ranges where $Q^2$ is not very
large compared to the average $\kt^2$, twist-4 contributions cannot be
neglected.  In particular, a twist-4 ``Cahn-like'' effect actually gives a large
contribution to the $\langle\cos 2\phi\rangle$ azimuthal moment, as was
explained in details in Ref.~\cite{Barone:2009hw}. 
This provides an additional term to the $F_{UU}^{\cos 2 \phi _h}$ structure
function, Eq.~(\ref{FUUcos2phi}), of the form
\be
2\sum_{q} e_q^2 \, \int d^2\bkt \!
\frac{\kt ^2}{Q^2}\left[2(\hat{\bfP}_T \cdot \hat{\bfk}_\perp)^2 - 1
\right] f_{q/p}(x, \kt)\,D_{h/q} (z,\pp) \,.
\label{twist4}
\ee

In Ref.~\cite{Barone:2009hw}, the presence of a twist-4 term proved to be
crucial to understand the available experimental data from
HERMES~\cite{Lamb:2009zza, Giordano:2009hi} and
COMPASS~\cite{Kafer:2008ud,Bressan:2009eu} results.
Here, a detailed study  inspired by
the fact that the HERMES $P_T$ spectrum can be reproduced
by Monte Carlo calculations with $\langle \kt ^2 \rangle =0.18$ GeV$^2$, pointed
out that slightly different values of TMD widths might be required for different 
experiments. 

\begin{figure}[t]
\includegraphics[width=0.24\textwidth,angle=-90]{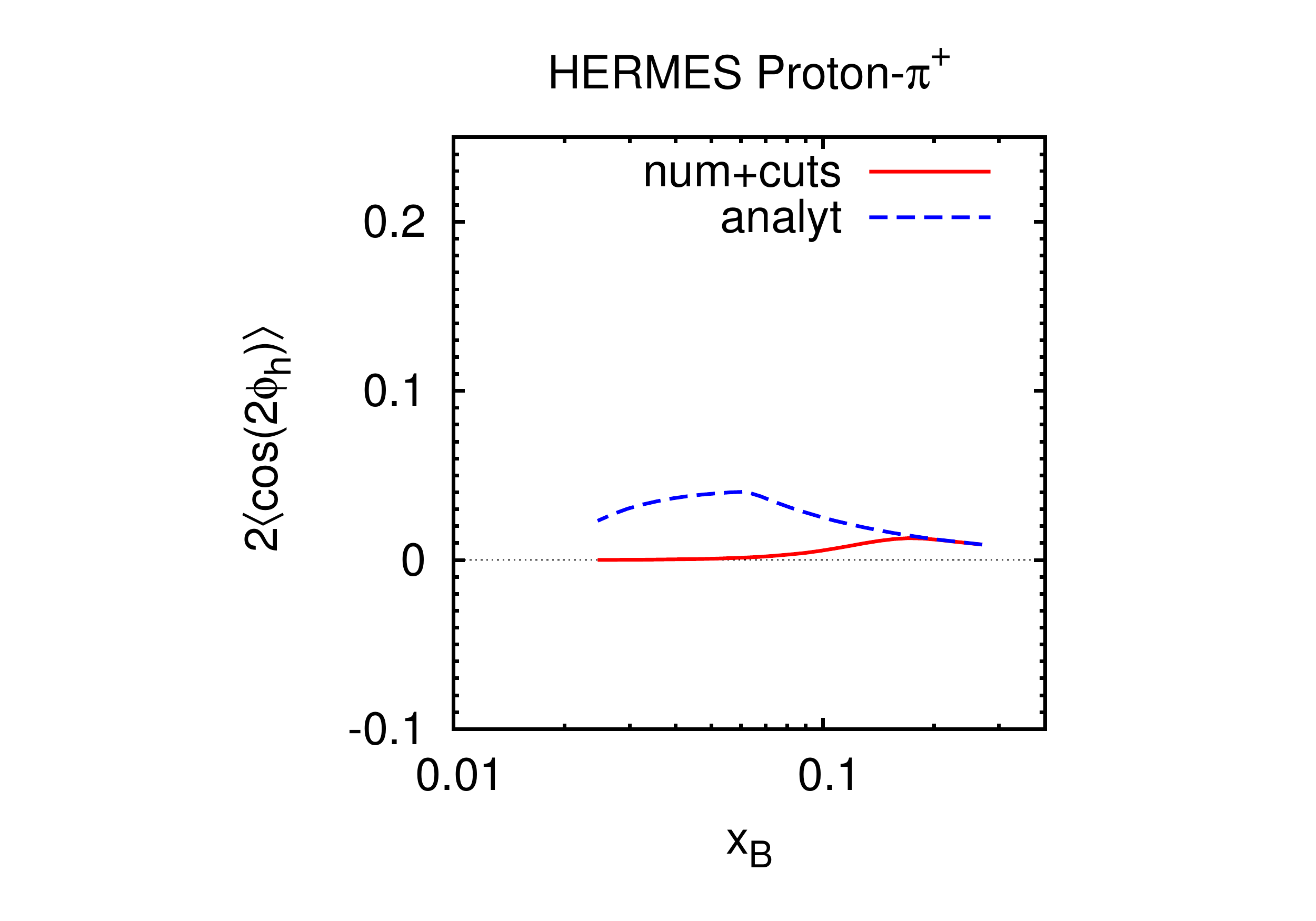}
\hspace*{-1.3cm}
\includegraphics[width=0.24\textwidth,angle=-90]{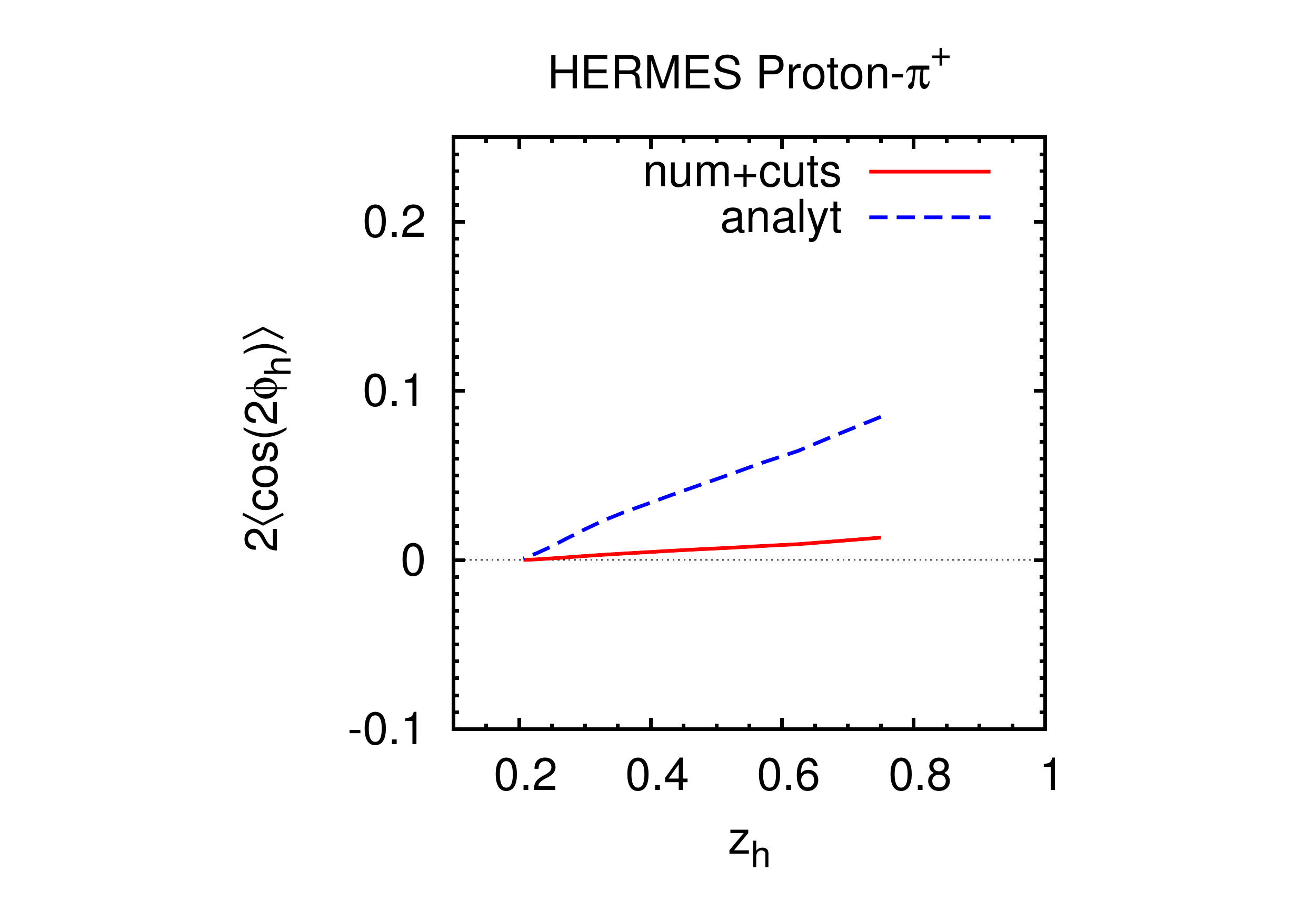}
\hspace*{-1.3cm}
\includegraphics[width=0.24\textwidth,angle=-90]{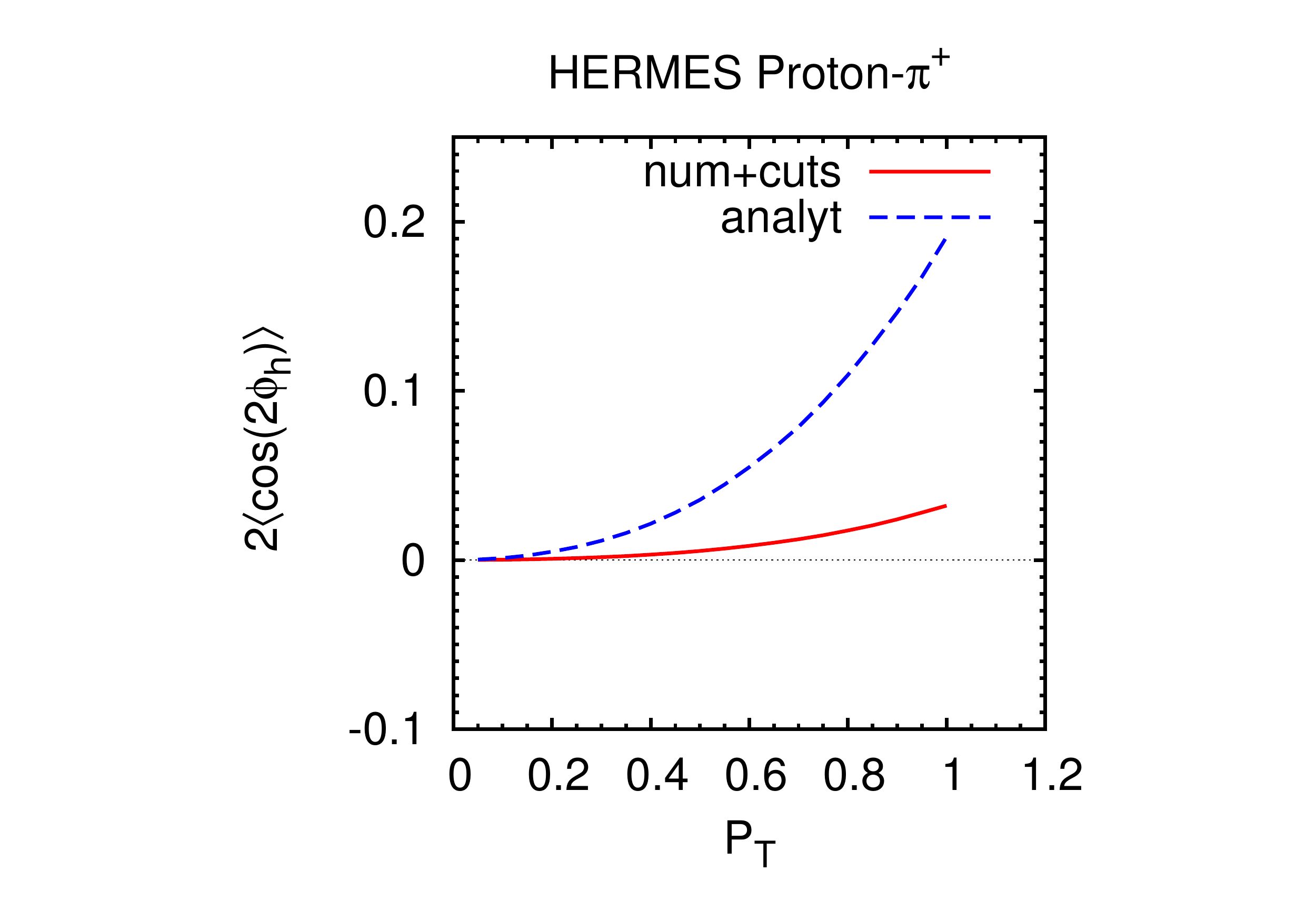}
\\
\includegraphics[width=0.24\textwidth,angle=-90]{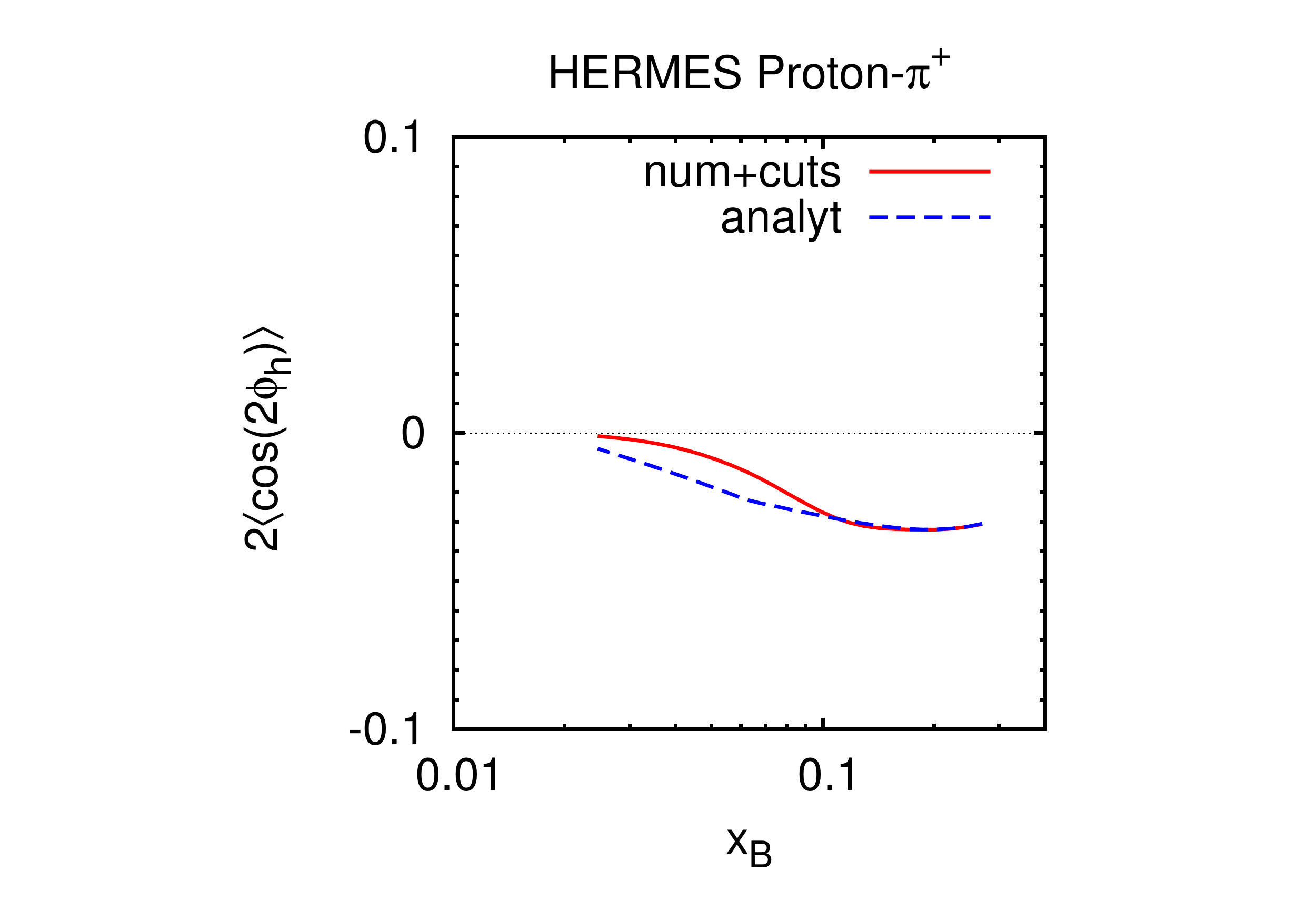}
\hspace*{-1.3cm}
\includegraphics[width=0.24\textwidth,angle=-90]{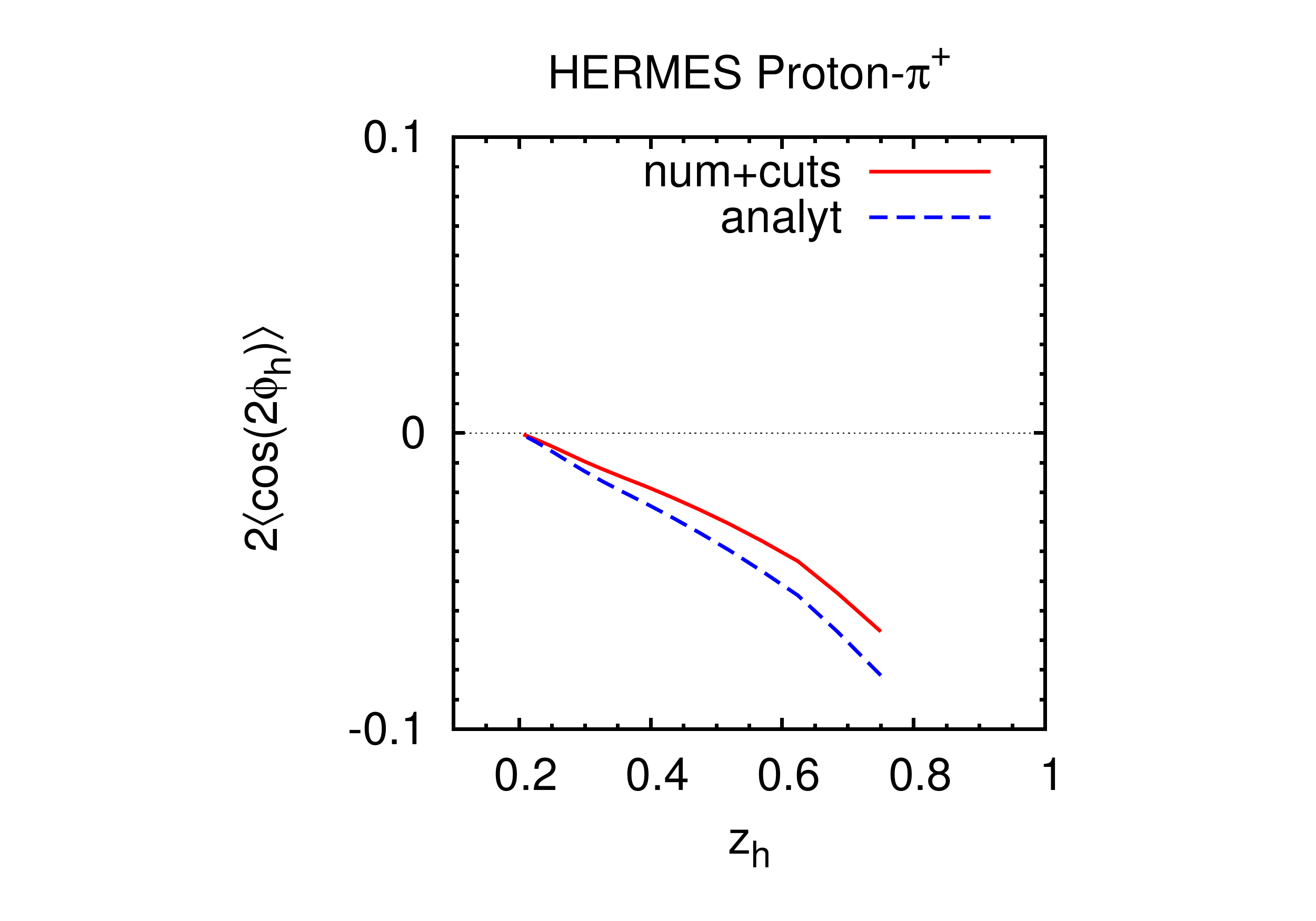}
\hspace*{-1.3cm}
\includegraphics[width=0.24\textwidth,angle=-90]{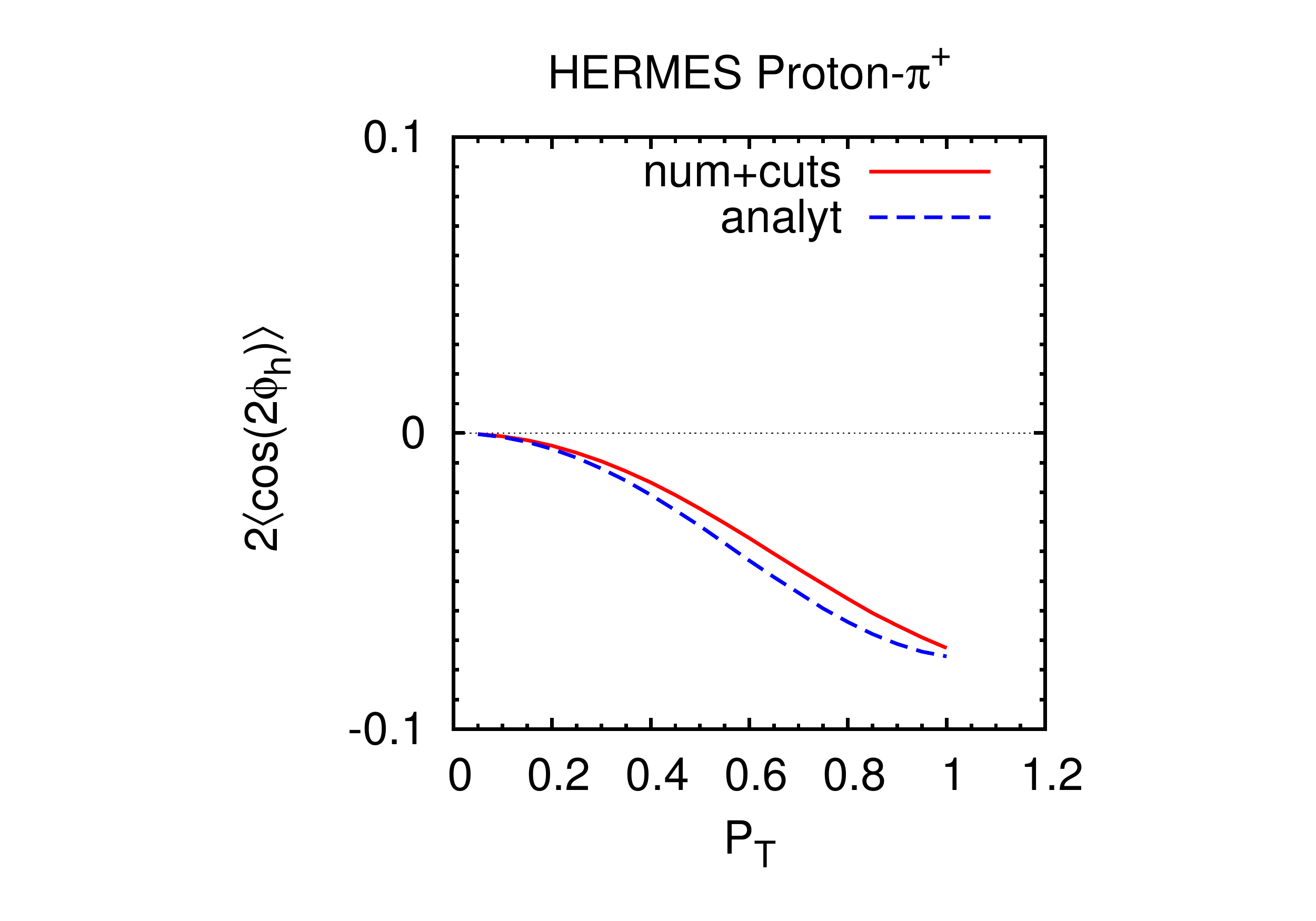}
\caption{\label{c2phi-hermes-ch-bm}
In the upper panels we show the twist-4 Cahn contribution to the $\langle\cos 2
\phi _h\rangle$ azimuthal modulation for $\pi ^+$ production at HERMES as a
function of $\xb$, $z_h$ and $P_T$, in the lower panels the twist-2 Boer-Mulders
contribution to the $\langle\cos 2 \phi _h\rangle$ azimuthal modulation for $\pi
^+$ production at HERMES, again as a function of $\xb$ (left plot), $z_h$
(central plot) and $P_T$ (right plot).
Ref.~\cite{Giordano:2009hi}.
}
\end{figure}

\begin{figure}[t]
\includegraphics[width=0.24\textwidth,angle=-90]{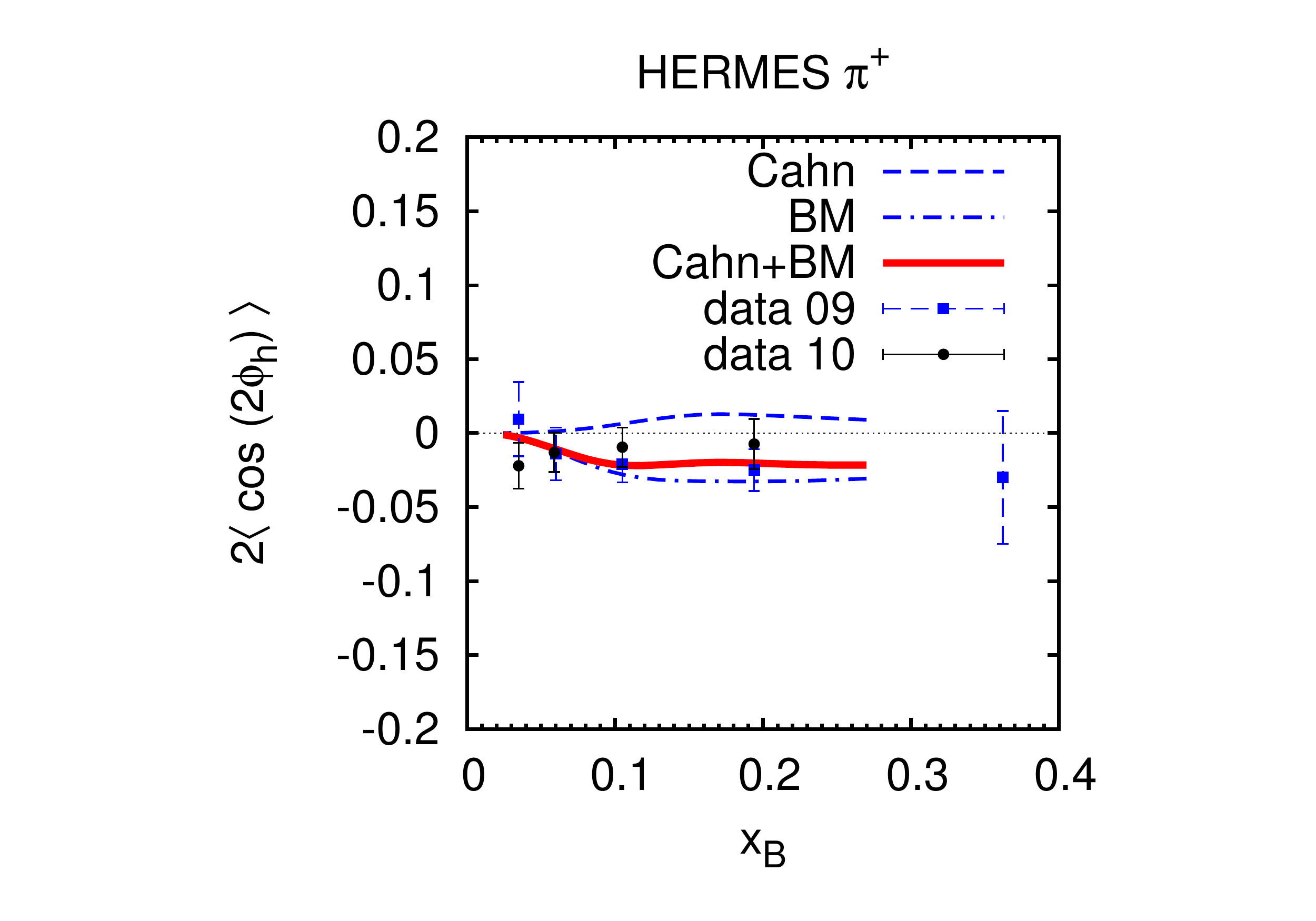}\hspace*
{-1.3cm}
\includegraphics[width=0.24\textwidth,angle=-90]{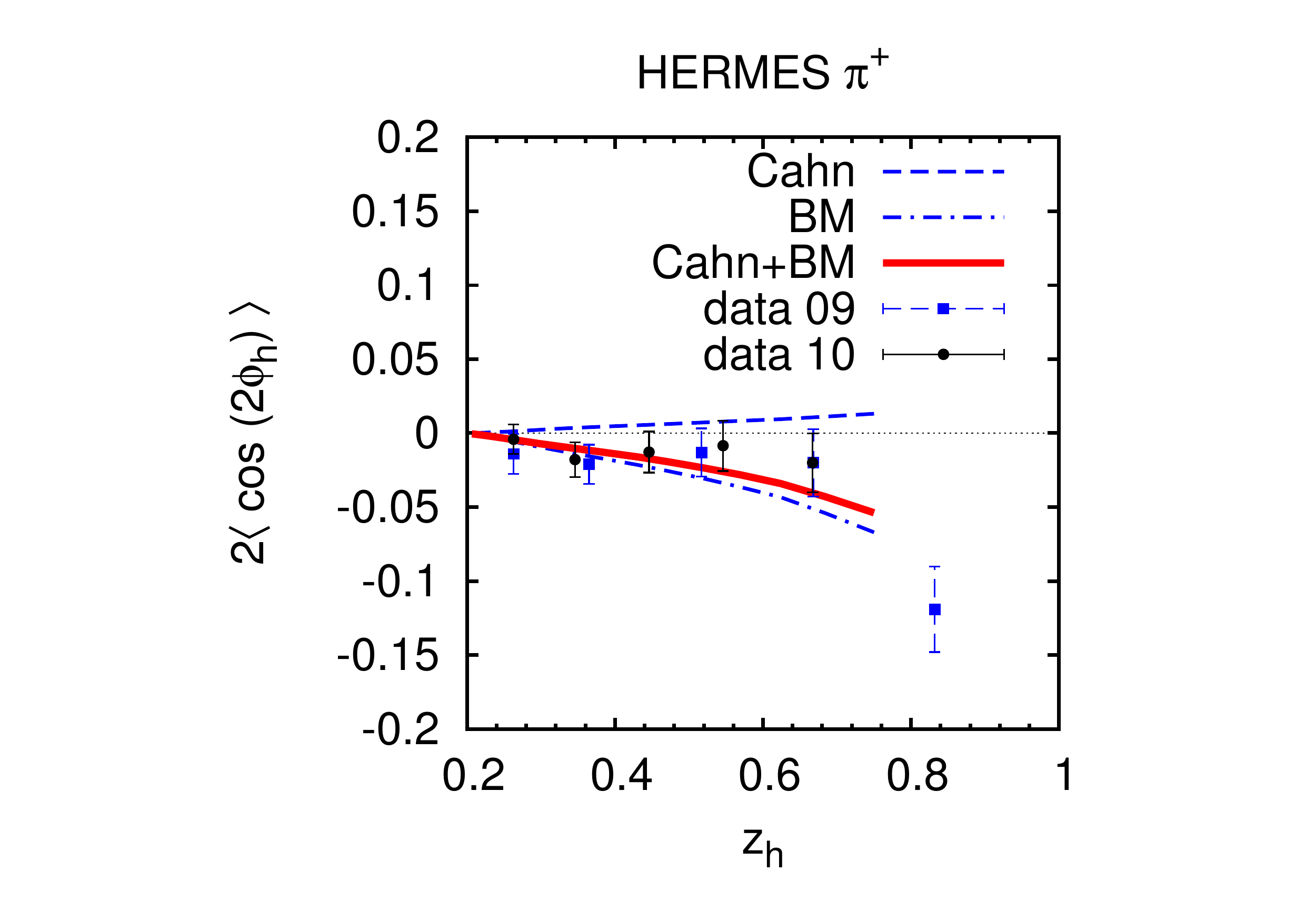}\hspace*
{-1.3cm}
\includegraphics[width=0.24\textwidth,angle=-90]{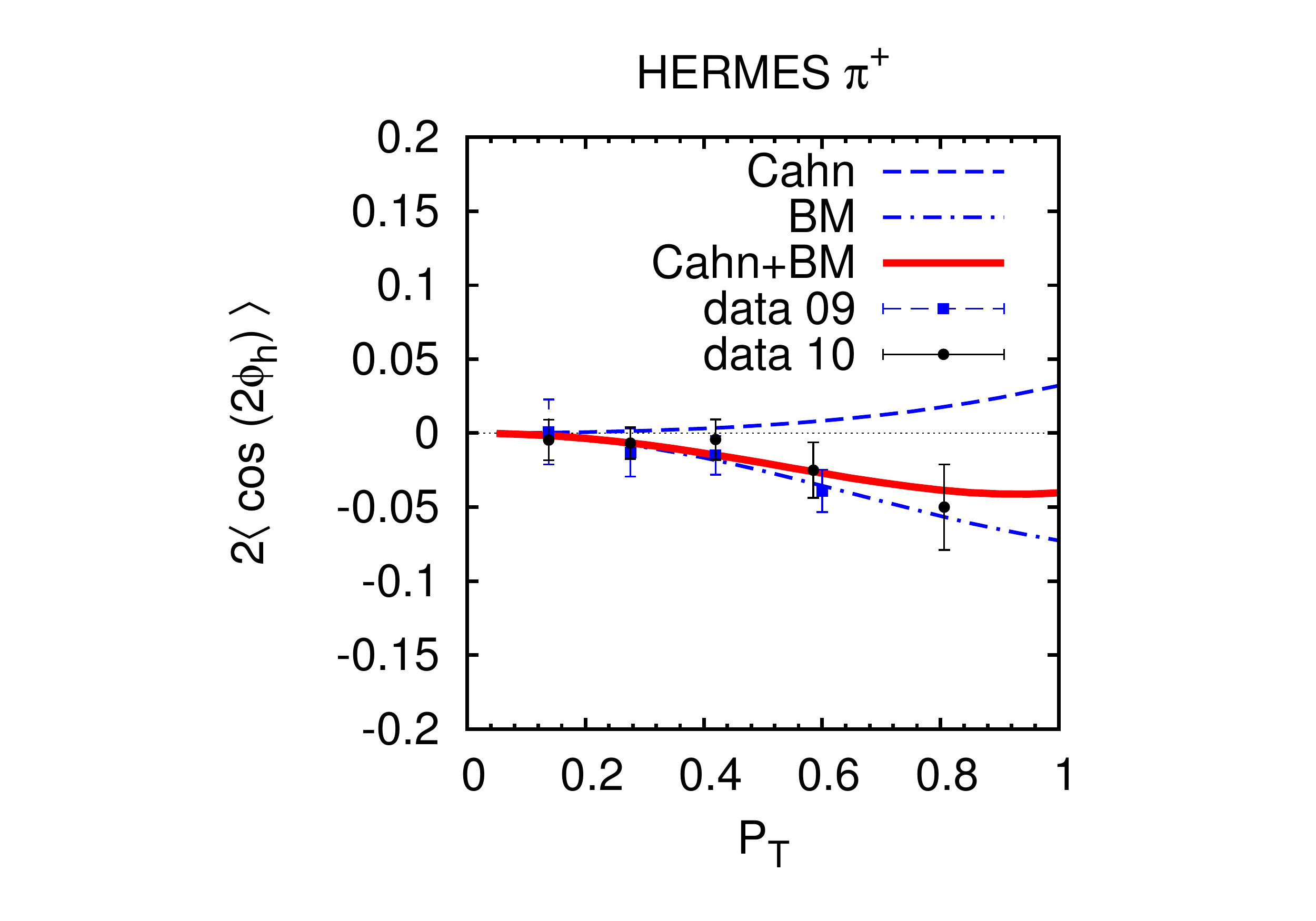}
\\
\includegraphics[width=0.24\textwidth,angle=-90]{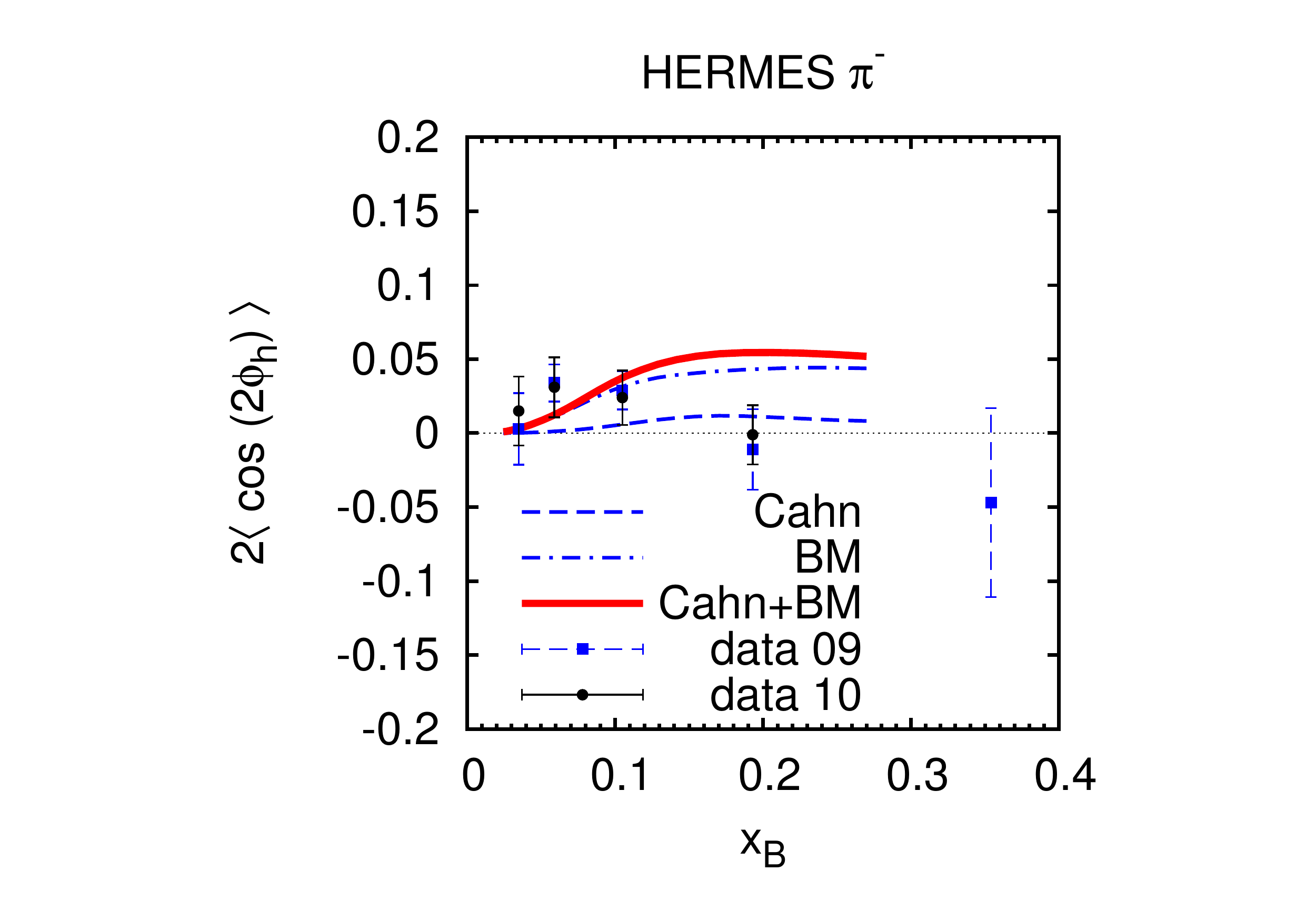}\hspace*
{-1.3cm}
\includegraphics[width=0.24\textwidth,angle=-90]{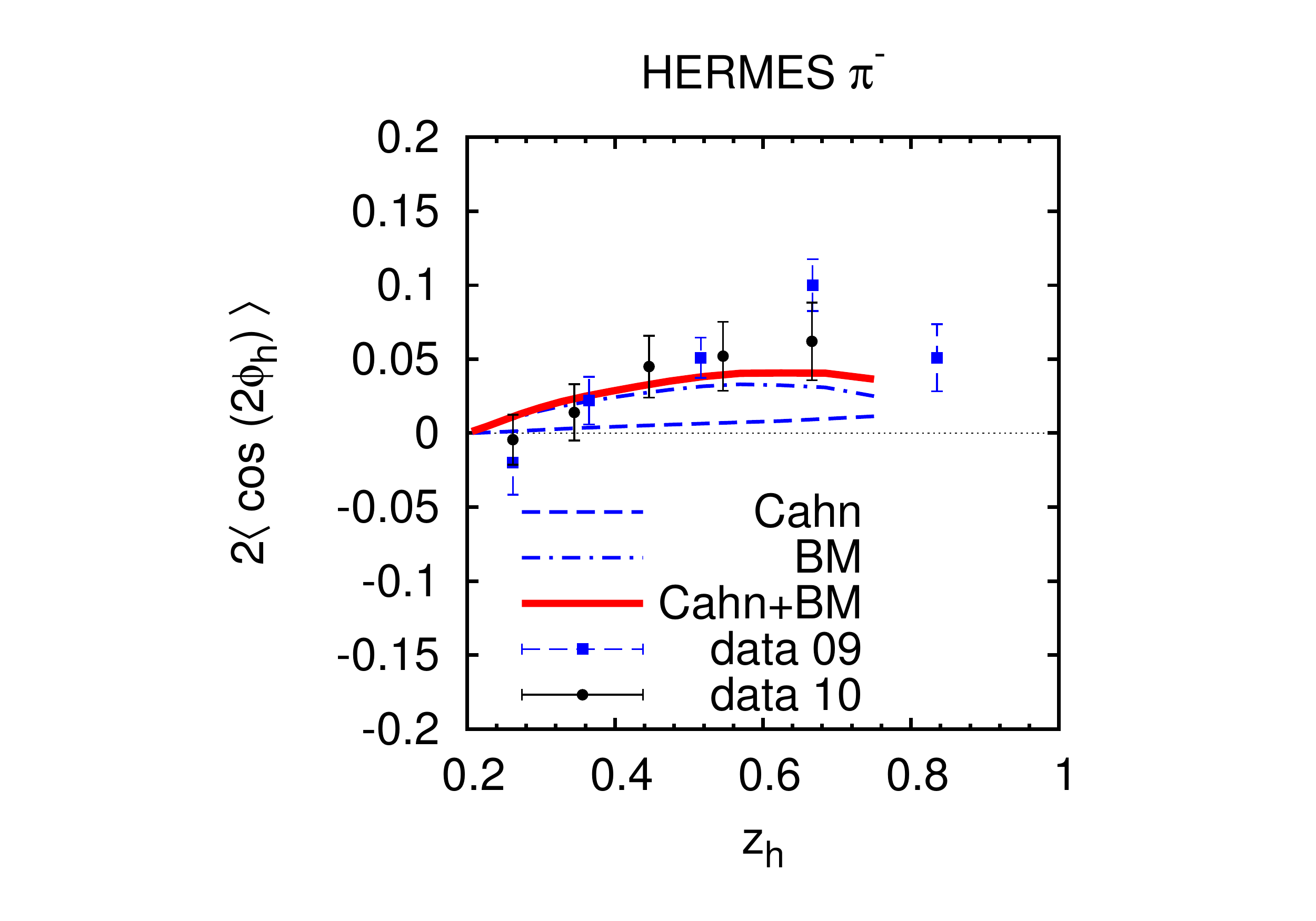}\hspace*
{-1.3cm}
\includegraphics[width=0.24\textwidth,angle=-90]{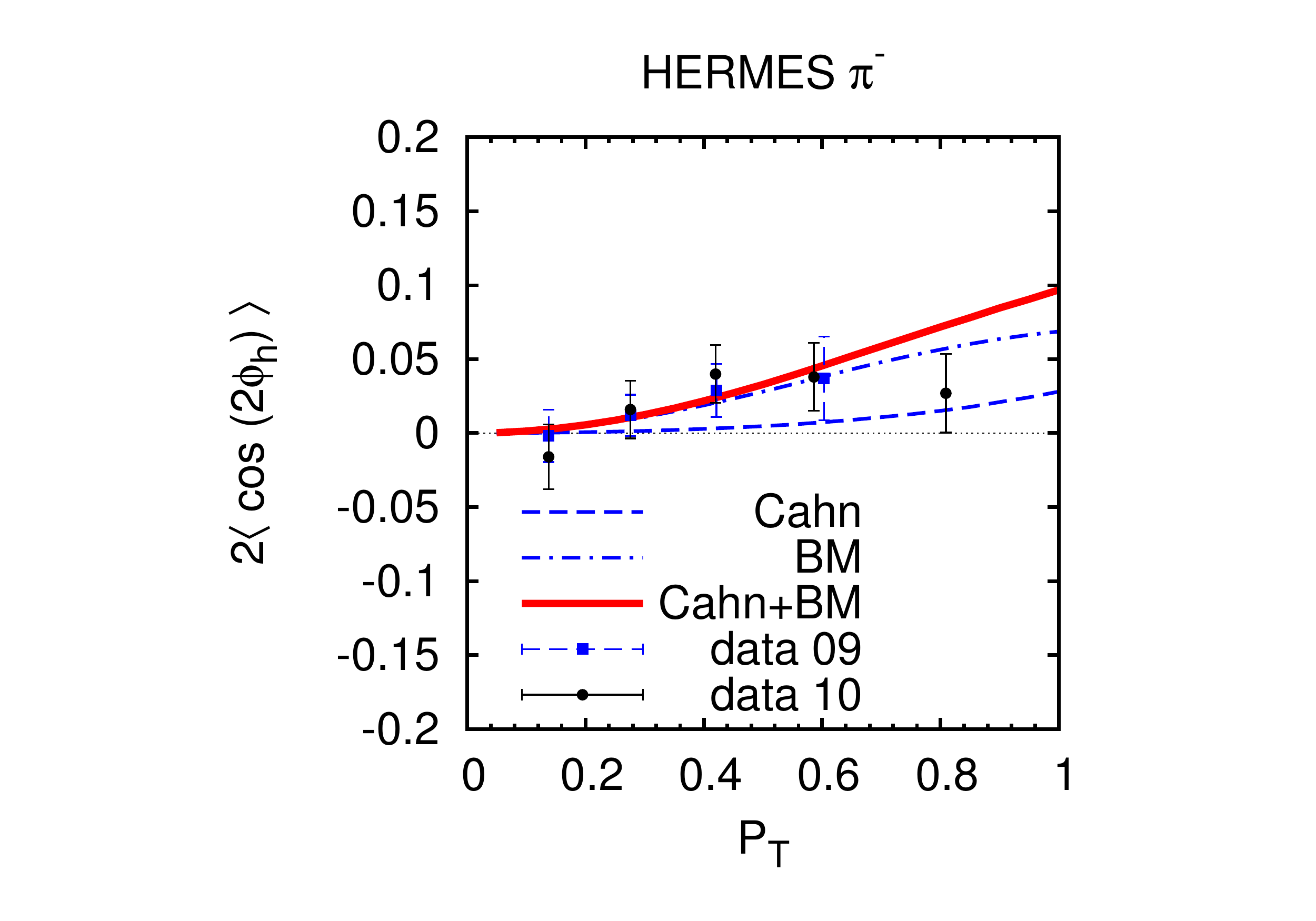}
\caption{\label{c2phi-hermes}
Boer-Mulders and Cahn contributions to the $\langle\cos 2 \phi _h\rangle$
azimuthal modulation for $\pi ^+$ (upper panel) and $\pi ^-$ (lower panel)
production at HERMES as a function of $\xb$ (left plot), $z_h$ (central plot)
and $P_T$ (right plot). Experimental data are from
Refs.~\cite{Lamb:2009zza,Giordano:2009hi,Giordano:2010zz}.}
\end{figure}

\begin{figure}[t]
\includegraphics[width=0.24\textwidth,angle=-90]{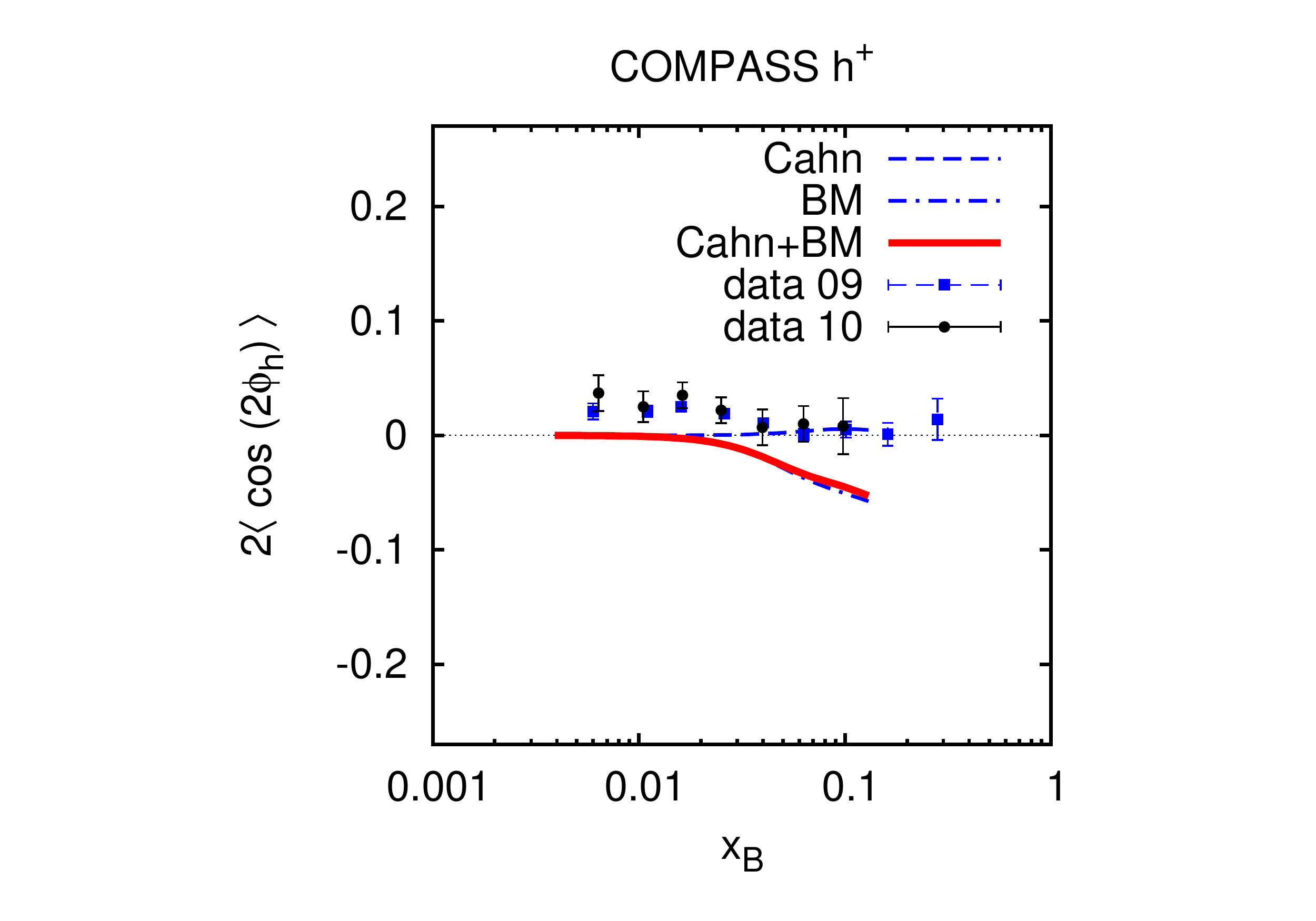}
\hspace*{-1.3cm}
\includegraphics[width=0.24\textwidth,angle=-90]{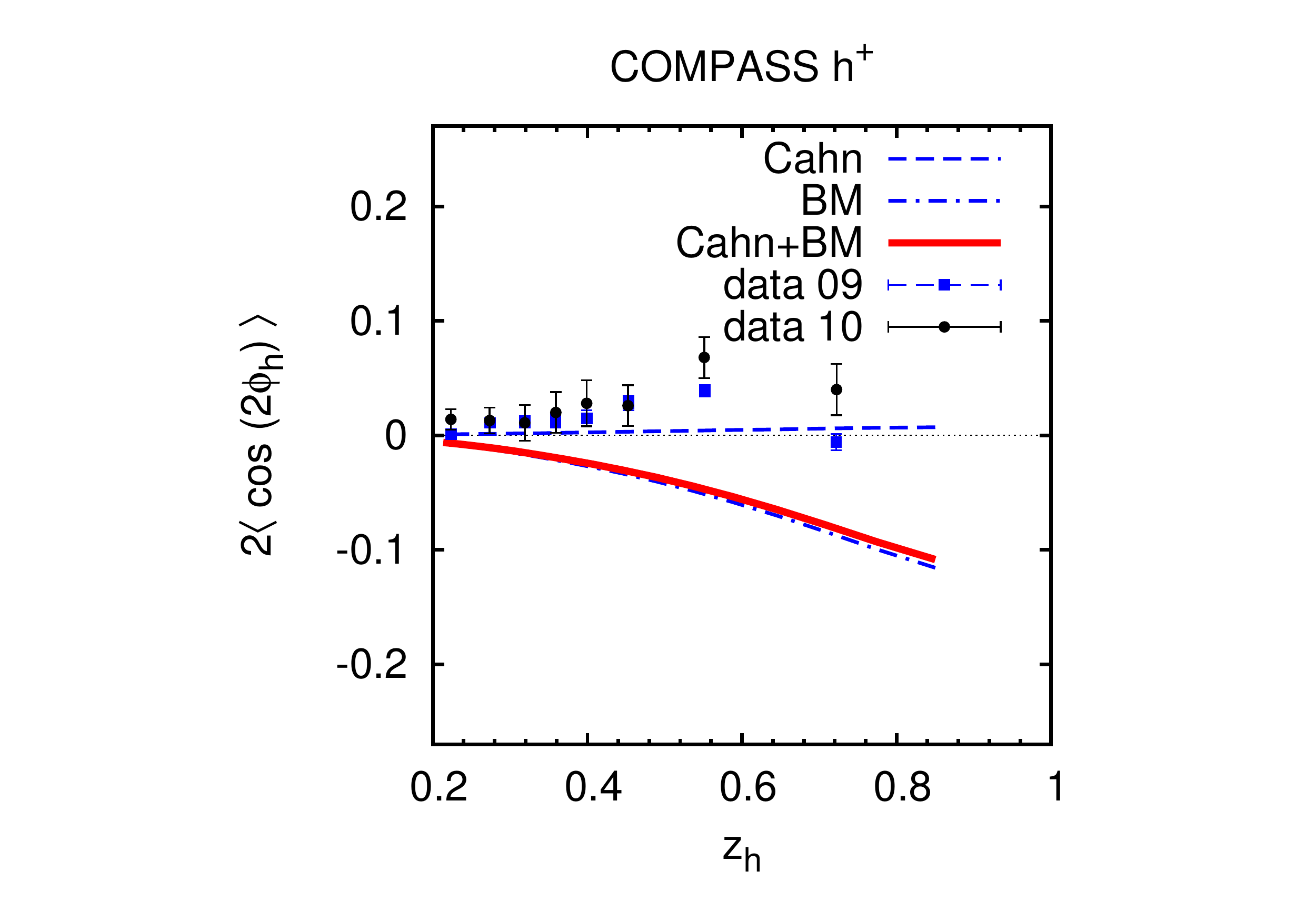}
\hspace*{-1.3cm}
\includegraphics[width=0.24\textwidth,angle=-90]{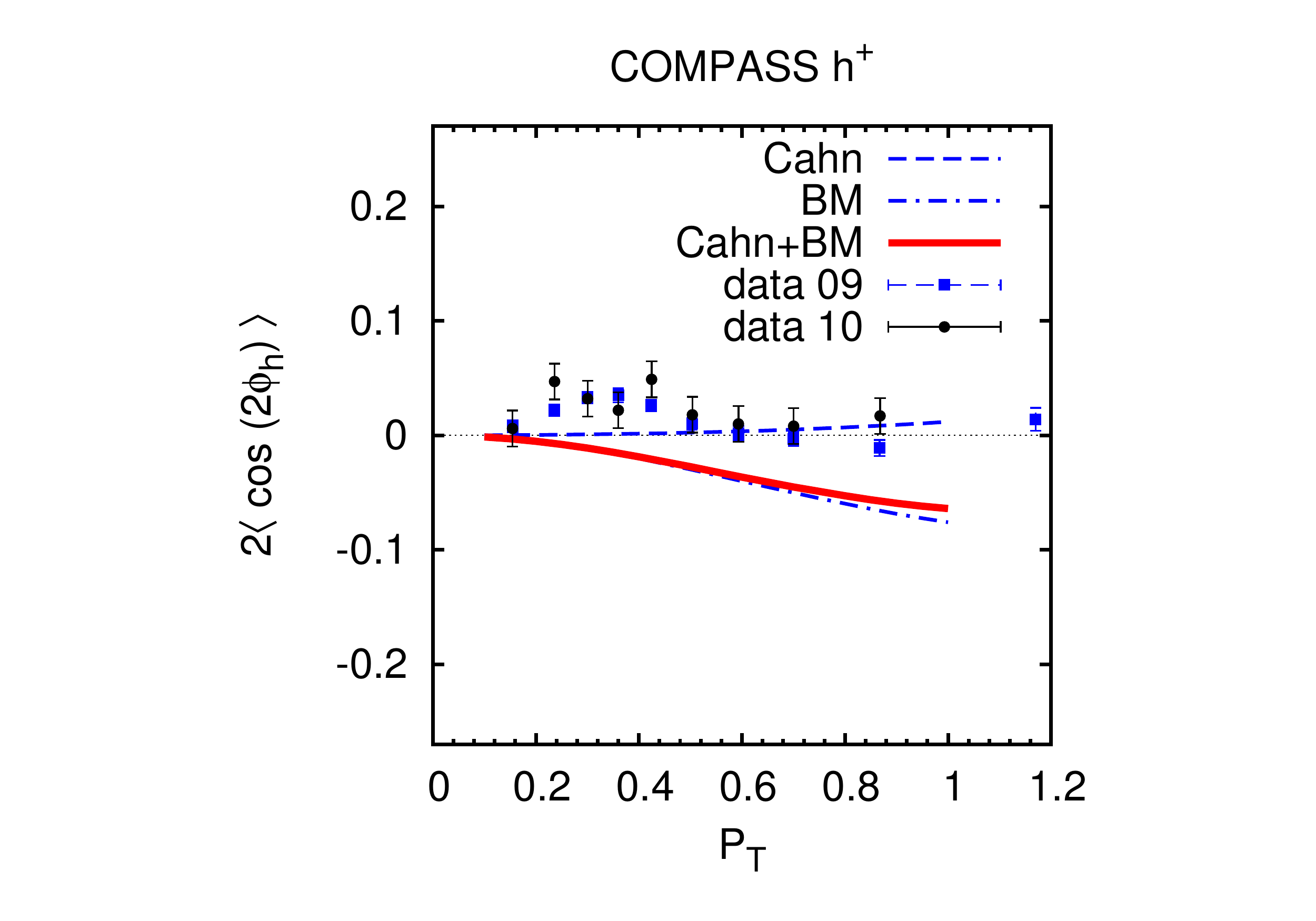}
\\
\includegraphics[width=0.24\textwidth,angle=-90]{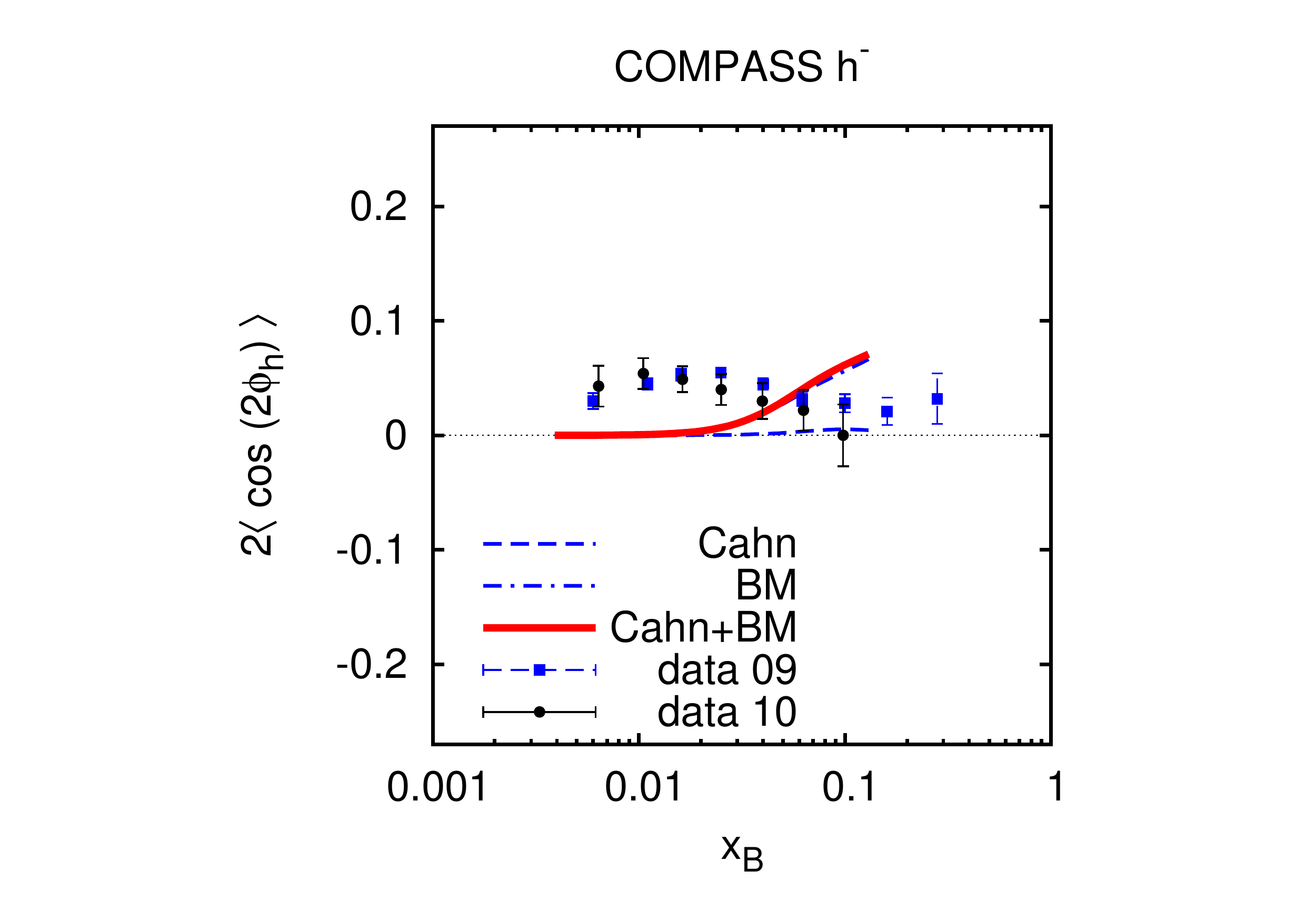}
\hspace*{-1.3cm}
\includegraphics[width=0.24\textwidth,angle=-90]{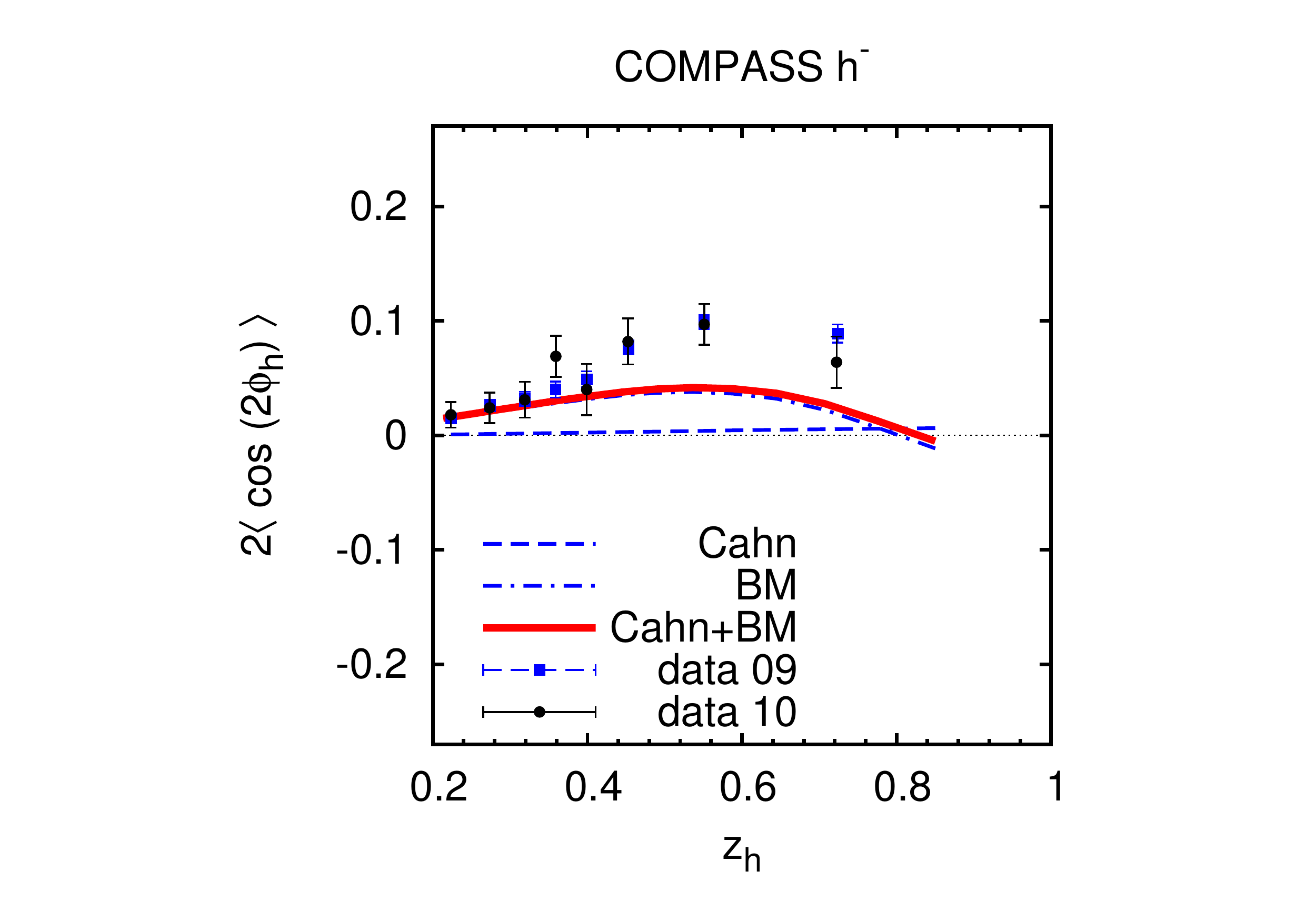}
\hspace*{-1.3cm}
\includegraphics[width=0.24\textwidth,angle=-90]{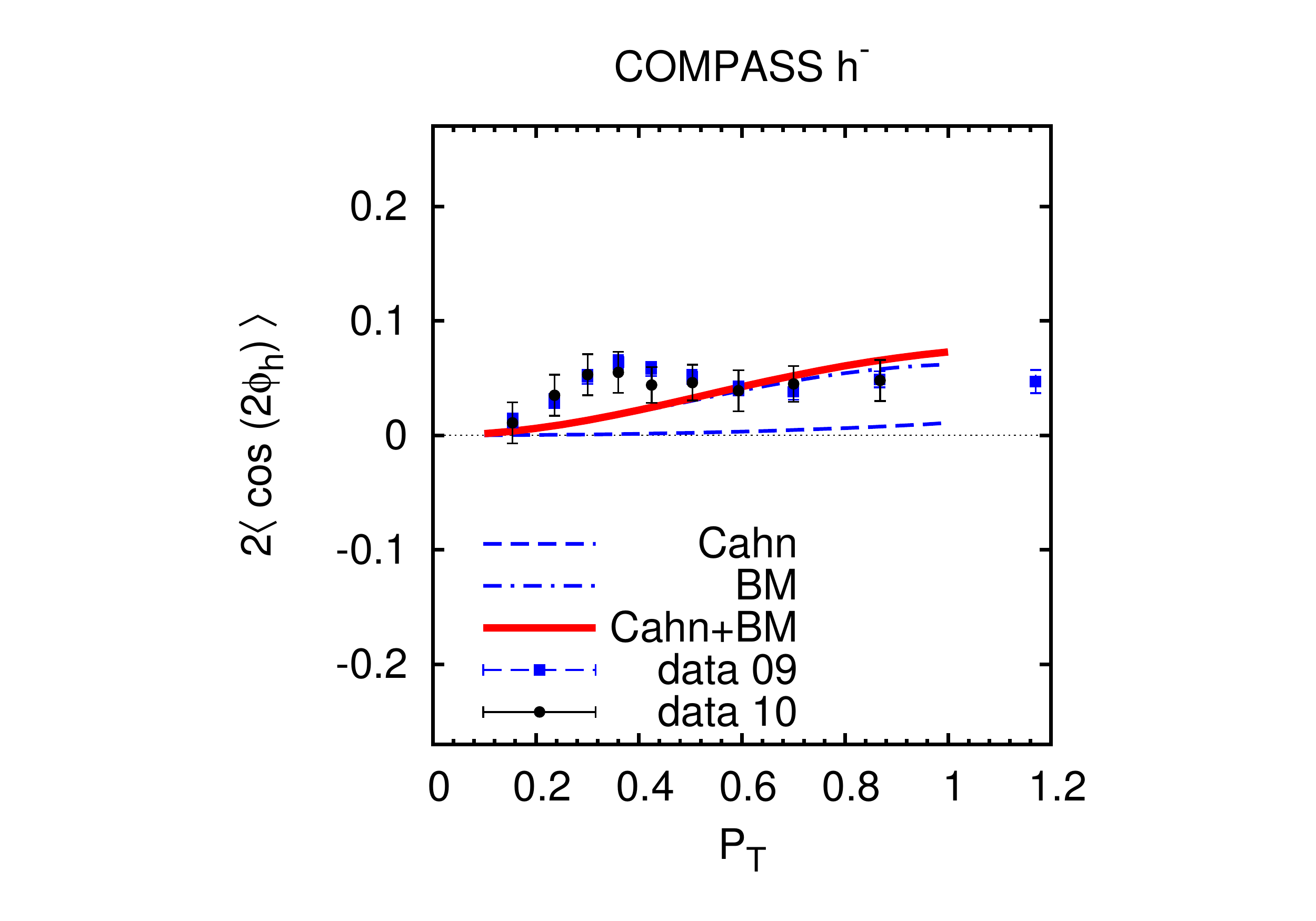}
\caption{\label{c2phi-compass}
Boer-Mulders and Cahn contributions to the $\langle\cos 2 \phi _h\rangle$
azimuthal modulation for $\pi ^+$ (upper panel) and $\pi ^-$ (upper panel)
production at COMPASS (deuteron target) as a function of $\xb$ (left plot),
$z_h$ (central plot) and $P_T$ (right plot). Experimental data are from
Ref.~\cite{Kafer:2008ud, Bressan:2009eu, Collaboration:2010fi}.}
\end{figure}

We revisit these calculations by applying the physical cuts on the partonic
transverse momenta, Eqs.~(\ref{cutenergy}) and (\ref{cutdirection}), 
and find that the value of the twist-4 Cahn effect is very sensitive to the
$\kt$ constraint,
as it can be seen in the upper panels of Fig.~\ref{c2phi-hermes-ch-bm}, while
the Boer-Mulders contribution does not exhibit such a strong dependence, as
shown in the lower panels of Fig.~\ref{c2phi-hermes-ch-bm}. This is explained by
the fact that, indeed, the accessed values of $(\kt^2/Q^2)$ are strongly
suppressed by limiting the range of $\kt$.
The sum of the Boer-Mulders and Cahn-like contributions which reproduces the
$\langle\cos 2 \phi _h\rangle$ azimuthal modulation is presented
in Fig.~\ref{c2phi-hermes}.
%
One can see that the description of the available data is very good. 
In Ref.~\cite{Barone:2009hw} a satisfactory description of the HERMES data was
achieved by adopting a smaller value of $\langle k_{\perp}^2\rangle=0.18$
GeV$^2$ for the HERMES data, while keeping $\langle k_{\perp}^2\rangle=0.25$
GeV$^2$ for fitting the COMPASS data (see FIT II).
Here a similar improvement is achieved by taking into account the physical cuts
on the partonic transverse momenta which, by cutting the range of allowed $\kt$
values, effectively reduces the average value of $\kt$  decreasing the
contribution generated by the Cahn effect.
Nevertheless, a slight puzzle still remains: while HERMES data seem to demand a
very small Cahn contribution, it can be seen from the analysis in
Ref.~\cite{Barone:2009hw} and from Fig.~\ref{c2phi-compass} that COMPASS data
seem to require a large Cahn contribution.
Large Cahn contributions can only be generated by large $k_{\perp}$ values, as
we have seen.
Since the COMPASS target is not a pure Deuterium target, but a $Li^6 D$
target, possible nuclear effects can enhance the values of $k_{\perp}$. 
Partonic transverse motion generated
by the nuclear smearing mechanisms  does not have to fulfill the bounds in
Eq.~\eqref{cutenergy} and \eqref{cutdirection}
and can be effectively simulated by a traditional Gaussian smearing, without any
restriction.
Future COMPASS data on pure hydrogen target will help our understanding.

%
%

%
%

%
\begin{figure}[t]
\includegraphics[width=0.24\textwidth,angle=-90]{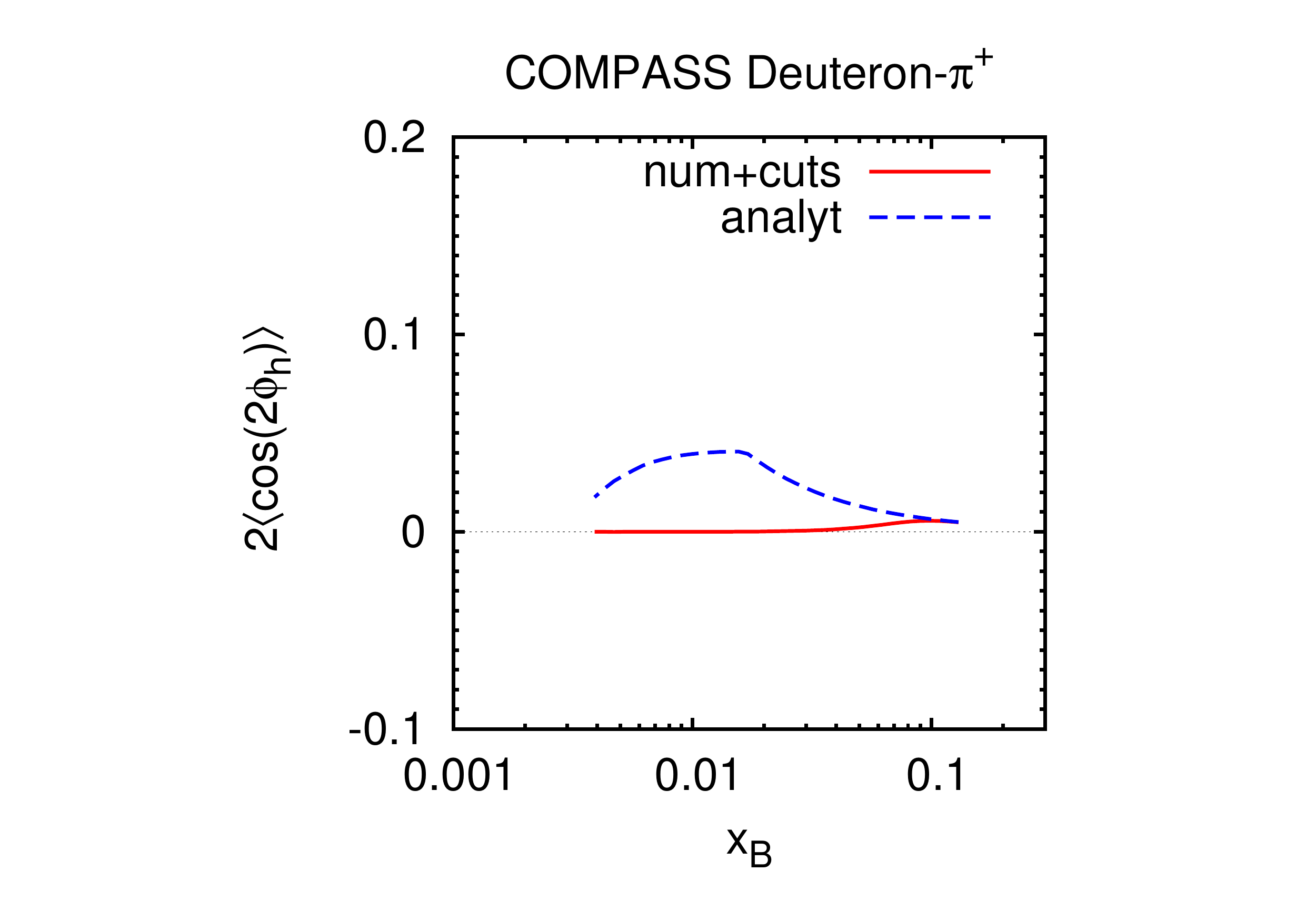}
\hspace*{-1.3cm}
\includegraphics[width=0.24\textwidth,angle=-90]{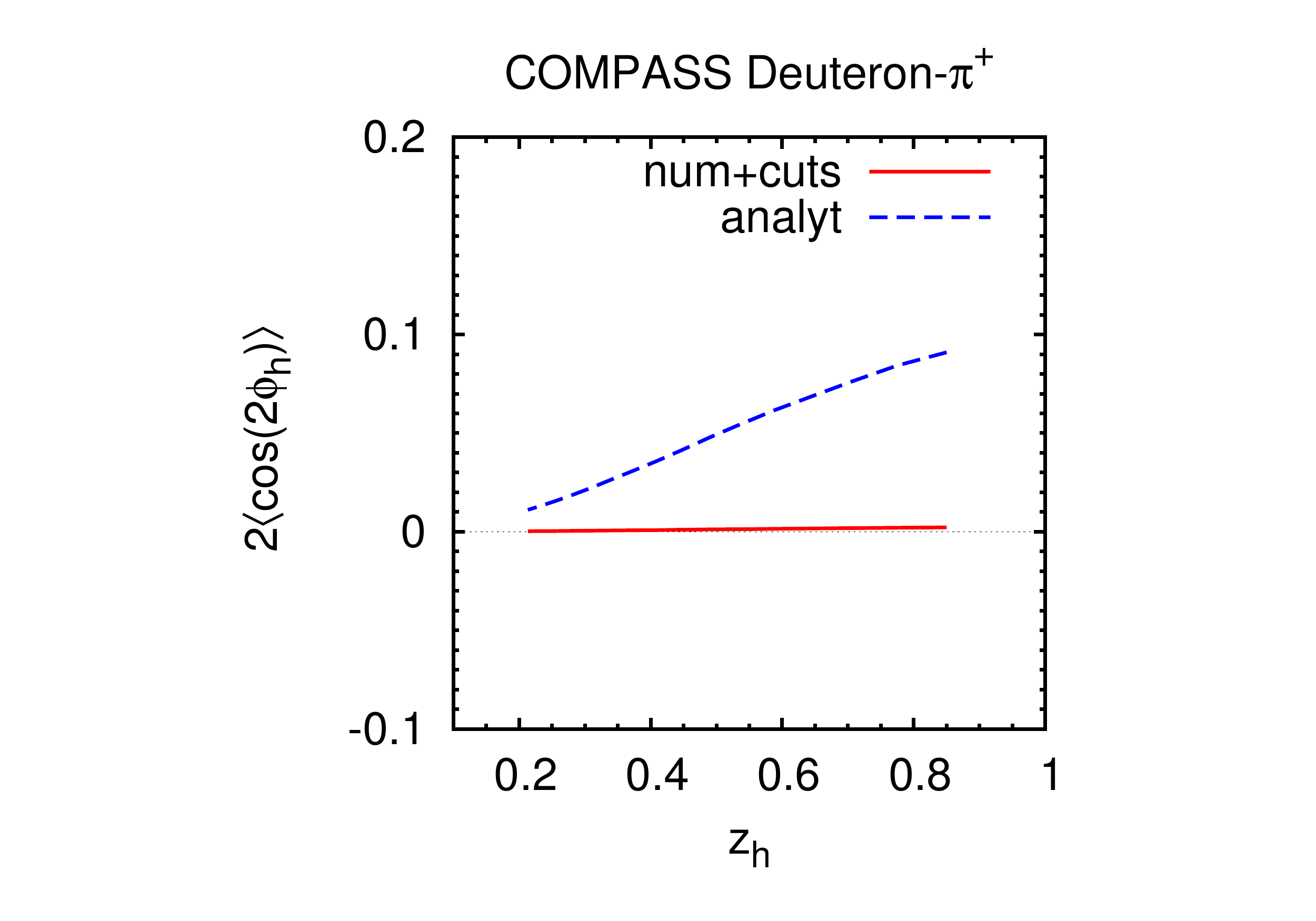}
\hspace*{-1.3cm}
\includegraphics[width=0.24\textwidth,angle=-90]{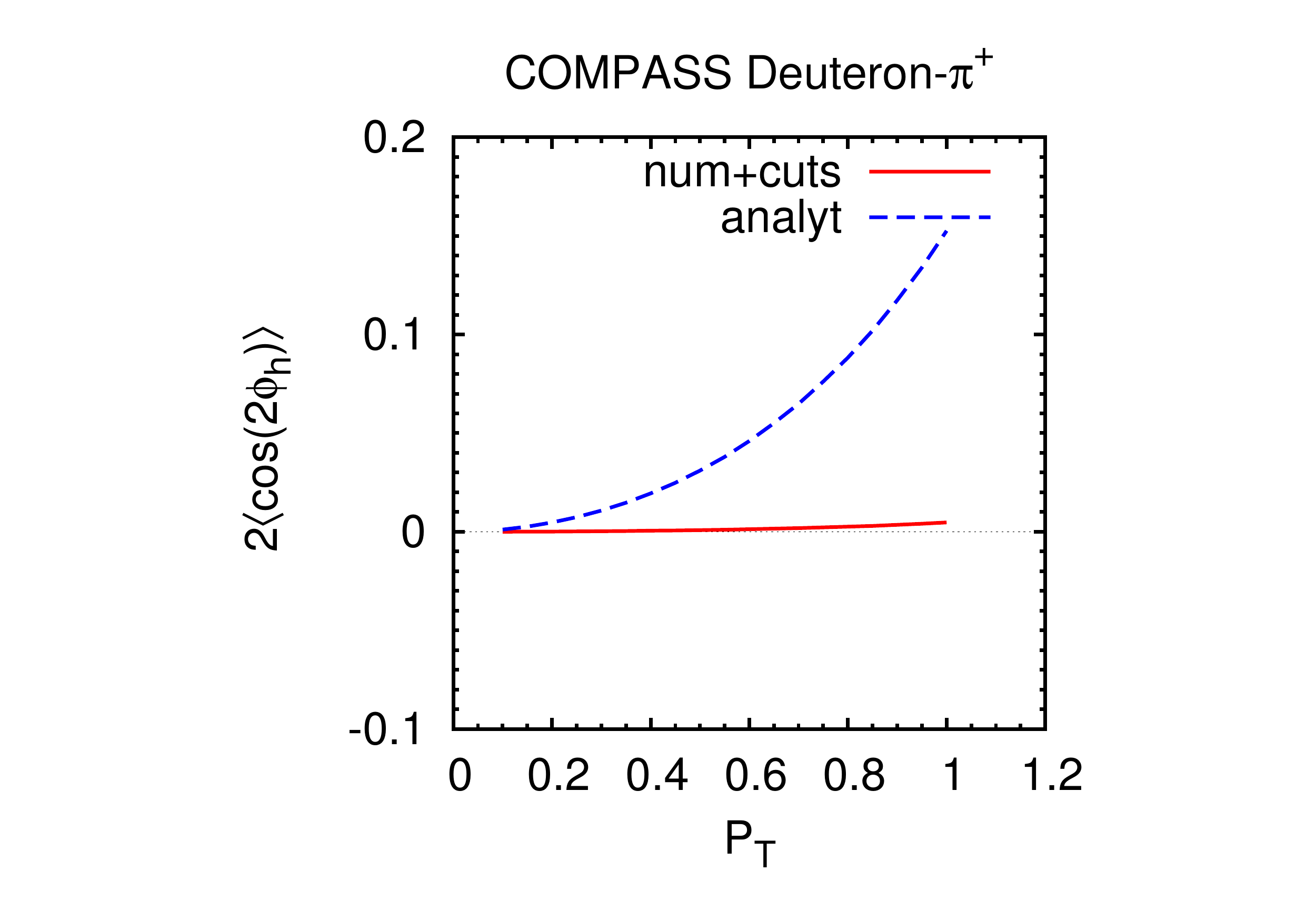}
\includegraphics[width=0.24\textwidth,angle=-90]{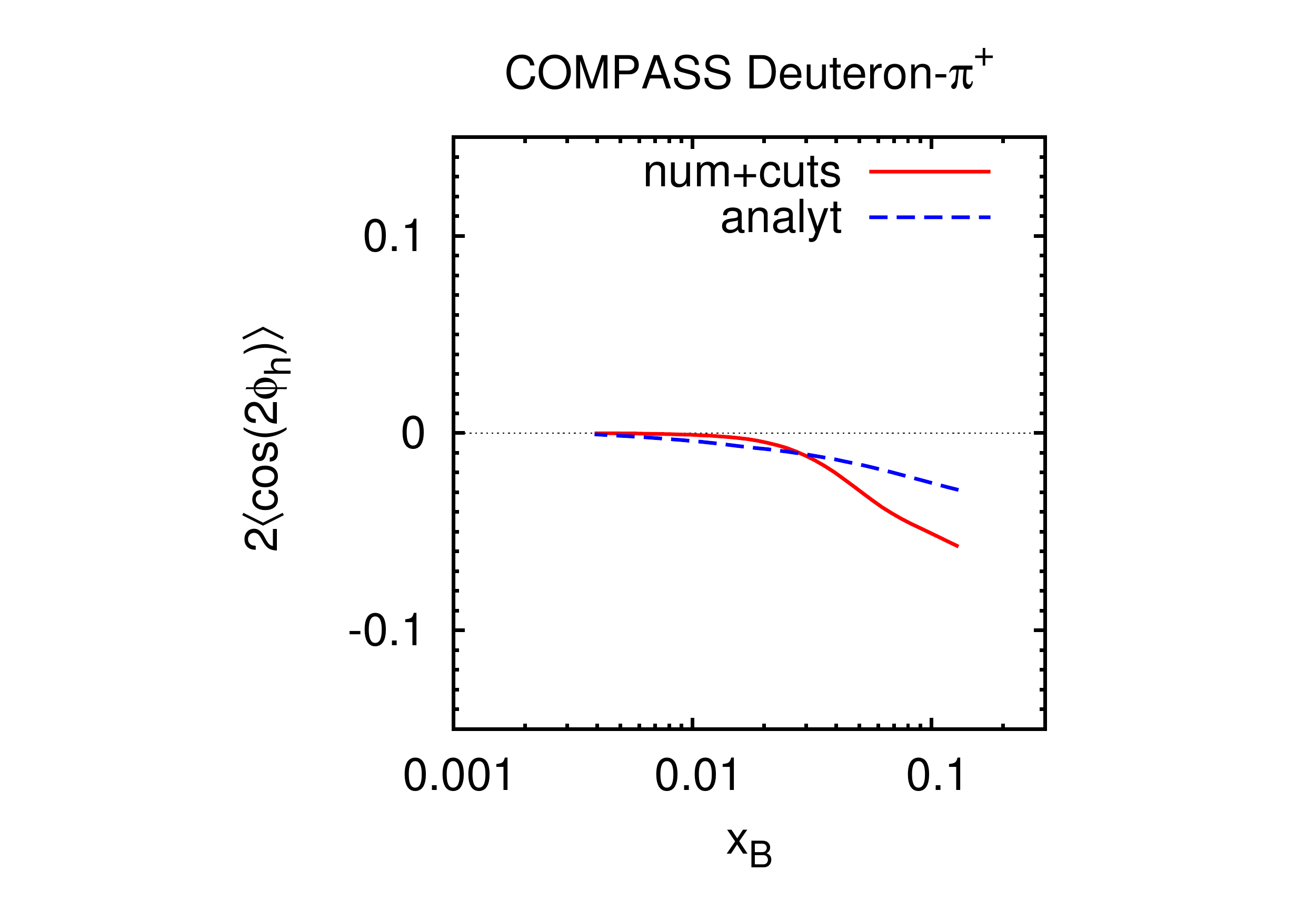}
\hspace*{-1.3cm}
\includegraphics[width=0.24\textwidth,angle=-90]{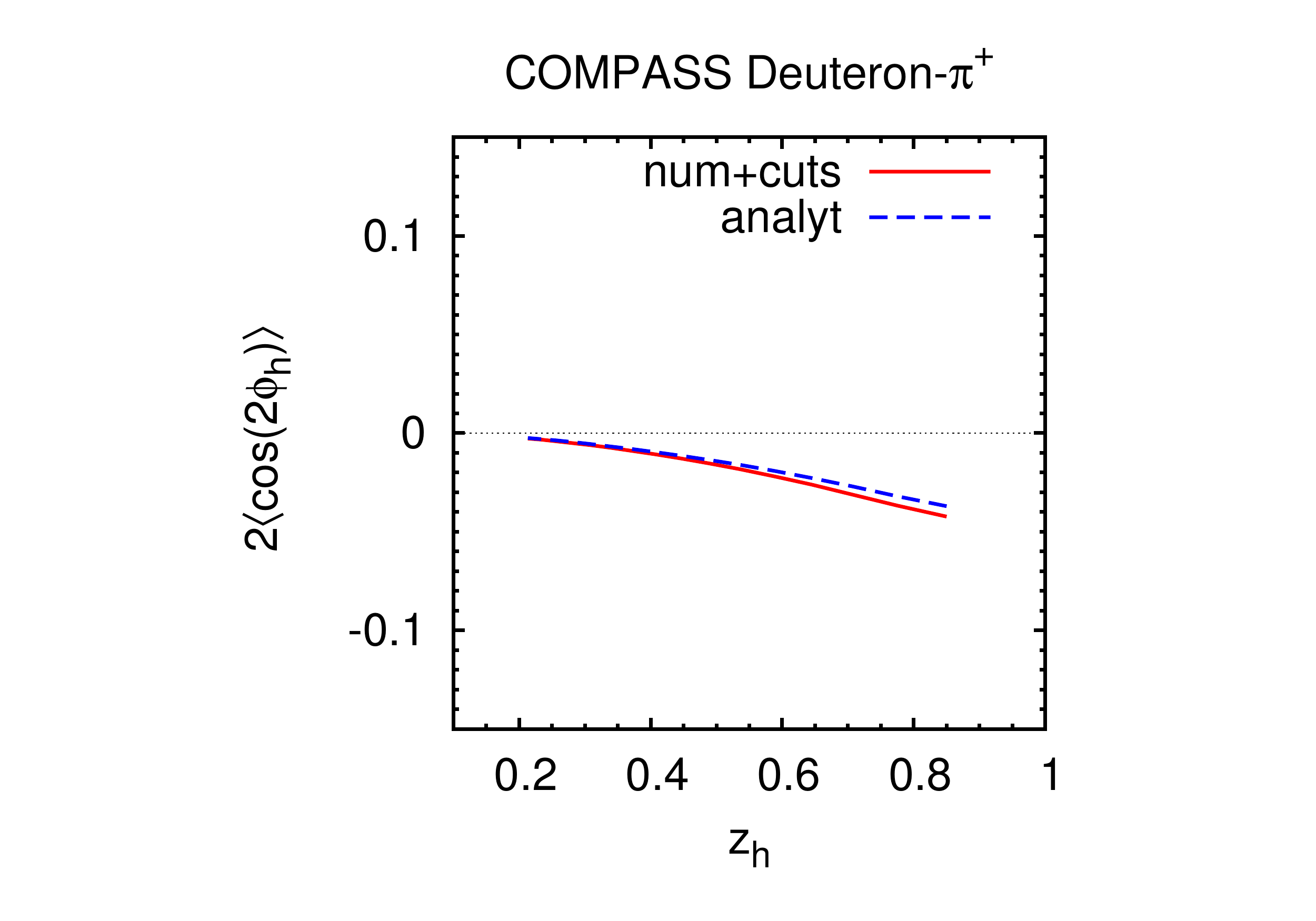}
\hspace*{-1.3cm}
\includegraphics[width=0.24\textwidth,angle=-90]{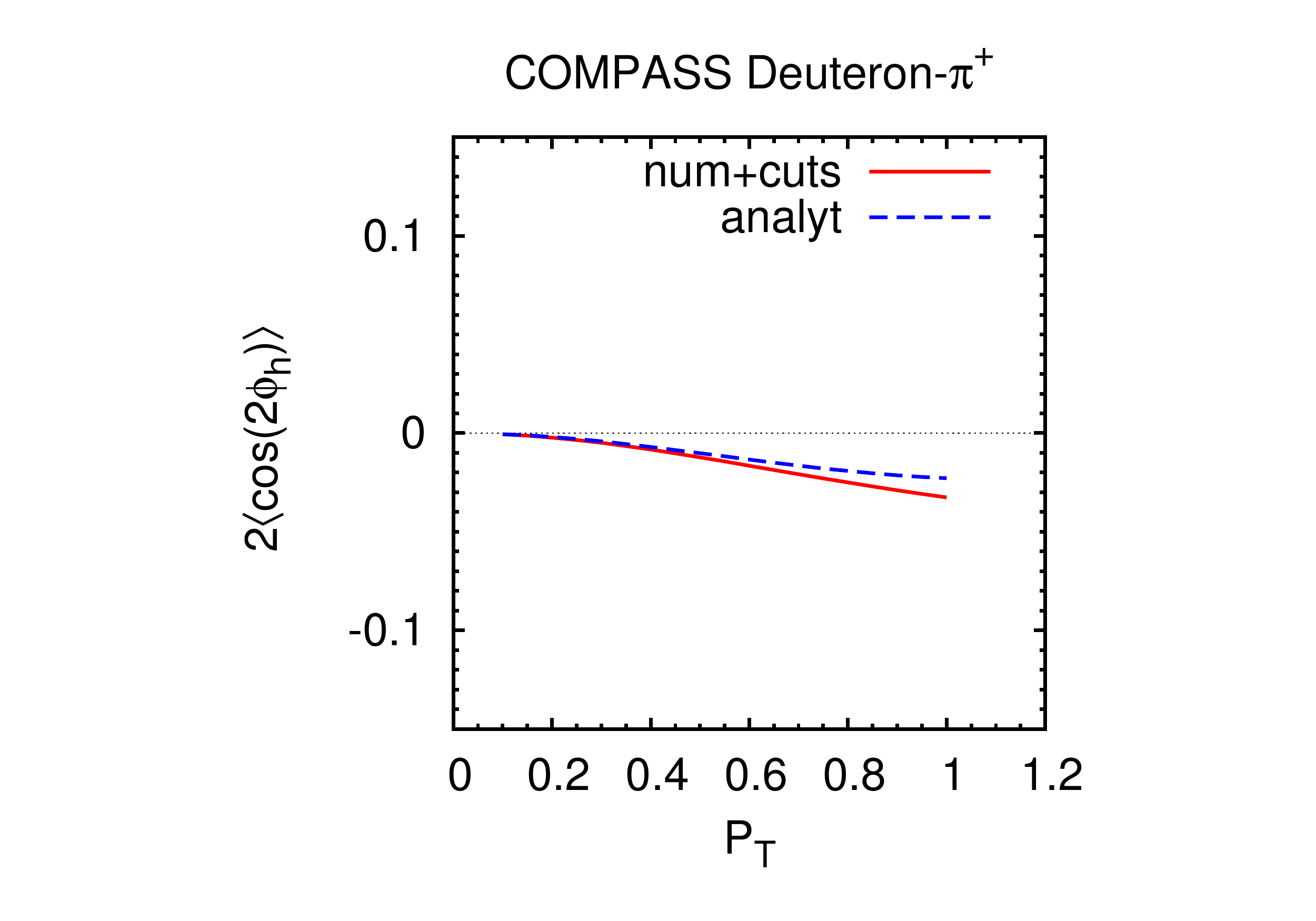}
\caption{\label{c2phi-compass-cahn-bm}
In the upper panels we show the twist-4 Cahn contribution to the $\langle\cos 2
\phi _h\rangle$ azimuthal modulation for $\pi ^+$ production at COMPASS
(deuteron target) as a function of $\xb$, $z_h$ and $P_T$, in the lower panels
the twist-2 Boer-Mulders contribution to the $\langle\cos 2 \phi _h\rangle$
azimuthal modulation for $\pi ^+$ production at HERMES again as a function of
$\xb$ (left plot), $z_h$ (central plot) and $P_T$ (right plot). 
data are from Ref.~\cite{Collaboration:2010fi}.
}
\end{figure}
%

%
%

\clearpage

\section{Conclusions}

In this paper we have studied SIDIS processes 
within a QCD parton model 
in the framework of TMD factorization;
the dependence on the parton instrinsic transverse momentum is modelled 
through a Gaussian parametrization. 
By requiring the parton energy to be smaller than the energy of
its parent hadron and preventing the parton to move backwards relatively to
its parent hadron, we were able to determine an upper bound,
$k_\perp^{max}(\xb,Q^2)$, to the range of allowed values of $\kt$. Under these
assumptions, we then re-calculated the three terms of the unpolarized SIDIS
cross section ($F_{UU}$, $F_{UU}^{\cos\phi_h}$ and $F_{UU}^{\cos2\phi_h}$), and
the detected hadron average transverse momentum, $\langle P_T^2\rangle$. Notice
that we made sure that the unpolarized TMD distribution function,
$f_{q/p}(\xb,\kt)$, respected the proper normalization condition by requiring
that the integral over $\kt$ in the restricted range
$[0,k_\perp^{max}(\xb,Q^2)]$ would still give the usual, collinear
$f_{q/p}(\xb)$.

Although the effects of our $\kt$ - cuts over the azimuthal-independent term
$F_{UU}$ are almost irrelevant (only a slight difference in the $F_{UU}$
dependence on $P_T$ can be appreciated), we realized that the detected hadron
average transverse momentum, $\langle P_T^2\rangle$, and the azimuthal moments
$\langle \cos\phi_h\rangle$ and $\langle \cos 2\phi_h\rangle$, are strongly
sensitive to the constraints on the  $\kt$ allowed values. 
In particular, by limiting the $\kt$ integration range, which effectively
reduces the Gaussian width 
$\langle \kt ^2\rangle$  suppressing the asymmetry at low $\xb$ (and
consequently low $Q^2$) values, we improve the description of the $\langle
\cos\phi_h\rangle$ azimuthal moment data from HERMES~\cite{Giordano:2010zz} and
COMPASS~\cite{Sbrizzai:2009fc} Collaborations, which were largely overestimated
by the predictions obtained with an analytical $\kt$ integration over the
unlimited $\kt$ range. Although the overall size of $\langle \cos\phi_h\rangle$,
a factor $2$ (ore more) smaller than that obtained without $\kt$ - cuts, is in
good agreement with the most recent experimental data,  some discrepancies
remain in the shape of our predictions, which can be interpreted as a signal of
the existence of higher twist contributions, that have been neglected in our
study. 

As far as the  $\langle \cos 2\phi_h\rangle$ azimuthal moment is concerned, the
situation remains slightly unresolved. Infact, while the effective reduction of
the Gaussian width $\langle \kt ^2\rangle$ helps in obtaining a satisfactory
agreement with HERMES measurements~\cite{Giordano:2009hi}, some considerable
inconsistencies remain in the description of the COMPASS
data~\cite{Collaboration:2010fi} which, instead,  seem to suggest the presence
of a much larger $\kt^2/Q^2$ Cahn contribution, that could only be achieved by
increasing the average $\kt$. This might suggest the presence of nuclear
smearing effects in the $Li^6 D$ target. COMPASS future data on a proton target
will help to clarify this issue.

Finally, we observed a significant deviation of the detected hadron average
transverse momentum, $\langle P_T^2\rangle$, from the theoretical value
$\avPT\G=\langle p_{\perp}^2\rangle+z^2_h \langle k_{\perp}^2\rangle$, obtained
by an analytical $\kt$ integration over the unlimited $\kt$ range. 
This is induced by two different mechanisms: on one side, the constraints we
applied on the  $\kt$ range of integration and, on the other side, the
(inevitable) $P_T$ cuts operated in the experimental analysis. In general, we
predict a much flatter behaviour than that of $\langle P^2_T\rangle_G$, but yet
quite far from that suggested by the COMPASS analysis of
Ref.~\cite{Rajotte:2010ir}. This is indeed an issue to be further studied in
future, as only very preliminary data are presently available from COMPASS,
HERMES and JLab.

In this study, higher twist contributions 
were neglected, 
together with different mechanisms to generate the intrinsic transverse momenta, 
like soft gluon emission and TMD QCD evolution effects~\cite{Aybat:2011zv,Collins:2011}. 
Therefore more refined phenomenological
descriptions are required to fix the details of such complex kinematics and
dynamics. However, we have shown that some extra care should be taken when
dealing with the present available experimental data, as they span a kinematical
region in which $\kt/Q$ contributions can be large and are not safely under
control, unless some limiting prescription over the allowed values of $\kt$ is
applied.  

Future experiments, like the Electron Ion Collider
(EIC)~\cite{Deshpande:2005wd,Horn:2009cu}, where the experimental cuts and the
$Q^2$ range would be easily adjustable in order to avoid unsafe kinematical
regions, will definitely help us to gain a much clearer understanding of the
three-dimensional structure of hadrons and, in particular, to disentangle higher
order corrections from leading twist contributions.

\section{Acknowledgement}

We would like to acknowledge useful discussions with Mauro Anselmino, Aram
Kotzinian, Enzo Barone, Barbara Pasquini, and Leonard Gamberg.
Authored by a Jefferson Science Associate, LLC under U.S. DOE Contract 
No. DE-AC05-06OR23177. The U.S. Government retains a non-exclusive, 
paid-up, irrevocable, 
world-wide license to publish
or reproduce this manuscript for U.S. Government purposes.

\appendix

\section{Sudakov decomposition\label{Sudakov}}

For the treatment of the SIDIS kinematics, we use the usual Sudakov
decomposition for four vectors:
\begin{equation}
v^\mu = v^+ n_+^\mu + v^- n_-^\mu + v_T^\mu
\end{equation}
where lightcone vectors $n_+^\mu$ and $n_-^\mu$ are
\begin{eqnarray}
&&   
n_+^\mu = \frac{1}{\sqrt{2}}\left(1,0,0,1\right) \;,\;\;\; 
n_-^\mu =\frac{1}{\sqrt{2}}\left(1,0,0,-1\right) , \\&&
n_+ \cdot n_- = 1\; , \;\;\;\;\;\;\;\;\;\;\;\;\;\;\;\; n_+^2 = n_-^2 = 0\; .
\end{eqnarray}
Using lightcone coordinates $a^\pm = 1/\sqrt{2} (a^0\pm a^3)$, and notations
$v^\mu = [v^-,v^+,\vect{v}_\perp]$ we can rewrite these vectors as
\begin{equation}
 n_+^\mu =  \left[0^-,1^+,\vect{0}_\perp\right] \; , n_-^\mu =
\left[1^-,0^+,\vect{0}_\perp\right]  \;
, v_T^{\mu}=(0,0,\vect{v}_{\perp}) .
\end{equation}
The product of two four vectors is then 
\begin{eqnarray}
   v\cdot w = v^+ w^- + v^- w^+ -  \vect{v}_\perp \cdot \vect{w}_\perp
\end{eqnarray}

Note that $v_T^\mu\cdot v_{T\mu}= - v_\perp^2$ and $n_\pm \cdot v_T = 0$.
The momentum four-vectors corresponding to the proton, the virtual photon and
the struck quark are:
\begin{eqnarray}
P^\mu &=& P^- n_{-}^\mu + \frac{M^2}{2 P^-} n_+^\mu \; ,  \nonumber \\ 
q^\mu &=& -\xi n_-^\mu + \frac{Q^2}{2 \xi} n_+^\mu \; ,\label{kin1}\\
k^\mu &=& x P^- n_-^\mu + \frac{k^2+k_\perp^2}{2xP^-} n_+^\mu + k_T^\mu \; ,
\nonumber
\end{eqnarray}
where the Sudakov vectors $n_+$, $n_-$ are defined in Appendix~\ref{Sudakov},
$x=k^-/P^-$ is the quark light-cone momentum fraction
and $\bfk_{\perp}$ is the quark intrinsic transverse momentum vector, see
Fig.~\ref{sidisfig}. 
Note that, according to the Trento conventions~\cite{Bacchetta:2004jz},  $P^-$
is the ``large'' component of the
proton's  momentum, 
i.e. along $n_{-}^\mu$. 

The variables of Eqs.~\eqref{kin1} then can be expressed as
\begin{equation}
\xi = \frac{Q^2}{\sqrt{2} W} \;,\> \qquad  \qquad 
P^- = \frac{1}{\sqrt{2}}\left ( W + \frac{Q^2}{W} \right )  \; ,
\end{equation}
and thus the proton and the virtual photon momenta can be written
in the $\gamma^* -p$ c.m. frame, as functions of the invariants $W$ and $Q$, in
this way:
\begin{eqnarray}
q^\mu &=&  \left[-\frac{Q^2}{\sqrt{2} W},
\frac{W}{\sqrt{2}},\vect{0}_\perp\right] = \left(P^0 -
\frac{Q^2}{W},\vect{0}_\perp,  P^0 \right) \; ,\\
P^\mu &=& \left[\frac{1}{\sqrt{2}}\left ( W + \frac{Q^2}{W} \right
),0^+,\vect{0}_\perp\right] = \left( P^0,\vect{0}_\perp, - P^0\right) \;,
\end{eqnarray}
where
\begin{equation}
P^0 = \frac{1}{2}\left ( W + \frac{Q^2}{W} \right )  \; .
\end{equation}
 
In the Generalised Parton Model the virtual photon scatters off an on-shell
quark, thus we will neglect the virtuality of the quark $k^2$ and write the
quark momentum in the $\gamma^*-p$ c.m. frame as:
\begin{equation}
k^\mu =
\left[x P^-,\frac{{k}_{\perp}^2}{2xP^-},\bfk_{\perp} 
\right]   = \left( x P^0 + \frac{{k}_{\perp}^2}{4xP^0},\bfk_{\perp},-x P^0 +
\frac{{k}_{\perp}^2}{4xP^0}\right)\,.
\end{equation}

\section{Kinematical cuts \label{exp-cuts}}
If not stated differently in the text, in our analysis we adopt the following
kinematical cuts. For the HERMES experiment:
\bea 
 && Q^2 \ge 1 \; {\rm GeV}^2\,, \quad W^2 > 10 \; {\rm GeV}^2 \,, \quad 0.05
<P_T < 1.0 \; {\rm GeV} 
\label{hermutcuts} \\ 
&& 0.023 < \xb < 0.27\,, \quad 0.2 < z_h < 0.75 \,, \quad
0.3 < y < 0.85 
\nonumber \>\,, 
\eea
and for COMPASS:
\bea 
 && Q^2 \ge 1 \; {\rm GeV}^2\,, \quad W^2 > 25 \; {\rm GeV}^2 \,, \quad 0.1 <P_T
< 1.0 \; {\rm GeV} 
\label{compasscuts} \\ 
&& 0.003 < \xb < 0.13\,, \quad 0.2 < z_h < 0.85 \,, \quad
0.2 \le y \le 0.9  \>\,.
\eea
Notice that hese kinematics correspond to the experimental cuts performed by
these collaborations in their most recent analysis of the unpolarized azimuthal
asymmetries~\cite{Giordano:2010zz,Collaboration:2010fi}.

For the EMC kinematics we used the following kinematical
cuts~\cite{Ashman:1991cj}: 
\bea 
 && Q^2 \ge 5 \; {\rm GeV}^2\,, \quad W^2 < 90 \; {\rm GeV}^2 \,, \quad 0.01
<P_T < 2.0 \; {\rm GeV} 
\label{emccuts} \\ 
&& E_h>5 {\rm GeV}\,, \quad 0.1 < z_h < 0.9 \,, \quad
0.2 \le y \le 0.8  \>\,.
\eea

The Electron Ion Collider (EIC) is a proposed
facility~\cite{Deshpande:2005wd,Horn:2009cu}
to provide further information on the proton structures. Thus we used the
following hypothetical kinematical configuration 
\bea 
 && Q^2 \ge 1 \;\;{\rm or}\;\;  Q^2 \ge 10 \; {\rm GeV}^2\,, \quad W^2 > 25 \; {\rm GeV}^2 \,, \quad 
0.05 < P_T < 1 \; {\rm GeV} 
\label{eiccuts} \\ 
&& 0.2 < z_h < 0.8 \,, \quad
0.05 \le y \le 0.8  \>.
\eea

\bibliographystyle{h-physrev5-ste}
\bibliography{newsample}

\end{document}